%% file: main.tex
\documentclass[journal,comsoc, 12pt,onecolumn,draftclsnofoot]{IEEEtran}
\usepackage{tikz}
\usepackage[T1]{fontenc}
\usetikzlibrary{patterns, arrows,backgrounds,fit,tikzmark,positioning,calc,decorations,snakes,shapes,matrix}

\usepackage{amsfonts}
\usepackage{amsmath,amssymb}
\usepackage{mathtools,hyperref}
\usepackage{amsthm}

\usepackage{subcaption} 
\usepackage{nicefrac} 
\usepackage{array}
\usepackage{pgfplots}
\usepackage[subnum]{cases} 
\usepackage{enumitem}
\usepackage{lscape}
\usepackage{cleveref}
\usepackage{multicol}
\usepackage{multirow}

\theoremstyle{remark}
\newtheorem{theorem}{ {Theorem}}
\newtheorem{corollary}{ {Corollary}}
\newtheorem{proposition}{{Proposition}}
\newtheorem{definition}{{Definition}}

\newtheorem{lemma}{ {Lemma}}
\newtheorem{remark}{ {Remark}}
\newtheorem{example}{{Example}}

\definecolor{OliveGreen}{rgb}{0,0.6,0}
\definecolor{RedBlue}{rgb}{0.8,0,0.5}
\definecolor{RedBlueGreen}{rgb}{0.8,0.6,0.5}
\definecolor{YellowOrange}{rgb}{0.4,0.4,0}

\DeclarePairedDelimiter\ceil{\lceil}{\rceil}

\input{content/tikz_drawing}
\ifCLASSINFOpdf
\else
\fi
\usepackage{amsmath}
\usepackage[cmintegrals]{newtxmath}
\begin{document}
%
\title{Fundamental Limits on Delivery Time in Cloud- and Cache-Aided Heterogeneous Networks}
%
%
%

\author{Jaber~Kakar,~\IEEEmembership{Member,~IEEE,}
        Soheil~Gherekhloo,~\IEEEmembership{Member,~IEEE,}
        and~Aydin~Sezgin,~\IEEEmembership{Senior Member,~IEEE}
\thanks{This paper was presented in part at the IEEE International Conference on Communications 2017 \cite{Kakar}.}
}

%
%

\markboth{Draft}%
{Shell \MakeLowercase{\textit{et al.}}: Bare Demo of IEEEtran.cls for IEEE Communications Society Journals}

%



\makeatletter
\newcommand*{\rom}[1]{\expandafter\@slowromancap\romannumeral #1@}
\makeatother

\maketitle


\begin{abstract}
A Fog radio access network (F-RAN) is considered as a network architecture candidate to meet the soaring demand in terms of reliability, spectral efficiency, and latency in next generation wireless networks.
This architecture combines the benefits associated with centralized cloud processing and wireless edge caching enabling primarily low-latency transmission under moderate fronthaul capacity requirements. The F-RAN we consider in this paper is composed of a centralized cloud server which is connected through fronthaul links to two edge nodes (ENs) serving two mobile users through a Z-shaped partially connected wireless network.  
We define an information-theoretic metric, the delivery time per bit (DTB), that captures the worst-case per-bit delivery latency for conveying any requested content to the users. For the cases when cloud and wireless transmission occur either sequentially or in parallel, we establish coinciding lower and upper bounds on the DTB as a function of cache size, backhaul capacity and wireless channel parameters.
Through optimized rate allocation, our achievability scheme determines the best combination of private, common signalling and interference neutralization that matches the converse. Our converse bounds use subsets of wireless, fronthaul and caching resources of the F-RAN as side information that enable a single receiver to decode either one or both users' requested files.
We show the optimality on the DTB for all channel regimes. In case of serial transmission, the functional DTB-behavior changes at fronthaul capacity thresholds. In this context, we combine multiple channel regimes to \emph{classes of channel regimes} which share the same fronthaul capacity thresholds and as such the same DTB-functional. In total, our analysis identifies four classes; in only three of those edge caching and cloud processing can provide nontrivial synergestic and non-synergestic performance gains. Interestingly, in these three classes, we show that \emph{only} under parallel fronthaul-edge transmission strategies edge caching becomes obsolete as long as a certain fronthaul capacity is exceeded.  
\end{abstract}


%
\IEEEpeerreviewmaketitle

\input{content/introduction}

\input{content/system_model}
\input{content/serial}

\input{content/parallel}
\input{content/conclusion}

\appendices
\input{content/appendixa}

\input{content/appendixb}

\input{content/appendixd}

\input{content/appendixc}

\input{content/appendixe}



\ifCLASSOPTIONcaptionsoff
  \newpage
\fi

\bibliographystyle{IEEEtran}
\bibliography{content/bibliography}

%

\end{document}

%% file: content/tikz_drawing.tex
\usepackage{ellipsis}
\usetikzlibrary{calc}
\usetikzlibrary{decorations.pathreplacing,decorations.markings,shapes.geometric}
\tikzset{naming/.style={align=center,font=\small}}
\tikzset{antenna/.style={insert path={-- coordinate (ant#1) ++(0,0.25) -- +(135:0.25) + (0,0) -- +(45:0.25)}}}
\tikzset{station/.style={naming,draw,shape=dart,shape border rotate=90, minimum width=10mm, minimum height=10mm,outer sep=0pt,inner sep=3pt}}
\tikzset{mobile/.style={naming,draw,shape=rectangle,minimum width=12mm,minimum height=6mm, outer sep=0pt,inner sep=3pt}}
\tikzset{radiation/.style={{decorate,decoration={expanding waves,angle=90,segment length=4pt}}}}

\newenvironment{customlegend}[1][]{%
    \begingroup
    \csname pgfplots@init@cleared@structures\endcsname
    \pgfplotsset{#1}%
}{%
    \csname pgfplots@createlegend\endcsname
    \endgroup
}%

\def\addlegendimage{\csname pgfplots@addlegendimage\endcsname}

\newcommand{\SymMod}[0]{
\begin{tikzpicture}[
roundnode/.style={circle, draw=green!60, fill=green!5, very thick, minimum size=7mm},
squarednode/.style={rectangle, draw=red!60, fill=red!5, very thick, minimum size=5mm},
]
\node [cloud, draw,cloud puffs=10,cloud puff arc=120, aspect=2, inner ysep=1em, scale=0.8] (c) at (0, 0) {Cloud};
\node[squarednode] (eNB) at (2.0, 0) {eNB};
\node[squarednode] (HeNB) at (2.0, 2) {HeNB};
\draw[thick] (c) -- (eNB);
\draw[->, thick] (c) -- (1.30, 2) node[pos=0.5,sloped,above] {$C_{F}$};
\node[squarednode] (Uone) at (5.5, 2) {$U_1$};
\node[squarednode] (Utwo) at (5.5, 0) {$U_2$};
\draw[->, dashed, thick] (eNB) -- (Utwo) node[pos=0.5,sloped,above] {$h_{3}$};
\draw[->, dashed, thick] (2.495, 0) -- (5.15, 1.9) node[pos=0.57,sloped,above] {$h_{2}$};
\draw[->, dashed, thick] (HeNB) -- (Uone) node[pos=0.5,sloped,above] {$h_{1}$};
\node (ch) at (2.0, 1.4) {\includegraphics[scale=.3]{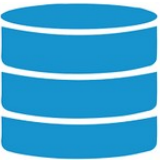}};
\draw[<->] (2.3, 1.15) -- (2.3, 1.62) node[pos=0.5,sloped, below] {$\scriptstyle\mu NL$};
\end{tikzpicture}
}

\newcommand{\Time}[0]{
\begin{tikzpicture}
[
squarednode/.style={rectangle, rounded corners, draw=red!60, fill=red!5, very thick},
]
\draw[line width=1mm] (0, 1.1) -- (13, 1.1);
\node[scale=2.5] at (13.7, 1.1) {$\ldots$};
\draw[->, line width=1mm] (14.4, 1.1) -- (16, 1.1);
\node[squarednode, anchor=north west, minimum width=0.15*\linewidth,minimum height=0.8cm] (time11) at (1.5,1.5) {};
\node[squarednode, draw=blue!60, fill=blue!5, right = 0 cm of time11, minimum width=0.15*\linewidth,minimum height=0.8cm] (time12) {};
\draw[-, dashed, thick] (1.5,1.5) -- (1.5,0) node[below, align=left] {Transmission \\ interval $j-1$};
\draw[<->, thick, color=red!60] (1.5, 1.5)++(0, 0.2) -- ([yshift=0.2cm]time11.north east) node[pos=0.5,sloped,above] {$T_{F}[j-1]$};
\draw[<->, thick, color=blue!60] ([yshift=0.2cm]time12.north west) -- ([yshift=0.2cm]time12.north east) node[pos=0.5,sloped,above] {$T_{E}[j-1]$};
\draw[<->, thick] (1.5, 1.5)++(0, 0.8) -- ([yshift=0.8cm]time12.north east) node[pos=0.5,sloped,above] {Latency Tx interval $j-1$};
\node[squarednode, anchor=north west, minimum width=0.075*\linewidth,minimum height=0.8cm] (time21) at ([xshift=1.5cm]time12.north east) {};
\node[squarednode, draw=blue!60, fill=blue!5, right = 0 cm of time21, minimum width=0.075*\linewidth,minimum height=0.8cm] (time22) {};
\draw[-, dashed, thick] ([xshift=1.5cm]time12.north east) -- ([xshift=1.5cm, yshift=-1.5cm]time12.north east) node[below, align=left] {Transmission \\ interval $j$};
\draw[<->, thick, color=red!60] (8, 1.5)++(0, 0.2) -- ([yshift=0.2cm]time21.north east) node[pos=0.5,sloped,above] {$T_{F}[j]$};
\draw[<->, thick, color=blue!60] ([yshift=0.2cm]time22.north west) -- ([yshift=0.2cm]time22.north east) node[pos=0.5,sloped,above] {$T_{E}[j]$};
\draw[<->, thick] ([yshift=0.8cm]time21.north west) -- ([yshift=0.8cm]time22.north east) node[pos=0.5,sloped,above] {Latency Tx interval $j$};
\node (tx1) at ([yshift=-1.2cm]time11.north east) {$\mathbf{d}[j-1],\mathbf{h}[j-1]$};
\node (tx2) at ([yshift=-1.2cm]time21.north east) {$\mathbf{d}[j],\mathbf{h}[j]$};
\end{tikzpicture}
}

\newcommand{\TimePar}[0]{
\begin{tikzpicture}
[
squarednode/.style={rectangle, rounded corners, draw=red!60, fill=red!5, very thick},
]
\draw[line width=1mm] (0, 1.1) -- (13, 1.1);
\node[scale=2.5] at (13.7, 1.1) {$\ldots$};
\draw[->, line width=1mm] (14.4, 1.1) -- (16, 1.1);
\node[squarednode, draw=black!60!green, fill=green!5, anchor=north west, minimum width=0.25*\linewidth,minimum height=0.8cm] (time11) at (1.5,1.5) {};
\draw[-, dashed, thick] (1.5,1.5) -- (1.5,0) node[below, align=left] {Transmission \\ interval $j-1$};
\draw[<->, thick, color=black!60!green] (1.5, 1.5)++(0, 0.2) -- ([yshift=0.2cm]time11.north east) node[pos=0.5,sloped,above] {$T_{P}[j-1]$};
\draw[<->, thick] (1.5, 1.5)++(0, 0.8) -- ([yshift=0.8cm]time11.north east) node[pos=0.5,sloped,above] {Latency Tx interval $j-1$};
\node[squarednode, draw=black!60!green, fill=green!5, anchor=north west, minimum width=0.10*\linewidth,minimum height=0.8cm] (time21) at ([xshift=1.5cm+0.05*\linewidth]time11.north east) {};
\draw[-, dashed, thick] (time21.north west) -- ([yshift=-1.5cm]time21.north west) node[below, align=left] {Transmission \\ interval $j$};
\draw[<->, thick, color=black!60!green] ([yshift=0.2cm]time21.north west) -- ([yshift=0.2cm]time21.north east) node[pos=0.5,sloped,above] {$T_{P}[j]$};
\draw[<->, thick] ([yshift=0.8cm]time21.north west) -- ([yshift=0.8cm]time21.north east) node[pos=0.5,sloped,above] {Latency Tx interval $j$};
\node (tx1) at ([yshift=-1.2cm, xshift=0.125\linewidth]time11.north west) {$\mathbf{d}[j-1],\mathbf{h}[j-1]$};
\node (tx2) at ([yshift=-1.2cm, xshift=0.075\linewidth]time21.north west) {$\mathbf{d}[j],\mathbf{h}[j]$};
\end{tikzpicture}
}

\newcommand{\ldmnewuser}[3]{
\begin{scope}[pipedot/.style={draw, circle, inner sep=0pt, minimum size=0.3cm}]
\def\username{#1}
\def\numPipes{#2}
\def\distanceDots{0.6}
\coordinate (position\username) at #3;

\foreach \x in {1,...,\numPipes} {
	\pgfmathparse{-(\x-1)*\distanceDots}
	\node[pipedot] (node\x\username) at ($(0, \pgfmathresult)+(position\username)$) {};
}
\node[draw,fit=(node1\username) (node\numPipes\username)] (group\username) {};
\end{scope}
} 

\newcommand{\ldmconnectusers}[4]{
\foreach \x in {1,...,#3} {
	\pgfmathtruncatemacro\mymacro{(\x+#4)}
	\draw[->, thick] (node\x #1)--(node\mymacro #2) ;
}
} 

\newcommand{\LDMblockone}[0]{
\begin{tikzpicture}[scale=1]

\ldmnewuser{HeNB}{5}{(0,0)}
\node[node distance=0.1cm, above=of groupHeNB] {HeNB};
\node[node distance=0.2cm, left=of node1HeNB] {$a_2\oplus b_3$};

\ldmnewuser{eNB}{5}{(0,-5)}
\node[node distance=0.1cm, above=of groupeNB] {eNB};
\node[node distance=0.2cm, left=of node1eNB] {$a_1$};
\node[node distance=0.2cm, left=of node2eNB] {$b_1$};
\node[node distance=0.2cm, left=of node3eNB] {$b_2$};
\node[node distance=0.2cm, left=of node4eNB] {$b_3$};
\node[node distance=0.2cm, left=of node5eNB] {$a_3$};

\ldmnewuser{fU}{5}{(5,0)}
\node[node distance=0.1cm, above=of groupfU] {$U_1$};
\node[node distance=0.2cm, right=of node1fU] {$a_1$};
\node[node distance=0.2cm, right=of node4fU] {$a_2$};
\node[node distance=0.2cm, right=of node5fU] {$a_3$};

\ldmnewuser{sU}{5}{(5,-5)}
\node[node distance=0.1cm, above=of groupsU] {$U_2$};
\node[node distance=0.2cm, right=of node3sU] {$b_1$};
\node[node distance=0.2cm, right=of node4sU] {$b_2$};
\node[node distance=0.2cm, right=of node5sU] {$b_3$};

\ldmconnectusers{HeNB}{fU}{2}{3}
\ldmconnectusers{eNB}{fU}{5}{0}
\ldmconnectusers{eNB}{sU}{4}{1}

\node[left=of node1HeNB.north, yshift=0.2cm, xshift=12pt] (HeNB1) {}; 
\node[left=of node1HeNB.south, yshift=-0.2cm, xshift=12pt] (HeNB2) {}; 
\draw[line width=1,<->] (-2.5,0.3) to (-2.5,-.3);
\node (a1) at (-3.1,0) [inner sep=0] {$R_{\text{IN}}^{n,*}$};

\node[left=of node1HeNB.south, yshift=-0.2cm, xshift=12pt] (HeNB3) {}; 
\node[left=of node5HeNB.south, yshift=-0.2cm, xshift=12pt] (HeNB4) {}; 
\draw[line width=1,<->] (-2.5,-0.3) to (-2.5,-2.7);
\node (a2) at (-3.1,-1.5) [inner sep=0] {$l_2^{*}$};

\node[left=of node1eNB.north, yshift=0.2cm, xshift=12pt] (eNB1) {}; 
\node[left=of node1eNB.south, yshift=-0.2cm, xshift=12pt] (eNB2) {}; 
\draw[line width=1,<->] (-2.5,-4.7) to (-2.5,-5.3);
\node (a3) at (-3.1,-5.0) [inner sep=0] {$R_{c}^{u,*}$};

\node[left=of node1eNB.south, yshift=-0.2cm, xshift=12pt] (eNB3) {}; 
\node[left=of node4eNB.south, yshift=-0.2cm, xshift=12pt] (eNB4) {}; 
\draw[line width=1,<->] (-2.5,-5.3) to (-2.5,-7.1);
\node (a4) at (-3.65,-6.2) [inner sep=0] {$R_{\text{IN}}^{v,*}+R_{c}^{v,*}$};
\draw[line width=1,<->] (-1.3,-5.3) to (-1.3,-6.5);
\node (b1) at (-1.9,-5.9) [inner sep=0] {$R_{c}^{v,*}$};
\draw[line width=1,<->] (-1.3,-6.5) to (-1.3,-7.1);
\node (b2) at (-1.9,-6.8) [inner sep=0] {$R_{\text{IN}}^{v,*}$};

\node[left=of node5eNB.north, yshift=0.2cm, xshift=12pt] (eNB5) {}; 
\node[left=of node5eNB.south, yshift=-0.2cm, xshift=12pt] (eNB6) {}; 
\draw[line width=1,<->] (-2.5,-7.1) to (-2.5,-7.7);
\node (a5) at (-3.75,-7.4) [inner sep=0] {$R_{1,p}^{w,*}=R_{1,p}^{u,*}$};

\begin{scope}[on background layer]
\node[pattern=north west lines, pattern color=red!80!black,fit=(node1HeNB) (node1HeNB)] (groupHeNBd1) {};
\node[pattern=north east lines, pattern color=black,fit=(node1HeNB) (node1HeNB)] (groupHeNBd2) {};

\node[pattern=north west lines, pattern color=red!80,fit=(node1eNB) (node1eNB)] (groupeNBd1) {};
\node[pattern=north east lines, pattern color=black,fit=(node2eNB) (node4eNB)] (groupeNBd3) {};
\node[pattern=north west lines, pattern color=red!80,fit=(node5eNB) (node5eNB)] (groupeNBd3) {};

\node[pattern=north west lines, pattern color=red!80,fit=(node1fU) (node1fU)] (groupfUd1) {};
\node[pattern=north east lines, pattern color=black,fit=(node2fU) (node3fU)] (groupfUd3) {};
\node[pattern=north west lines, pattern color=red!80,fit=(node4fU) (node5fU)] (groupfUd3) {};

\node[pattern=north west lines, pattern color=red!80,fit=(node2sU) (node2sU)] (groupsUd1) {};
\node[pattern=north east lines, pattern color=black,fit=(node3sU) (node5sU)] (groupsUd3) {};
\end{scope}

\end{tikzpicture}
}

\newcommand{\LDMblockthree}[0]{
\begin{tikzpicture}[scale=1]

\ldmnewuser{HeNB}{6}{(0,0)}
\node[node distance=0.1cm, above=of groupHeNB] {HeNB};
\node[node distance=0.2cm, left=of node1HeNB] {$a_3\oplus b_3$};
\node[node distance=0.2cm, left=of node2HeNB] {$a_4\oplus b_2$};

\ldmnewuser{eNB}{6}{(0,-5)}
\node[node distance=0.1cm, above=of groupeNB] {eNB};
\node[node distance=0.2cm, left=of node1eNB] {$a_1$};
\node[node distance=0.2cm, left=of node2eNB] {$a_2$};
\node[node distance=0.2cm, left=of node3eNB] {$b_4$};
\node[node distance=0.2cm, left=of node4eNB] {$b_3$};
\node[node distance=0.2cm, left=of node5eNB] {$b_2$};
\node[node distance=0.2cm, left=of node6eNB] {$b_1$};

\ldmnewuser{fU}{6}{(5,0)}
\node[node distance=0.1cm, above=of groupfU] {$U_1$};
\node[node distance=0.2cm, right=of node2fU] {$a_1$};
\node[node distance=0.2cm, right=of node3fU] {$a_2$};
\node[node distance=0.2cm, right=of node5fU] {$a_3$};
\node[node distance=0.2cm, right=of node6fU] {$a_4$};

\ldmnewuser{sU}{6}{(5,-5)}
\node[node distance=0.1cm, above=of groupsU] {$U_2$};
\node[node distance=0.2cm, right=of node3sU] {$b_4$};
\node[node distance=0.2cm, right=of node4sU] {$b_3$};
\node[node distance=0.2cm, right=of node5sU] {$b_2$};
\node[node distance=0.2cm, right=of node6sU] {$b_1$};

\ldmconnectusers{HeNB}{fU}{2}{4}
\ldmconnectusers{eNB}{fU}{5}{1}
\ldmconnectusers{eNB}{sU}{6}{0}

\node[left=of node1HeNB.north, yshift=0.2cm, xshift=12pt] (HeNB1) {}; 
\node[left=of node1HeNB.south, yshift=-0.2cm, xshift=12pt] (HeNB2) {}; 
\draw[line width=1,<->] (-4.5,0.3) to (-4.5,-.9);
\node (a1) at (-5.1,-.3) [inner sep=0] {$R_{\text{IN}}^{n,*}$};

\node[left=of node1HeNB.south, yshift=-0.2cm, xshift=12pt] (HeNB3) {}; 
\node[left=of node5HeNB.south, yshift=-0.2cm, xshift=12pt] (HeNB4) {}; 
\draw[line width=1,<->] (-4.5,-0.9) to (-4.5,-3.3);
\node (a2) at (-5.1,-2.1) [inner sep=0] {$l_2^{*}$};

\node[left=of node1eNB.north, yshift=0.2cm, xshift=12pt] (eNB1) {}; 
\node[left=of node1eNB.south, yshift=-0.2cm, xshift=12pt] (eNB2) {}; 
\draw[line width=1,<->] (-4.5,-4.7) to (-4.5,-5.9);
\node (a3) at (-5.1,-5.3) [inner sep=0] {$R_{c}^{u,*}$};

\node[left=of node1eNB.south, yshift=-0.2cm, xshift=12pt] (eNB3) {}; 
\node[left=of node4eNB.south, yshift=-0.2cm, xshift=12pt] (eNB4) {}; 
\draw[line width=1,<->] (-4.5,-5.9) to (-4.5,-8.3);
\node (a4) at (-6.25,-7.1) [inner sep=0] {$R_{1,p}^{w,*}+R_{\text{IN}}^{v,*}+R_{c}^{v,*}$};
\draw[line width=1,<->] (-1.3,-5.9) to (-1.3,-6.5);
\node (b1) at (-1.9,-6.2) [inner sep=0] {$R_{c}^{v,*}$};
\draw[line width=1,<->] (-1.3,-6.5) to (-1.3,-7.7);
\node (b2) at (-1.9,-7.1) [inner sep=0] {$R_{\text{IN}}^{v,*}$};
\draw[line width=1,<->] (-1.3,-7.7) to (-1.3,-8.3);
\node (b3) at (-2.55,-8.0) [inner sep=0] {$R_{1,p}^{w,*}=R_{1,p}^{v,*}$};

\node[left=of node5eNB.north, yshift=0.2cm, xshift=12pt] (eNB5) {}; 
\node[left=of node5eNB.south, yshift=-0.2cm, xshift=12pt] (eNB6) {}; 

\begin{scope}[on background layer]
\node[pattern=north west lines, pattern color=red!80!black,fit=(node1HeNB) (node2HeNB)] (groupHeNBd1) {};
\node[pattern=north east lines, pattern color=black,fit=(node1HeNB) (node2HeNB)] (groupHeNBd2) {};

\node[pattern=north west lines, pattern color=red!80,fit=(node1eNB) (node2eNB)] (groupeNBd1) {};
\node[pattern=north east lines, pattern color=black,fit=(node3eNB) (node6eNB)] (groupeNBd3) {};

\node[pattern=north west lines, pattern color=red!80,fit=(node2fU) (node3fU)] (groupfUd1) {};
\node[pattern=north east lines, pattern color=black,fit=(node4fU) (node4fU)] (groupfUd3) {};
\node[pattern=north west lines, pattern color=red!80!black,fit=(node5fU) (node6fU)] (groupfUd4) {};

\node[pattern=north west lines, pattern color=red!80,fit=(node1sU) (node2sU)] (groupsUd1) {};
\node[pattern=north east lines, pattern color=black,fit=(node3sU) (node6sU)] (groupsUd3) {};
\end{scope}

\end{tikzpicture}
}

\edef\defaultpgflinewidth{\the\pgflinewidth}

\newlength{\mylabelwidth}

\newcommand*\circled[1]{\tikz[baseline=(char.base)]{
            \node[shape=circle,draw,inner sep=2pt] (char) {#1};}}

\pgfplotsset{every tick label/.append style={font=\footnotesize}}
 
\newcommand{\Resulttwo}[0]{
\begin{axis}[
        xmin=-0.2,
        xlabel=$\mu$,
        ylabel={DTB -- ${\Delta^{*}_{\text{det}}(\mu,n_F,\mathbf{n}})$}, xtick={0, 0.75, 1}, ytick={1, 1.6},
        xticklabels={$0$,${\mu^{\prime}(\mathbf{n})}$, $1$}, 
        yticklabels={${\Delta_{\text{LB}}^{\prime}(\mathbf{n})}$, $\Delta_0(\mathbf{n})$}, grid,
        legend cell align=left]
    \addplot[mark=*,blue, ultra thick] plot coordinates {
        (0,1.6)
        (0.75,1)
        (1,1)
    };
\addlegendentry{$0\leq n_{F}\leq \max\{n_{d2},n_{d3}\}$}
        \addplot[mark=*,red, ultra thick] plot coordinates {
        (0,1.2)
        (0.75,1)
        (1,1)
    };
    \addlegendentry{$n_{F}>\max\{n_{d2},n_{d3}\}$}
    
    \node[scale=0.70,label={185:{\circled{$A_1$}}},circle,fill,inner sep=2pt] at (axis 				cs:0,1.6) {};
    \node[scale=0.70,label={180:{\circled{$A_2$}}},circle,fill,inner sep=2pt] at (axis 				cs:0,1.2) {};
    \node[scale=0.70,label={5:{\circled{$B_2$}}},circle,fill,inner sep=2pt] at (axis 					cs:0.75,1) {};
\end{axis}
}

\newcommand{\Resultthree}[0]{
\begin{axis}[
        xmin=-0.2,
        xlabel=$\mu$,
        ylabel={DTB -- ${\Delta^{*}_{\text{det}}(\mu,n_F,\mathbf{n}})$}, xtick={0, 0.25, 0.75, 1}, ytick={1, 1.8}, extra y ticks={1.25},
        xticklabels={$0$,${\mu^{\prime\prime}(\mathbf{n})}$, ${\mu^{\prime}	(\mathbf{n})}$, $1$}, 
        yticklabels={${\Delta_{\text{LB}}^{\prime}(\mathbf{n})}$, $\Delta_0(\mathbf{n})$}, 
        extra y tick labels={$\frac{1}{n_{d3}-n_{d2}}$}, 
        grid,
            extra y tick style={%
            ,grid=major, tick label style={rotate=90},
            },        
        legend cell align=left]
    \addplot[mark=*,blue, ultra thick] plot coordinates {
        (0,1.8)
        (0.25,1.25)
        (0.75,1)
        (1,1)
    };

\addlegendentry{$0\leq n_{F}\leq n_{d2}$}
        \addplot[mark=*,black!60!green, ultra thick] plot coordinates {
        (0,1.48)
        (0.25,1.25)
        (0.75,1)
        (1,1)
    };
    \addlegendentry{$n_{d2}< n_{F}\leq n_{d3}$}
    
       	\addplot[mark=*,red, ultra thick] plot coordinates {
        (0,1.2)
        (0.75,1)
        (1,1)
    };
    \addlegendentry{$n_{F}> n_{d3}$}
    
    \addplot[dashed, black, thick] plot coordinates {
        (0,1.375)
        (0.25,1.25)
    };
        
	\node[scale=0.70, label={185:{\circled{$A_1$}}},circle,fill,inner sep=2pt] at (axis 				cs:0,1.8) {};
	\node[scale=0.70,label={180:{\circled{$A_3$}}},circle,fill,inner sep=2pt] at (axis 				cs:0,1.48) {};
	\node[scale=0.70,label={183:{\circled{$A_2$}}},circle,fill,inner sep=2pt] at (axis 					cs:0,1.2) {};
	\node[scale=0.70,label={5:{\circled{$B_1$}}},circle,fill,inner sep=2pt] at (axis 					cs:0.25,1.25) {};
	\node[scale=0.70,label={5:{\circled{$B_2$}}},circle,fill,inner sep=2pt] at (axis 					cs:0.75,1) {};
\end{axis}
}

\newcommand{\Resultfour}[0]{
\begin{axis}[
        xmin=-0.2,
        xlabel=$\mu$,
        ylabel={DTB -- ${\Delta^{*}_{\text{det}}(\mu,n_F,\mathbf{n}})$}, 	                xtick={0, 0.5, 0.75, 1}, ytick={1, 1.8},
        xticklabels={$0$,${\mu^{\prime\prime\prime}(\mathbf{n})}$, ${\mu^{\prime\prime}(\mathbf{n})}$, $1$}, 
        yticklabels={${\Delta_{\text{LB}}^{\prime}(\mathbf{n})}$, $\Delta_0(\mathbf{n})$}, grid,
        legend cell align=left]
    \addplot[mark=*,blue, ultra thick] plot coordinates {
        (0,1.8)
        (0.5,1)
        (0.75,1)
        (1,1)
    };
\addlegendentry{$0\leq n_{F}\leq n_{d2}$}
        \addplot[mark=*,red, ultra thick] plot coordinates {
        (0,1.5)
        (0.5,1)
        (0.75,1)
        (1,1)
    };
    \addlegendentry{$n_{F}> n_{d2}$}  

	\node[scale=0.70,label={185:{\circled{$A_1$}}},circle,fill,inner sep=2pt] at (axis 				cs:0,1.8) {};
	\node[scale=0.70,label={180:{\circled{$A_4$}}},circle,fill,inner sep=2pt] at (axis 				cs:0,1.5) {};
	\node[scale=0.70,label={5:{\circled{$C_1$}}},circle,fill,inner sep=2pt] at (axis 					cs:0.5,1) {};
\end{axis}
}

\newcommand{\Resultfive}[0]{
\begin{axis}[
        ymax=1.1,
        ymin=0.99,
        xmin=-0.2,
        xlabel=$\mu$,
        ylabel={DTB -- ${\Delta^{*}_{\text{det}}(\mu,n_F,\mathbf{n}})$}, 	                xtick={0, 1}, ytick={1},
        xticklabels={$0$,$1$}, 
        yticklabels={${\Delta^{\prime}_{\text{LB}}}(\mathbf{n})$}, 
        grid,
        legend cell align=left,
         ]
    \addplot[mark=*,blue, ultra thick] plot coordinates {
        (0,1)
        (1,1)
    };
	\addlegendentry{$\forall n_{F}$}
	\node[scale=0.7,label={175:{\circled{$A_1$}}},circle,fill,inner sep=2pt] at (axis 				cs:0,1) {};
\end{axis}
}

\newcommand{\ParBlockMod}[0]{
\begin{tikzpicture}
[
squarednode/.style={rectangle, rounded corners, draw=red!60, fill=red!5, very thick},
]
\coordinate (A0) at (0,2.9);
\node[squarednode, anchor=north west, minimum width=0.125*\linewidth,minimum height=0.8cm] (time11) at ([yshift=-0.5cm, xshift=0cm]A0) {};
\node[squarednode, draw=blue!60, fill=blue!5, anchor=north west, minimum width=0.125*\linewidth,minimum height=0.8cm] (time21) at ([xshift=0.025*\linewidth, yshift=-3.0cm]A0) {};

\node[squarednode, anchor=north west, minimum width=0.125*\linewidth,minimum height=0.8cm] (time12) at ([xshift=0.05cm]time11.north east) {};
\node[squarednode, draw=blue!60, fill=blue!5, anchor=north west, minimum width=0.125*\linewidth,minimum height=0.8cm] (time22) at ([xshift=0.05cm]time21.north east) {};

\node[squarednode, anchor=north west, minimum width=0.125*\linewidth,minimum height=0.8cm] (time13) at ([xshift=0.05cm]time12.north east) {};
\node[squarednode, draw=blue!60, fill=blue!5, anchor=north west, minimum width=0.125*\linewidth,minimum height=0.8cm] (time23) at ([xshift=0.05cm]time22.north east) {};

\node[squarednode, anchor=north west, minimum width=0.125*\linewidth,minimum height=0.8cm] (time14) at ([xshift=0.125*\linewidth+0.05cm]time13.north east) {};
\node[squarednode, draw=blue!60, fill=blue!5, anchor=north west, minimum width=0.125*\linewidth,minimum height=0.8cm] (time24) at ([xshift=0.125*\linewidth+0.05cm]time23.north east) {};

\path (time13) -- (time14) node [midway, sloped, scale=1.5] {$\dots$};
\path (time23) -- (time24) node [midway, sloped, scale=1.5] {$\dots$};

\node[squarednode, anchor=north west, minimum width=0.125*\linewidth,minimum height=0.8cm] (time15) at ([xshift=0.05cm]time14.north east) {};

\node[squarednode, draw=blue!60, fill=blue!5, anchor=north west, minimum width=0.125*\linewidth,minimum height=0.8cm] (time25) at ([xshift=0.05cm]time24.north east) {};

\draw ($(A0)+(0,5pt)$) -- ($(A0)-(0,5pt)$);
\node[scale=0.75] at ($(A0)+(0,3ex)$) {$0$};
\coordinate (A1) at ([xshift=1+0.025*\linewidth, yshift=0cm]A0);
\draw ($(A1)+(0,5pt)$) -- ($(A1)-(0,5pt)$);
\node[scale=0.75]  at ($(A1)+(0,3ex)$) {$T_O$};
\coordinate (A2) at ([xshift=1+0.025*\linewidth+0.125*\linewidth+0.05cm, yshift=0cm]A0);
\draw ($(A2)+(0,5pt)$) -- ($(A2)-(0,5pt)$);
\node[scale=0.75]  at ($(A2)+(0,3ex)$) {$T_{B}+T_O$};
\foreach \x in {3,4}
    {        
      \coordinate (A\x) at ([xshift=1+0.025*\linewidth+(\x-1)*0.125*\linewidth+\x*0.05cm, yshift=0cm]A0) {};
      \draw ($(A\x)+(0,5pt)$) -- ($(A\x)-(0,5pt)$);
      \pgfmathtruncatemacro{\y}{int(\x-1)}
      \node[scale=0.75]  at ($(A\x)+(0,3ex)$) {\y $T_B+T_O$};
    }
\coordinate (A5) at ([xshift=1+0.025*\linewidth+5*0.125*\linewidth+7*0.05cm, yshift=0cm]A0);
\draw ($(A5)+(0,5pt)$) -- ($(A5)-(0,5pt)$);
\node[scale=0.75]  at ($(A5)+(0,3ex)$) {$(B-1)T_{B}+T_O$};
\coordinate (A6) at ([xshift=1+0.025*\linewidth+6*0.125*\linewidth+8*0.05cm, yshift=0cm]A0);
\draw ($(A6)+(0,5pt)$) -- ($(A6)-(0,5pt)$);
\node[scale=0.75]  at ($(A6)+(0,3ex)$) {$BT_{B}+T_O$};
\draw[ultra thick,arrows=->] (A0) -- (A6) -- ($(A6)+1.5*(0.5,0)$);

\node[color=red!100,scale=1.25, anchor=south west] at ([xshift=1cm, yshift=-0.75cm]time11.south west) {Fronthaul Transmission};
\node[color=blue!100,scale=1.25, anchor=south west] at ([xshift=1cm, yshift=0.15cm]time21.north west) {Wireless Transmission};
\draw[-, dashed, thick] ([xshift=0.01cm, yshift=-3.05cm]A0) -- ([xshift=-0.025*\linewidth, yshift=-0.025cm]time21.south west);
\draw[-, dashed, thick] ([xshift=-0.05cm, yshift=-3.05cm]A1) -- ([xshift=0cm, yshift=-0.025cm]time21.south west);
\draw[<->, thick] ([xshift=0.01cm, yshift=-3.0cm]A0) -- ([xshift=-0.05cm, yshift=-3cm]A1) node[above, midway, scale=0.75] {$T_0$};

\end{tikzpicture}
}

\newcommand{\Resultonepar}[0]{
\begin{axis}[
        xmin=-0.2,
        xlabel=$\mu$,
        ylabel={DTB -- ${\Delta^{*}_{\text{det}}(\mu,n_F,\mathbf{n}})$}, 					xtick={0, 0.25, 0.75, 1}, 
        ytick={0.25},
        xticklabels={$0$,${\mu^{\prime}_{P}(n_{F_2},\mathbf{n})}$,${\mu^{\prime}_{P}(n_{F_1},\mathbf{n})}$, $1$},  
        yticklabels={${\Delta_{\text{LB}}^{\prime}(\mathbf{n})}$}, 
        grid,
        legend cell align=left]
    
   \addplot[mark=*,blue, ultra thick] plot coordinates {
        (0,2/5) (6/8,2/8) (1,2/8)
    };
   \addlegendentry{}
   \addplot[mark=*,black!60!green, ultra thick] plot coordinates {
        (0,2/7) (2/8,2/8) (1,2/8)
    };
    \addlegendentry{\smash{\raisebox{-0.3ex}{$0\leq n_{F}< n_{F,\max}(\mathbf{n})$}}}
    \addplot[mark=*,red, ultra thick] plot coordinates {
        (0,2/8) (1,2/8)
    };
    \addlegendentry{$n_{F}\geq n_{F,\max}(\mathbf{n})$}
    \node[circle,fill,inner sep=2pt] (p1) at (axis cs:0,0.4) {};
    \node[circle,fill,inner sep=2pt] (p2) at (axis cs:0,0.285714) {}; 
    \node[circle,fill,inner sep=2pt] (p3) at (axis cs:0,0.25) {};
    \node (c1) at (axis cs: -0.10, 0.325) {\circled{$A_1$}};
    \draw [->] (c1) -- (p1);
    \draw [->] (c1) -- (p2);
    
    \node[circle,fill,inner sep=2pt] (t1) at (axis cs:0.75,0.25) {};
    \node[circle,fill,inner sep=2pt] (t2) at (axis cs:0.25,0.25) {}; 
	\node (c2) at (axis cs: 0.5, 0.27) {\circled{$B_2$}}; 
	\draw [->] (c2) -- (t1);
    \draw [->] (c2) -- (t2);   

	\node[scale=0.85] (a) at (axis cs: 0.70, 0.34) {\circled{$A_1$}$\:=\Big(0,\frac{2}{n_{F}+\max\{n_{d2},n_{d3}\}}\Big)$};
\end{axis}
}

\newcommand{\Resulttwopar}[0]{
\begin{axis}[
        xmin=-0.2,
        xlabel=$\mu$,
        ylabel={DTB -- ${\Delta^{*}_{\text{det}}(\mu,n_F,\mathbf{n}})$}, 		
        xtick={0,0.2,0.6,0.8,1},
        ytick={0.2},
        extra y ticks={0.25},
        extra x ticks={0.25, 0.5},
		xticklabels={$0$, ${\mu^{\prime}_{P}(n_{F_3},\mathbf{n})}$, ${\mu^{\prime}_{P}(n_{F_2},\mathbf{n})}$, $\qquad\:\:\:{\mu^{\prime}_{P}(n_{F_1},\mathbf{n})}$, $\qquad 1$}, 
        yticklabels={${\Delta_{\text{LB}}^{\prime}(\mathbf{n})}$}, 
        extra y tick labels={$\quad\:\frac{1}{n_{d3}-n_{d2}}$},
        extra x tick labels={${\mu^{\prime\prime}_{P}(n_{F_2},\mathbf{n})}$, $\qquad\:\:{\mu^{\prime\prime}_{P}(n_{F_1},\mathbf{n})}$},
extra x tick style={%
            ,grid=major
            ,ticklabel pos=upper
            },
            extra y tick style={%
            ,grid=major, tick label style={rotate=90},
            },
        grid,
        legend cell align=left]
    
   \addplot[mark=*,blue, ultra thick] plot coordinates {
        (0,1/2) (2/4,1/4) (4/5,1/5) (1,1/5)
    };
   \addlegendentry{}
   \addplot[mark=*,black!60!brown, ultra thick] plot coordinates {
        (0,1/3) (1/4,1/4) (3/5,1/5) (1,1/5)
    };
    \addlegendentry{\smash{\raisebox{-0.3ex}{$0\leq n_{F}< n_{F,\text{IM}}(\mathbf{n})$}}}
    \addplot[mark=*,black!60!green, ultra thick] plot coordinates {
        (0,2/9) (1/5, 1/5) (1, 1/5)
    };
    \addlegendentry{$n_{F,\text{IM}}(\mathbf{n})\leq n_{F}< n_{F,\max}(\mathbf{n})$}
    \addplot[mark=*,red, ultra thick] plot coordinates {
        (0,1/5) (1, 1/5)
    };
    \addlegendentry{$n_{F}\geq n_{F,\max}(\mathbf{n})$}
	\node[circle,fill,inner sep=2pt] (p1) at (axis cs:0,0.5) {};
	\node[circle,fill,inner sep=2pt] (p2) at (axis cs:0,0.3333333) {};
	\node (c1) at (axis cs: -0.10, 0.41666666) {\circled{$A_2$}};
	\draw [->] (c1) -- (p1);
	\draw [->] (c1) -- (p2);
    
	\node[circle,fill,inner sep=2pt] (t1) at (axis cs:0,0.2222222) {};
	\node (c2) at (axis cs: -0.10, 0.23) {\circled{$A_1$}}; 
	
	\node[circle,fill,inner sep=2pt] (s1) at (axis cs:0.25,0.25) {}; 
	\node[circle,fill,inner sep=2pt] (s2) at (axis cs:0.5,0.25) {}; 
	\node (c3) at (axis cs: 0.25, 0.3) {\circled{$B_1$}};
	\draw [->] (c3) -- (s1);
	\draw [->] (c3) -- (s2);

	\node[circle,fill,inner sep=2pt] (r1) at (axis cs:0.2,0.2) {}; 
	\node[circle,fill,inner sep=2pt] (r2) at (axis cs:0.6,0.2) {}; 
	\node[circle,fill,inner sep=2pt] (r3) at (axis cs:0.8,0.2) {}; 
	\node (c4) at (axis cs: 0.8, 0.29) {\circled{$B_2$}}; 
	\draw [->] (c4) -- (r1);
	\draw [->] (c4) -- (r2);  
	\draw [->] (c4) -- (r3); 
	
	\node[scale=0.85] (a) at (axis cs: 0.85, 0.35) {\circled{$A_2$}$\:=\Big(0,\frac{1}{n_{F}+n_{d2}}\Big)$}; 
\end{axis}
}

\newcommand{\Resulthreepar}[0]{
\begin{axis}[
        xmin=-0.2,
        xlabel=$\mu$,
        ylabel={DTB -- ${\Delta^{*}_{\text{det}}(\mu,n_F,\mathbf{n}})$}, 					xtick={0, 0.4, 0.6, 1}, 
        ytick={0.2},
        xticklabels={$0$,${\mu^{\prime\prime\prime}_{P}(n_{F_2},\mathbf{n})}$,$\qquad\:\:\:\:\:\:{\mu^{\prime\prime\prime}_{P}(n_{F_1},\mathbf{n})}$, $1$},  
        yticklabels={${\Delta_{\text{LB}}^{\prime}(\mathbf{n})}$}, 
        grid,
        legend cell align=left]
    
   \addplot[mark=*,blue, ultra thick] plot coordinates {
        (0,1/2) (3/5,1/5) (1,1/5)
    };
   \addlegendentry{}
   \addplot[mark=*,black!60!brown, ultra thick] plot coordinates {
        (0,1/3) (2/5,1/5) (1,1/5)
    };
    \addlegendentry{\smash{\raisebox{-0.3ex}{$0\leq n_{F}< n_{F,\max}(\mathbf{n})$}}}
    \addplot[mark=*,red, ultra thick] plot coordinates {
        (0,1/5) (1,1/5)
    };
    \addlegendentry{$n_{F}\geq n_{F,\max}(\mathbf{n})$}
    \node[circle,fill,inner sep=2pt] (p1) at (axis cs:0,0.5) {};
    \node[circle,fill,inner sep=2pt] (p2) at (axis cs:0,0.333333333) {};
    \node[circle,fill,inner sep=2pt] (p3) at (axis cs:0,0.2) {};
    \node (c1) at (axis cs: -0.10, 0.416666666) {\circled{$A_2$}};
    \draw [->] (c1) -- (p1);
    \draw [->] (c1) -- (p2);
    
    \node[circle,fill,inner sep=2pt] (t1) at (axis cs:0.6,0.2) {};
    \node[circle,fill,inner sep=2pt] (t2) at (axis cs:0.4,0.2) {}; 
	\node (c2) at (axis cs: 0.7, 0.27) {\circled{$C_1$}}; 
	\draw [->] (c2) -- (t1);
    \draw [->] (c2) -- (t2);   

\end{axis}
}

\newcommand{\Resultfourpar}[0]{
\begin{axis}[
        ymax=1.1,
        ymin=0.99,
        xmin=-0.2,
        xlabel=$\mu$,
        ylabel={DTB -- ${\Delta^{*}_{\text{det}}(\mu,n_F,\mathbf{n}})$}, 	                xtick={0, 1}, ytick={1},
        xticklabels={$0$,$1$}, 
        yticklabels={${\Delta_{\text{LB}}^{\prime}}(\mathbf{n})$}, grid,
        legend cell align=left,
         ]
    \addplot[mark=*,red, ultra thick] plot coordinates {
        (0,1)
        (1,1)
    };
    \addlegendentry{$n_{F}\geq n_{F,\max}(\mathbf{n})$}
\end{axis}
}

\newcommand{\SCHEMERxoneSCL}[0]{

\node (Zero) at (-4.5,-4) [inner sep=0] {$0$};
\draw[line width=1,-] (-4,-4) to (1,-4);

\node (zerotx2rx1) at (-3,-4.3) [inner sep=0] {$\boldsymbol S^{q-n_{d2}}\boldsymbol x_2$};
\node (zerotx3rx1) at (0,-4.3) [inner sep=0] {$\boldsymbol S^{q-n_{d1}}\boldsymbol x_1$};

\draw[line width=1,-] (-1,3.0) to (1,3.0);
\node (zerotx1rx1) at (1.3,3.0) [inner sep=0] {$n_{d1}$};

\draw[line width=1] (-0.5,2.4) rectangle (0.5,3.0);
\node (zerotx1rx1) at (0,2.7) [inner sep=0] {$\boldsymbol{0}_{l_{4}}$};

\draw[line width=1] (-0.5,1.4) rectangle (0.5,2.4);
\node (zerotx1rx1) at (0,1.9) [inner sep=0] {$\boldsymbol{u}_{2,p}$};

\draw[line width=1] (-0.5,-1.5) rectangle (0.5,1.4);
\node (zerotx1rx1) at (0,-0.05) [inner sep=0] {$\boldsymbol{0}_{l_{3}}$};

\draw[line width=1] (-0.5,-3.5) rectangle (0.5,-1.5);
\node (zerotx1rx1) at (0,-2.5) [inner sep=0] {$\boldsymbol{n}_{\text{IN}}$};

\draw[line width=1] (-0.5,-4) rectangle (0.5,-3.5);
\node (zerotx1rx1) at (0,-3.75) [inner sep=0] {$\boldsymbol{0}$};

\draw[line width=1,-] (-4,1) to (-2,1);
\node (zerotx1rx1) at (-4.3,1) [inner sep=0] {$n_{d2}$};

\draw[line width=1] (-3.5,0.5) rectangle (-2.5,1);
\node (zerotx1rx1) at (-3,0.75) [inner sep=0] {$\boldsymbol{u}_{c}$};

\draw[line width=1] (-3.5,-1.5) rectangle (-2.5,0.5);
\node (zerotx1rx1) at (-3,-0.5) [inner sep=0] {$\boldsymbol{v}_{c}$};

\draw[line width=1] (-3.5,-3.5) rectangle (-2.5,-1.5);
\node (zerotx1rx1) at (-3,-2.5) [inner sep=0] {$\boldsymbol{v}_{\text{IN}}$};

\draw[line width=1] (-3.5,-4) rectangle (-2.5,-3.5);
\node (u1prx1) at (-3,-3.75) [inner sep=0] {$\boldsymbol{u}_{1,p}$};

\draw[line width=1,dashed,-] (-2.5,-3.5) to (-0.5,-3.5);
\draw[line width=1,dashed,-] (-2.5,-1.5) to (-0.5,-1.5);
\draw[line width=1,dashed,-] (-2,1) to (-0.5,1);

\draw[line width=1,<->] (-1,1) to (-1,3);
\node (zerotx1rx1) at (-1.9,2) [inner sep=0] {$n_{d1}-n_{d2}$};

}

\newcommand{\SCHEMERxtwoSCL}[0]{

\node (Zero) at (-4.5,-4) [inner sep=0] {$0$};
\draw[line width=1,-] (-4,-4) to (-2,-4);

\node (zerotx2rx1) at (-3,-4.3) [inner sep=0] {$\boldsymbol S^{q-n_{d3}}\boldsymbol x_2$};

\draw[line width=1,-] (-4,0.5) to (-2,0.5);
\node (zerotx1rx1) at (-4.3,0.5) [inner sep=0] {$n_{d3}$};

\draw[line width=1] (-3.5,0) rectangle (-2.5,0.5);
\node (zerotx1rx1) at (-3,0.25) [inner sep=0] {$\boldsymbol{u}_{c}$};

\draw[line width=1] (-3.5,-2) rectangle (-2.5,0);
\node (zerotx1rx1) at (-3,-1) [inner sep=0] {$\boldsymbol{v}_{c}$};

\draw[line width=1] (-3.5,-4) rectangle (-2.5,-2);
\node (zerotx1rx1) at (-3,-3) [inner sep=0] {$\boldsymbol{v}_{\text{IN}}$};

}

\newcommand{\SCHEMERxoneWCL}[0]{

\node (Zero) at (-4.5,-4) [inner sep=0] {$0$};
\draw[line width=1,-] (-4,-4) to (1,-4);

\node (zerotx2rx1) at (-3,-4.3) [inner sep=0] {$\boldsymbol S^{q-n_{d2}}\boldsymbol x_2$};
\node (zerotx3rx1) at (0,-4.3) [inner sep=0] {$\boldsymbol S^{q-n_{d1}}\boldsymbol x_1$};

\draw[line width=1,-] (-1,2.35) to (1,2.35);
\node (zerotx1rx1) at (1.3,2.35) [inner sep=0] {$n_{d1}$};

\draw[line width=1] (-0.5,1.8) rectangle (0.5,2.35);
\node (zerotx1rx1) at (0,2.075) [inner sep=0] {$\boldsymbol{0}_{l_{4}}$};

\draw[line width=1] (-0.5,0.8) rectangle (0.5,1.8);
\node (zerotx1rx1) at (0,1.3) [inner sep=0] {$\boldsymbol{u}_{2,p}$};

\draw[line width=1] (-0.5,-2.3) rectangle (0.5,0.8);
\node (zerotx1rx1) at (0,-0.75) [inner sep=0] {$\boldsymbol{0}_{l_{3}}$};

\draw[line width=1] (-0.5,-3.8) rectangle (0.5,-2.3);
\node (zerotx1rx1) at (0,-3.05) [inner sep=0] {$\boldsymbol{n}_{\text{IN}}$};

\draw[line width=1] (-0.5,-4) rectangle (0.5,-3.8);
\node (zerotx1rx1) at (0,-3.9) [inner sep=0,scale=0.75] {$\boldsymbol{0}$};

\draw[line width=1,-] (-4,0.7) to (-2,0.7);
\node (zerotx1rx1) at (-4.3,0.7) [inner sep=0] {$n_{d2}$};

\draw[line width=1] (-3.5,-1.3) rectangle (-2.5,0.7);
\node (zerotx1rx1) at (-3,-0.3) [inner sep=0] {$\boldsymbol{u}_{c}$};

\draw[line width=1] (-3.5,-2.3) rectangle (-2.5,-1.3);
\node (zerotx1rx1) at (-3,-1.8) [inner sep=0] {$\boldsymbol{v}_{c}$};

\draw[line width=1] (-3.5,-3.8) rectangle (-2.5,-2.3);
\node (zerotx1rx1) at (-3,-3.05) [inner sep=0] {$\boldsymbol{v}_{\text{IN}}$};

\draw[line width=1] (-3.5,-4) rectangle (-2.5,-3.8);

\draw[line width=1,dashed,-] (-2.5,-3.8) to (-0.5,-3.8);
\draw[line width=1,dashed,-] (-2.5,-2.3) to (-0.5,-2.3);
\draw[line width=1,dashed,-] (-2,0.7) to (-0.5,0.7);

\draw[line width=1,<->] (-1,0.7) to (-1,2.35);
\node (zerotx1rx1) at (-1.9,1.525) [inner sep=0] {$n_{d1}-n_{d2}$};

}

\newcommand{\SCHEMERxtwoWCL}[0]{

\node (Zero) at (-4.5,-4) [inner sep=0] {$0$};
\draw[line width=1,-] (-4,-4) to (-2,-4);

\node (zerotx2rx1) at (-3,-4.3) [inner sep=0] {$\boldsymbol S^{q-n_{d3}}\boldsymbol x_2$};

\draw[line width=1,-] (-4,1.5) to (-2,1.5);
\node (zerotx1rx1) at (-4.3,1.5) [inner sep=0] {$n_{d3}$};

\draw[line width=1] (-3.5,-0.5) rectangle (-2.5,1.5);
\node (zerotx1rx1) at (-3,0.5) [inner sep=0] {$\boldsymbol{u}_{c}$};

\draw[line width=1] (-3.5,-1.5) rectangle (-2.5,-0.5);
\node (zerotx1rx1) at (-3,-1) [inner sep=0] {$\boldsymbol{v}_{c}$};

\draw[line width=1] (-3.5,-3.0) rectangle (-2.5,-1.5);
\node (zerotx1rx1) at (-3,-2.25) [inner sep=0] {$\boldsymbol{v}_{\text{IN}}$};

\draw[line width=1] (-3.5,-4) rectangle (-2.5,-3.0);
\node (zerotx1rx1) at (-3,-3.5) [inner sep=0] {$\boldsymbol{v}_{1,p}$};
}

\newcommand{\SCHEMERxoneSCLPar}[0]{

\node (Zero) at (-4.5,-4) [inner sep=0] {$0$};
\draw[line width=1,-] (-4,-4) to (1,-4);

\node (zerotx2rx1) at (-3,-4.3) [inner sep=0] {$\boldsymbol S^{q-n_{d2}}\boldsymbol x_2[t]$};
\node (zerotx3rx1) at (0,-4.3) [inner sep=0] {$\boldsymbol S^{q-n_{d1}}\boldsymbol x_1[t]$};

\draw[line width=1,-] (-1,3.0) to (1,3.0);
\node (zerotx1rx1) at (1.3,3.0) [inner sep=0] {$n_{d1}$};

\draw[line width=1] (-0.5,2.4) rectangle (0.5,3.0);
\node (zerotx1rx1) at (0,2.7) [inner sep=0,scale=0.85] {$\boldsymbol{0}_{l_{4}[t]}$};

\draw[line width=1] (-0.5,1.4) rectangle (0.5,2.4);
\node (zerotx1rx1) at (0,1.9) [inner sep=0,scale=0.85] {$\boldsymbol{u}_{2,p}[t]$};

\draw[line width=1] (-0.5,-2.5) rectangle (0.5,1.4);
\node (zerotx1rx1) at (0,-0.55) [inner sep=0,scale=0.85] {$\boldsymbol{0}_{l_{3}[t]}$};

\draw[line width=1] (-0.5,-3.5) rectangle (0.5,-2.5);
\node (zerotx1rx1) at (0,-3.0) [inner sep=0,scale=0.85] {$\boldsymbol{n}_{\text{IN}}[t]$};

\draw[line width=1] (-0.5,-4) rectangle (0.5,-3.5);
\node (zerotx1rx1) at (0,-3.75) [inner sep=0,scale=0.85] {$\boldsymbol{0}$};

\draw[line width=1,-] (-4,1) to (-2,1);
\node (zerotx1rx1) at (-4.3,1) [inner sep=0,scale=0.85] {$n_{d2}$};

\draw[line width=1] (-3.5,0.5) rectangle (-2.5,1);
\node (zerotx1rx1) at (-3,0.75) [inner sep=0,scale=0.85] {$\boldsymbol{u}_{2,c}[t]$};

\draw[line width=1] (-3.5,0) rectangle (-2.5,0.5);
\node (zerotx1rx1) at (-3,0.25) [inner sep=0,scale=0.85] {$\boldsymbol{v}_{2,c}[t]$};

\draw[line width=1] (-3.5,-0.5) rectangle (-2.5,0);
\node (zerotx1rx1) at (-3,-0.25) [inner sep=0,scale=0.85] {$\boldsymbol{u}_{1,c}[t]$};

\draw[line width=1] (-3.5,-2.5) rectangle (-2.5,-0.5);
\node (zerotx1rx1) at (-3,-1.5) [inner sep=0,scale=0.85] {$\boldsymbol{v}_{1,c}[t]$};

\draw[line width=1] (-3.5,-3.5) rectangle (-2.5,-2.5);
\node (zerotx1rx1) at (-3,-3) [inner sep=0,scale=0.85] {$\boldsymbol{v}_{\text{IN}}[t]$};

\draw[line width=1] (-3.5,-4) rectangle (-2.5,-3.5);
\node (u1prx1) at (-3,-3.75) [inner sep=0,scale=0.85] {$\boldsymbol{u}_{1,p}[t]$};

\draw[line width=1,dashed,-] (-2.5,-3.5) to (-0.5,-3.5);
\draw[line width=1,dashed,-] (-2.5,-2.5) to (-0.5,-2.5);
\draw[line width=1,dashed,-] (-2,1) to (-0.5,1);

\draw[line width=1,<->] (-1,1) to (-1,3);
\node (zerotx1rx1) at (-1.9,2) [inner sep=0] {$n_{d1}-n_{d2}$};

\draw[thick, dashed, red] (-3.75,-3.5) rectangle (-2.25,0);

}

\newcommand{\SCHEMERxtwoSCLPar}[0]{

\node (Zero) at (-4.5,-4) [inner sep=0] {$0$};
\draw[line width=1,-] (-4,-4) to (-2,-4);

\node (zerotx2rx1) at (-3,-4.3) [inner sep=0] {$\boldsymbol S^{q-n_{d3}}\boldsymbol x_2[t]$};

\draw[line width=1,-] (-4,0.7) to (-2,0.7);
\node (zerotx1rx1) at (-4.3,0.7) [inner sep=0] {$n_{d3}$};

\draw[line width=1] (-3.5,0.2) rectangle (-2.5,0.7);
\node (zerotx1rx1) at (-3,0.45) [inner sep=0,scale=0.85] {$\boldsymbol{u}_{2,c}[t]$};

\draw[line width=1] (-3.5,-0.3) rectangle (-2.5,0.2);
\node (zerotx1rx1) at (-3,-0.05) [inner sep=0,scale=0.85] {$\boldsymbol{v}_{2,c}[t]$};

\draw[line width=1] (-3.5,-0.8) rectangle (-2.5,-0.3);
\node (zerotx1rx1) at (-3,-0.55) [inner sep=0,scale=0.85] {$\boldsymbol{u}_{1,c}[t]$};

\draw[line width=1] (-3.5,-2.8) rectangle (-2.5,-0.8);
\node (zerotx1rx1) at (-3,-1.8) [inner sep=0,scale=0.85] {$\boldsymbol{v}_{1,c}[t]$};

\draw[line width=1] (-3.5,-3.8) rectangle (-2.5,-2.8);
\node (zerotx1rx1) at (-3,-3.3) [inner sep=0,scale=0.85] {$\boldsymbol{v}_{\text{IN}}[t]$};

\draw[line width=1] (-3.5,-4) rectangle (-2.5,-3.8);

\draw[thick, dashed, red] (-3.75,-3.8) rectangle (-2.25,-0.3);

}

\newcommand{\SCHEMERxoneWCLPar}[0]{

\node (Zero) at (-4.5,-4) [inner sep=0] {$0$};
\draw[line width=1,-] (-4,-4) to (1,-4);

\node (zerotx2rx1) at (-3,-4.3) [inner sep=0] {$\boldsymbol S^{q-n_{d2}}\boldsymbol x_2[t]$};
\node (zerotx3rx1) at (0,-4.3) [inner sep=0] {$\boldsymbol S^{q-n_{d1}}\boldsymbol x_1[t]$};

\draw[line width=1,-] (-1,2.35) to (1,2.35);
\node (zerotx1rx1) at (1.3,2.35) [inner sep=0] {$n_{d1}$};

\draw[line width=1] (-0.5,1.8) rectangle (0.5,2.35);
\node (zerotx1rx1) at (0,2.075) [inner sep=0,scale=0.85] {$\boldsymbol{0}_{l_{4}[t]}$};

\draw[line width=1] (-0.5,0.8) rectangle (0.5,1.8);
\node (zerotx1rx1) at (0,1.3) [inner sep=0,scale=0.85] {$\boldsymbol{u}_{2,p}[t]$};

\draw[line width=1] (-0.5,-2.3) rectangle (0.5,0.8);
\node (zerotx1rx1) at (0,-0.75) [inner sep=0,scale=0.85] {$\boldsymbol{0}_{l_{3}[t]}$};

\draw[line width=1] (-0.5,-3.8) rectangle (0.5,-2.3);
\node (zerotx1rx1) at (0,-3.05) [inner sep=0,scale=0.85] {$\boldsymbol{n}_{\text{IN}}[t]$};

\draw[line width=1] (-0.5,-4) rectangle (0.5,-3.8);
\node (zerotx1rx1) at (0,-3.9) [inner sep=0,scale=0.75,scale=0.85] {$\boldsymbol{0}$};

\draw[line width=1,-] (-4,0.7) to (-2,0.7);
\node (zerotx1rx1) at (-4.3,0.7) [inner sep=0] {$n_{d2}$};

\draw[line width=1] (-3.5,0.2) rectangle (-2.5,0.7);
\node (zerotx1rx1) at (-3,0.45) [inner sep=0,scale=0.85] {$\boldsymbol{u}_{2,c}[t]$};

\draw[line width=1] (-3.5,-0.3) rectangle (-2.5,0.2);
\node (zerotx1rx1) at (-3,-0.05) [inner sep=0,scale=0.85] {$\boldsymbol{v}_{2,c}[t]$};

\draw[line width=1] (-3.5,-1.3) rectangle (-2.5,-0.3);
\node (zerotx1rx1) at (-3,-0.8) [inner sep=0,scale=0.85] {$\boldsymbol{u}_{1,c}[t]$};

\draw[line width=1] (-3.5,-2.3) rectangle (-2.5,-1.3);
\node (zerotx1rx1) at (-3,-1.8) [inner sep=0,scale=0.85] {$\boldsymbol{v}_{1,c}[t]$};

\draw[line width=1] (-3.5,-3.8) rectangle (-2.5,-2.3);
\node (zerotx1rx1) at (-3,-3.05) [inner sep=0,scale=0.85] {$\boldsymbol{v}_{\text{IN}}[t]$};

\draw[line width=1] (-3.5,-4) rectangle (-2.5,-3.8);

\draw[line width=1,dashed,-] (-2.5,-3.8) to (-0.5,-3.8);
\draw[line width=1,dashed,-] (-2.5,-2.3) to (-0.5,-2.3);
\draw[line width=1,dashed,-] (-2,0.7) to (-0.5,0.7);

\draw[line width=1,<->] (-1,0.7) to (-1,2.35);
\node (zerotx1rx1) at (-1.9,1.525) [inner sep=0] {$n_{d1}-n_{d2}$};

\draw[thick, dashed, red] (-3.75,-3.8) rectangle (-2.25,-0.3);

}

\newcommand{\SCHEMERxtwoWCLPar}[0]{

\node (Zero) at (-4.5,-4) [inner sep=0] {$0$};
\draw[line width=1,-] (-4,-4) to (-2,-4);

\node (zerotx2rx1) at (-3,-4.3) [inner sep=0] {$\boldsymbol S^{q-n_{d3}}\boldsymbol x_2$};

\draw[line width=1,-] (-4,1.5) to (-2,1.5);
\node (zerotx1rx1) at (-4.3,1.5) [inner sep=0] {$n_{d3}$};

\draw[line width=1] (-3.5,1) rectangle (-2.5,1.5);
\node (zerotx1rx1) at (-3,1.25) [inner sep=0,scale=0.85] {$\boldsymbol{u}_{2,c}[t]$};

\draw[line width=1] (-3.5,0.5) rectangle (-2.5,1);
\node (zerotx1rx1) at (-3,0.75) [inner sep=0,scale=0.85] {$\boldsymbol{v}_{2,c}[t]$};

\draw[line width=1] (-3.5,-0.5) rectangle (-2.5,0.5);
\node (zerotx1rx1) at (-3,0.0) [inner sep=0,scale=0.85] {$\boldsymbol{u}_{1,c}[t]$};

\draw[line width=1] (-3.5,-1.5) rectangle (-2.5,-0.5);
\node (zerotx1rx1) at (-3,-1) [inner sep=0,scale=0.85] {$\boldsymbol{v}_{1,c}[t]$};

\draw[line width=1] (-3.5,-3.0) rectangle (-2.5,-1.5);
\node (zerotx1rx1) at (-3,-2.25) [inner sep=0,scale=0.85] {$\boldsymbol{v}_{\text{IN}}[t]$};

\draw[line width=1] (-3.5,-4) rectangle (-2.5,-3.0);
\node (zerotx1rx1) at (-3,-3.5) [inner sep=0,scale=0.85] {$\boldsymbol{v}_{1,p}[t]$};

\draw[thick, dashed, red] (-3.75,-3.0) rectangle (-2.25,0.5);

}

\newcommand{\FronthaulRegime}[0]{
\draw[->, line width=1mm] (0, 1.1) -- (8, 1.1);
\draw[-, dashed, thick] (0.5,1.9) -- (0.5,0.9) node[below, align=left] {\small $0$};
\draw[-, dashed, thick] (1.5,1.9) -- (1.5,0.9) node[below, align=left] {\small $n_{d2}$};
\draw[-, dashed, thick] (4.5,1.9) -- (4.5,0.9) node[below, align=left] {\small $\max\{n_{d2},n_{d3}\}$};
\draw[<->, thick, red] (0.5, 1.5)--  (1.5, 1.5) node[pos=0.5,sloped,above] {\small \textcolor{red}{Low}};
\draw[<->, thick, YellowOrange] (1.5, 1.5)--  (4.5, 1.5) node[pos=0.5,sloped,above] {\small \textcolor{YellowOrange}{Intermediate}};
\draw[<-, thick, OliveGreen] (4.5, 1.5)--  (8, 1.5) node[pos=0.5,sloped,above] {\small \textcolor{OliveGreen}{High}};
\node at (8.3, 1.1) {\small $n_F$};

\node[squarednode, draw=red, anchor=north west, minimum width=4.0cm,minimum height=0.15cm] (CA_1) at (0.5,0.25) {\footnotesize \textcolor{red}{\textbf{Enlarged} Low}};
\node[squarednode, draw=OliveGreen, anchor=north west, minimum width=3.5cm,minimum height=0.15cm] (CA_3) at (4.5,0.25) {\footnotesize \textcolor{OliveGreen}{High}};

\node[squarednode, draw=red, anchor=north west, minimum width=1cm,minimum height=0.15cm] (CB_1) at (0.5,-0.75) {\footnotesize \textcolor{red}{Low}};
\node[squarednode, draw=YellowOrange, anchor=north west, minimum width=3cm,minimum height=0.15cm] (CB_2) at (1.5,-0.75) {\footnotesize \textcolor{YellowOrange}{Intermediate}};
\node[squarednode, draw=OliveGreen, anchor=north west, minimum width=3.5cm,minimum height=0.15cm] (CB_3) at (4.5,-0.75) {\footnotesize \textcolor{OliveGreen}{High}};

\node[squarednode, draw=red, anchor=north west, minimum width=1cm,minimum height=0.15cm] (CC_1) at (0.5,-1.75) {\footnotesize \textcolor{red}{Low}};
\node[squarednode, draw=OliveGreen, anchor=north west, minimum width=6.5cm,minimum height=0.15cm] (CC_3) at (1.5,-1.75) {\footnotesize \textcolor{OliveGreen}{\textbf{Enlarged} High}};

\node[squarednode, draw=red, anchor=north west, minimum width=7.5cm,minimum height=0.15cm] (CD_1) at (0.5,-2.75) {\footnotesize \textcolor{red}{\textbf{Only} Low}};

\node (CA) at ([yshift=-0.3cm,xshift=-3.90cm]CA_1.north west) {\footnotesize Class \rom{1} (\textit{"High Fronthaul Threshold"})};
\node (CB) at ([yshift=-0.3cm,xshift=-3.65cm]CB_1.north west) {\footnotesize Class \rom{2} (\textit{"Multiple Fronthaul Thresholds"})};
\node (CC) at ([yshift=-0.3cm,xshift=-3.95cm]CC_1.north west) {\footnotesize Class \rom{3} (\textit{"Low Fronthaul Threshold"})};
\node (CD) at ([yshift=-0.3cm,xshift=-4.0cm]CD_1.north west) {\footnotesize Class \rom{4} (\textit{"No Fronthaul Threshold"})};
}

%% file: content/introduction.tex
\section{Introduction}
\label{sec:intro}

In recent years, mobile usage characteristics in wireless networks have changed profoundly from conventional connection-centric (e.g., phone calls) to content-centric (e.g, HD video) behaviors. This shift is mainly driven by the rapid growth in multimedia content, particularly by video \cite{cisco, Bastug}. Over the last decade, however, while the demand for rich
multimedia content has increased tremendously, the capacity of the mobile radio and backhaul network, could not cope at the same pace with the exponentially growing mobile traffic (despite PHY and MAC layer improvement) due to the \emph{centralized} nature of mobile network architectures \cite{Wang12}. As part of standardizing next generation (5G) mobile networks, two major solutions that have great potential to facilitate this shift are content \emph{in-network} caching \cite{Wang14} and multi-tier networks \cite{Andrews13} in the form of \emph{heterogenous} networks. These solutions go hand in hand with the design of more \emph{decentralized} network (HetNet) architectures. 

In-network caching prefetches popular content during off-peak traffic hours in intermediate servers potentially belonging to various hierarchical network layers of the mobile network. Two main places where caches can be deployed are at the core network and/or at the radio access network (RAN) \cite{Wang14}. In this regard, placing caches to the very edge of the network is a RAN-based caching approach which is generally known as \emph{edge caching} and in the small-cell scenario as \emph{femto caching} \cite{Shanmugam13}. It has the advantage that it brings popular content very close to destinations; thus, reducing the usage of expensive backhaul connections from edge nodes (EN) to remote cloud servers and thereby lessening the latency to address the increasing demand in content retrieval. 
Recent trends in the immensely growing number of base stations \cite{Malladi12}, suggest that future networks will be highly heterogeneous in which both small and macro ENs coexist in a HetNet\footnote{Henceforth, we call small and macro ENs as eNB and HeNB (Home eNB).}. Thus, recently research interests have shifted towards the investigation of edge caching of HetNets \cite{Bastug}.    
However, deploying solely cache-based HetNets prevents the eNBs from joint centralized baseband processing; thus, avoiding to some extent advantages of \emph{cooperative} communication strategies \cite{Lozano,Soheil}. Typically, joint processing enables, amongst others, enhanced interference management, flexibility, scalability and power efficiency \cite{Checko}; all of which are key factors in the design of HetNets \cite{Andrews13}.  

Cache and cloud-aided architectures (e.g., Fog radio access networks (F-RAN)) are new hybrid solutions that bring together the advantages of both centralized cloud processing and edge caching \cite{Peng16}. F-RAN is particularly relevant for HetNets. A simplistic HetNet involving both cloud and edge processing is shown in Fig. \ref{fig:HetNet}, which has been first introduced in \cite{Azimi}. We aim at understanding the synergestic benefits of cloud and edge processing for HetNets from a 
delivery time perspective \cite{Liu2011}. To this end, we focus in this work on characterizing the fundamental trade-off on delivery time 
in cloud and cache-aided HetNets of the model as shown in Fig. \ref{fig:HetNet}.

Recently, the impact of caching on the 
delivery time for cache-aided networks has been investigated \cite{Maddah-Ali2,Maddah_Ali,avik,Xu16}. In this regard, receiver (Rx) \cite{Maddah-Ali2, Hachem16, Amiri16, Wan16} and transmitter (Tx) caching \cite{Maddah_Ali,avik, Liu15} as well as simultaneous Tx/Rx caching \cite{Xu16,Naderializadeh17} offer great potential for reducing the induced delivery time for file retrieval. 
Rx caching, on the one hand, was first studied in \cite{Maddah-Ali2} for a shared link with one server and multiple cache-enabled receivers. The authors show that caching can exploit multicast opportunities and as such significantly reduces the delivery latency over the shared link. On the other hand, the impact of Tx caching on the delivery time has mainly been investigated by analyzing the inverse degrees-of-freedom (DoF) metric for Gaussian networks. 
To this end, the authors of \cite{Maddah_Ali} developed a novel interference alignment achievability scheme characterizing the metric as a function of the cache storage capability for a 3-user Gaussian interference network. The cache placement was designed to facilitate transmitter cooperation such that interference coordination techniques can be applied. A converse on this metric was developed in \cite{avik} for a network with arbitrary number of edge nodes and users showing the optimality of schemes presented in \cite{Maddah_Ali} for certain regimes of cache sizes. Extensions of this work include the characterization of the latency-memory tradeoff to cloud and cache-assisted F-RAN \cite{Tandon, Azimi}. 
Tx-caching through cache-enabled helpers can also significantly reduce the delivery time than a system without cachnig as was shown in \cite{SoheilTWC} for a MIMO broadcast channel. 
As opposed to \cite{Maddah_Ali, avik, Tandon}, authors of \cite{Azimi} modeled the wireless channel by a binary fading channel \cite{Vahid17}. Through this simplification, their results gave first insight on the delivery time as a function of the cache size and \emph{probabilistic} parameters on the wireless channel. Related papers that study the influence of cloud and edge processing on achievable rates, backhaul costs and power consumption for non-uniform file requests are, amongst others, \cite{Yigit, Park, Tao16, Chen16}.

In this paper, we study the fundamental limits on the delivery time 
for the cloud and cache-aided HetNet introduced in \cite{Azimi} and shown in Fig. \ref{fig:HetNet} which consists of two eNBs and two users. For this network, two distinct cloud-edge transmission policies are feasible. In the first policy, the so-called \emph{serial} policy, the cloud transmission terminates before the wireless transmission initiates, whereas in the second policy, the so-called \emph{parallel} policy, cloud and wireless transmissions are executed simultaneously. For both policies, we measure the performance through the latency-centric metric \emph{delivery time per bit} (DTB) \cite{avik}. As opposed to the results established in \cite{Azimi}, where a binary fading channel is used to determine the DTB for serial transmission only, we instead use the linear deterministic model (LDM) \cite{Avestimehr, Anas} as a model that comes closest to the Gaussian system model of the proposed network. 
Under this particular channel which takes into account distinct channel strengths, we characterize the DTB of the network through means of lower bounds (converse) and upper bounds (achievability) on the DTB. 
For serial and parallel transmissions, we establish five lower bounds that utilize five distinct combinations of cached information, fronthaul and wireless signals that enable reliable decoding of either one or two requested files of the users at any arbitrary decoder, respectively. 
All bounds are tight and required for the characterization of the DTB. As far as the achievability is concerned, we propose a generalized scheme that is based on rate splitting of private, common and interference-neutralizing information. We formulate a linear \emph{max-min} optimization problem to maximize the least number of desired bits conveyable to the users per block of channel uses. Hereby, the least number of desired bits conveyable corresponds to the \emph{per-user rate} of the \emph{weaker} user. Thus, \emph{maximizing} this quantity optimizes the weaker users' rate. This optimization problem intrinsically captures the importance of treating each user equally from a rate perspective. Since the per-user rate of the weaker user is inverse proportional to the DTB, 
we are able to minimize the achievable DTB and determine the optimal rate splitting which achieves the lower bound. To the best of the author's knowledge, existing work \cite{Maddah_Ali,avik,Hachem16,Tandon} treat all channel links as equally strong and use the inverse DoF as their 
delivery time metric. However, these results do not capture the inherent 
dependency of the time for file delivery on the wireless channel strength. Instead, apart from cloud and cache parameters, our DTB metric also captures the influence of channel strength on the latency. 
Through the additional perspective on channel strength, we are able to broadly identify channel regimes for which edge caching and cloud processing can provide nontrivial synergestic and non-synergestic performance gains.

The rest of the paper is organized as follows.  In  Section \ref{sec:Sym_Model}, we  introduce the system model. The main results on the DTB, including achievability and converse, for serial and parallel cloud-edge transmissions are  presented in sections \ref{sec:ser_trans} through \ref{sec:ser_trans_ub} and \ref{sec:par_trans} through \ref{sec:par_upp_bound}, respectively. Finally, Section \ref{sec:conclusion} concludes the paper. The appendix of this paper is
devoted to give further details on lower and upper bounds.

\textbf{Notation:} Throughout the paper, we use $\mathbb{F}_2$ to denote the binary field and $\oplus$ to denote the modulo 2 addition. We use normal lower-case, normal upper-case, boldface lower-case, and boldface upper-case letters to denote scalars, scalar random variables, vectors, and matrices, respectively. $\text{Bern}(a)$ is a Bernoulli distribution with probability $a$. The vector $\mathbf{0}_q$ denotes the zero-vector of length $q$ and the matrix $I_q$ is the $q\times q$ identity matrix. We use the superscript $(\cdot)^{T}$ to represent the transpose of a matrix. Furthermore, for any two integers $a$ and $b$ with $a\leq b$, we define $[a:b]\triangleq\{a,a+1,\ldots,b\}$. When $a=1$, we simply write $[b]$ for $\{1,\ldots,b\}$. Furthermore, we define the function $(x)^{+}\triangleq\max\{0,x\}$.

%% file: content/system_model.tex
\section{System Model and Latency Metric}
\label{sec:Sym_Model}
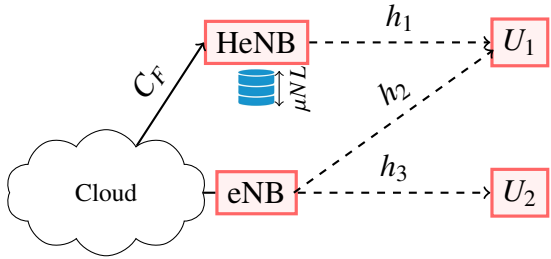
\begin{figure}[t]
\begin{center}
        \begin{tikzpicture}[scale=0.8]
		\SymMod
		\end{tikzpicture}
	\caption{\small System Model of Cache- and Cloud-aided HetNet}	
	\label{fig:HetNet}
	\end{center}
\end{figure}

In  this  section,  we  first  outline the system model for the cloud and  cache-aided  F-RAN in Fig. \ref{fig:HetNet}. Then, we introduce the delivery time per bit (DTB) metric, along with its operational meaning to provide additional context on the adopted model and performance metric. In sub-section \ref{subsec:sym_model_LDM}, we introduce the linear-determinstic model (LDM) as an approximation method of the Gaussian system model of sub-section \ref{subsec:sym_model_gauss} and further alter the DTB metric of sub-section \ref{subsec:sym_model_LDM} to make it applicable for the LDM.    

\subsection{System Model}
\label{subsec:sym_model_gauss}
We study the downlink of a cache and cloud-aided HetNet as shown in Fig. \ref{fig:HetNet}. The HetNet consists of a HeNB and a macro eNB which serves two users -- a small-cell user ($U_1$) and a macro cell user ($U_2$) -- over a wireless channel. As the HeNB transmits at much lower power than the eNB, we model the wireless channel by a partially connected network in the spirit of a Z-channel. At every transmission interval $j$, both users request any file $W_i$, all of which are of $L$ bits in size, from a library of $N$ popular files. The request pattern of both users is revealed to the cloud and the edge nodes (eNB and HeNB) prior to the transmission which seek to satisfy the users' demands at the lowest possible latency. Hereby, cloud and edge transmission can, amongst others, be carried out in two ways: (a) \emph{serial/sequential} transmission, where the cloud transmission over the fronthaul link terminates before the edge transmission over the wireless channel initiates; (b) \emph{parallel/pipelined} transmission, where cloud and edge transmission are executed simultaneously in which the HeNB operates as a causal full-duplex cache-aided relay (cf. Fig. \ref{fig:Lat}). The transmission scheme of interval $j$ terminates when the requested files have been delivered. This induces a total 
delivery time consisting of the sum of fronthaul and edge latencies ($T_{F}[j]+T_{E}[j]$) for (a) serial and $T_{P}[j]$ for (b) parallel transmissions.     
The system model, notation and main assumptions are summarized as follows:

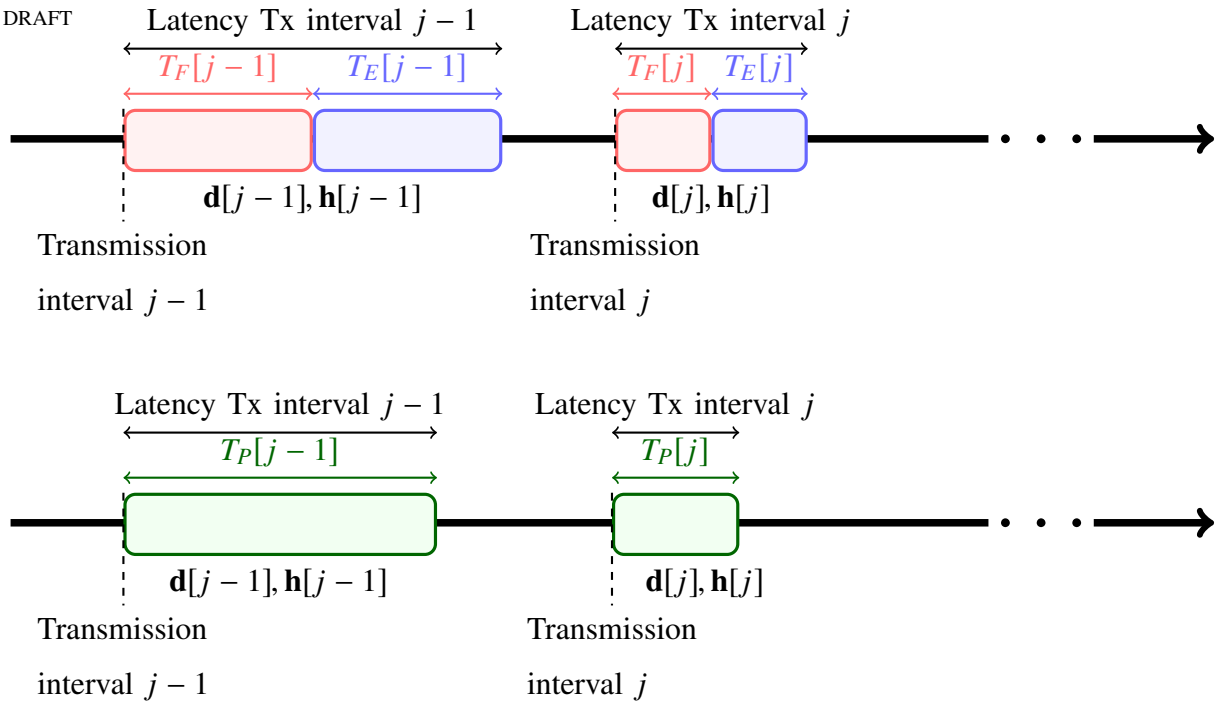
\begin{figure*}[t]
    \centering
    	\begin{subfigure}[t]{1\textwidth}
		\begin{tikzpicture}[scale=1]
			\Time
		\end{tikzpicture}
    	\end{subfigure}%
    	\newline
    	\vfill
    	\begin{subfigure}[t]{1\textwidth}
		\begin{tikzpicture}[scale=1]
			\TimePar
		\end{tikzpicture}
    	\end{subfigure}
\caption{\small Illustration of the delivery latency within each transmission interval for the F-RAN of Fig. \ref{fig:HetNet} with (a) serial and (b) parallel fronthaul-edge transmission.}
\label{fig:Lat}
\end{figure*} 
%

\begin{itemize}
\item Let $\mathcal{W}=\{W_1,\ldots,W_{N}\}$ denote the library of popular files, where each file $W_i$ is of size $L$ bits. Each file $W_i$ is chosen uniformly at random from $[2^{L}]\triangleq\{1,2,\ldots,2^{L}\}$. We define the demand vector $\mathbf{d}[j]=(d_1[j],d_2[j])^{T}\in[N]^{2}$ to denote the request pattern of both users. Thus, at every transmission interval $j$, $U_1$ and $U_2$ request files $W_{d_1[j]}$ and $W_{d_2[j]}$ from the library $\mathcal{W}$, respectively. 
\item The HeNB is endowed with a cache capable of storing $\mu NL$ bits, where $\mu\in[0,1]$ corresponds to the \emph{fractional cache size}. It denotes how much content can be stored at the HeNB relatively to the entire library $\mathcal{W}$.
\item The cloud server has access to all $N$ files. The cloud server and the eNB are co-located whereas the HeNB is connected to the cloud via a fronthaul link of fixed capacity $C_{F}$.
For $C_F= 0$, we explicitly assume that the link from HeNB-to-cloud is absent only in the file delivery phase; and, the user can still prefetch some content in its local cache in the off-peak hours. In doing so, the HeNB may still have some useful contents even though the link between HeNB and cloud may not be present while the files ought to be delivered.
\item Global channel state information (CSI) at transmission interval $j$ is summarized by the channel vector $\mathbf{h}[j]=(h_1[j],h_2[j],h_3[j])^{T}\in\mathbb{C}^{3}$, where $h_1[j],h_2[j]$ and $h_3[j]$ represent the complex channel coefficients from HeNB and eNB to $U_1$ and $U_2$, respectively. Hereby, the channel coefficients in $\mathbf{h}[j]$ are constant over a single transmission interval $j$. The channel coefficients are assumed to be drawn i.i.d. from a continuous distribution.  
\end{itemize} 
Communication over the wireless channel occurs in two consecutive phases, \emph{placement phase} followed by the \emph{delivery phase}.  
In the following, we will describe the modeling of placement and delivery phase for both serial and parallel fronthaul-edge transmission. 

\begin{enumerate}
\item \vspace{.5em} \emph{Placement phase}: During this phase, the HeNB is given full access to the database of $N$ files. \emph{Irrespective} of serial and parallel fronthaul-edge transmission, the cached content at the HeNB is generated through its caching function. 
\vspace{.5em}
\begin{definition}[Caching function]\label{def_cache_fct} The HeNB maps each file $W_i\in\mathcal{W}$ to its local \emph{file cache content} 
\begin{equation}
S_{i}=\phi_{i}(W_{i}),\qquad\forall i=1,\ldots,N\nonumber.
\end{equation} All $S_{i}$ are concatenated to form the total cache content 
\begin{equation}
S=(S_{1},S_{2},\ldots,S_{N})\nonumber
\end{equation} at the HeNB. Hereby, due to the assumption of symmetry in caching, the entropy $H(S_{i})$ of each component $S_i$, $i=1,\ldots,N$, is upper bounded by $\nicefrac{\mu NL}{N}=\mu L$. 
The definition of the caching function presumes that every file $W_i$ is subjected to \emph{individual} caching functions. Thus, permissible caching policies allow for intra-file coding but avoid inter-file coding. Moreover, the caching policy is kept fixed over multiple transmission intervals. Thus, it is indifferent to the user's request pattern and of channel realizations. \end{definition}
\item \vspace{.5em} \emph{Delivery phase}: In this phase, a fronthaul and edge transmission policy at the cloud as well as at HeNB and eNB 
is applied in each transmission interval $j$ to satisfy the given user's requests $\mathbf{d}[j]$ under the current channel realizations $\mathbf{h}[j]$. In the sequel, we will focus on a \emph{single} transmission interval and therefore omit indexing the transmission interval explicitly. Hereafter, the main differences between serial and parallel fronthaul-edge transmission are outlined.  
\vspace{.5em}
\begin{definition}[Encoding functions]\label{def_enc_fct} For $C_{F}\geq 0$, the cloud encoding function 
\begin{equation}\label{eq:cloud_enc}
\psi_{C}:\begin{cases}&[2^{NL}]\times [N]^{2}\times\mathbb{C}^{3}\rightarrow \mathbb{C}^{T_F}\text{ for serial transmission}\\&[2^{NL}]\times [N]^{2}\times\mathbb{C}^{3}\rightarrow \mathbb{C}^{T_P}\text{ for parallel transmission}\end{cases}
\end{equation} determines the fronthaul messages $\mathbf{x}_{F}^{T}$ as a function of $\mathcal{W},\mathbf{d}$ and $\mathbf{h}$. Depending on the edge-fronthaul transmission, these messages are transmitted in $T\in\{T_F,T_P\}$ channel uses. Note that in $T$ channel uses, the fronthaul message cannot exceed $TC_{F}$ bits. 
The encoding function of the HeNB for \emph{serial} fronthaul-edge transmission on the one hand is defined by
\begin{equation}\label{eq:henb_enc_ser}
\psi_{1}:[2^{\mu NL}]\times \mathbb{C}^{T_F}\times [N]^{2}\times\mathbb{C}^{3}\rightarrow \mathbb{C}^{T_E}. 
\end{equation} 
This encoding function $\psi_{1}$ maps the cached content $S$, the demand vector $\mathbf{d}$, the fronthaul messages $\mathbf{x}_{F}^{T_F}$ and global CSI given by $\mathbf{h}$ to the codeword $\mathbf{x}_{1}^{T_E}$ of $T_{E}$ channel uses in duration while satisfying the average power constraint given by the parameter $P$. On the other hand, to account for parallel fronthaul-edge transmission, we modify \eqref{eq:henb_enc_ser} as follows:   
\begin{equation}\label{eq:henb_enc_par}
\psi_{1}^{[t]}:[2^{\mu NL}]\times \mathbb{C}^{t-1}\times [N]^{2}\times\mathbb{C}^{3}\rightarrow \mathbb{C},\quad t\in[T_p] 
\end{equation}
For any time instant $t$, \eqref{eq:henb_enc_par} accounts for the simultaneous reception and transmission through fronthaul and wireless links at the HeNB. More precisely, the transmitted signal at the $t-$th channel use is generated by $\mathbf{x}_{1}[t]=\psi_{1}^{[t]}\big(S,\mathbf{x}_{F}^{[t-1]},\mathbf{d},\mathbf{h}\big)$.  
Similarly to \eqref{eq:henb_enc_ser} and \eqref{eq:henb_enc_par}, the macro eNB uses the encoding function \begin{equation}
\psi_{2}:\begin{cases}&[2^{NL}]\times [N]^{2}\times\mathbb{C}^{3}\rightarrow \mathbb{C}^{T_E}\text{ for serial transmission}\\&[2^{NL}]\times [N]^{2}\times\mathbb{C}^{3}\rightarrow \mathbb{C}^{T_P}\text{ for parallel transmission}\end{cases}
\end{equation} to construct the codeword $\mathbf{x}_{2}^{T}=\psi_{2}(\mathcal{W},\mathbf{d},\mathbf{h})$ for $T\in\{T_{E},T_{P}\}$ subjected to an average power constraint of $P$.  
\end{definition}
\end{enumerate}
\vspace{.5em}
\begin{definition}[Decoding function]\label{def_dec_fct} The decoding operation at $U_k$, $k\in\{1,2\}$, follows the mapping
\begin{equation}
\eta_{k}:\begin{cases}&\mathbb{C}^{T_E}\times [N]^{2}\times\mathbb{C}^{3}\rightarrow [2^{L}]\text{ for serial transmission}\\&\mathbb{C}^{T_P}\times [N]^{2}\times\mathbb{C}^{3}\rightarrow [2^{L}]\text{ for parallel transmission}\end{cases}. 
\end{equation} 
The decoding function $\eta_k$ takes as its arguments $\mathbf{h}$, the available demand pattern $\mathbf{d}$ and the channel outputs $\mathbf{y}_{k}^{T}$ for $T\in\{T_{E},T_{P}\}$ given by
\begin{equation}\label{eq:Gaus_mod}
\mathbf{y}_k^{T}=\begin{cases}
h_{1}\mathbf{x}_{1}^{T}+h_{2}\mathbf{x}_{2}^{T}+\mathbf{z}_{1}^{T}&\text{ for }k=1\\
h_{3}\mathbf{x}_{2}^{T}+\mathbf{z}_{2}^{T}&\text{ for } k=2
\end{cases}
\end{equation} to provide an estimate $\hat{W}_{d_k}=\eta_{k}\big(\mathbf{y}_k^{T},\mathbf{d},\mathbf{h}\big)$ of the requested file $W_{d_k}$. $\mathbf{z}_{k}^{T}$ denotes complex Gaussian noise of zero mean and unit power which is i.i.d. across time and users. \end{definition} 
A proper choice of a caching, encoding and decoding function that satisfies the reliability condition; that is the worst-case error probability \begin{equation}\label{eq:error_prob}
P_e=\max_{\mathbf{d}\in [N]^{2}}\max_{k\in\{1,2\}}\mathbb{P}(\hat{W}_{d_k}\neq W_{d_k}) 
\end{equation} approaches $0$ as $L\rightarrow\infty$, is called a \emph{feasible policy}. 
\subsection{Latency Metric: Delivery Time per Bit}
\label{subsec:lat_metric_DTB}
We next define the proposed performance metric of delivery time per bit (DTB) for a serial and pipelined fronthaul-edge transmission scheme. 
\begin{definition}[Delivery time per bit] The DTB for $\mathbf{d}$ and $\mathbf{h}$ is defined as \begin{equation}\label{eq:DTB}
\Delta(\mu, C_F, \mathbf{h}, P)=\max_{\mathbf{d}\in [N]^{2}}\limsup_{L\rightarrow\infty}\begin{cases}&\frac{T_{F}(\mathbf{d},\mathbf{h})+T_{E}(\mathbf{d},\mathbf{h})}{L}\text{ for serial transmission}\\&\frac{T_{P}(\mathbf{d},\mathbf{h})}{L}\qquad\:\:\:\:\text{ for parallel transmission}\end{cases}.
\end{equation} 
The minimum DTB $\Delta^{*}$ is the infimum of the DTB of all achievable schemes.
\end{definition}  
\begin{remark}
The DTB measures the overall per-bit latency within a transmission interval for the worst-case request pattern of $U_1$ and $U_2$. In case of serial transmission, the latency consists of the fronthaul latency incurred from cloud to HeNB and the latency of the wireless channel from (H)eNB to the users. Note that parallel transmission naturally outperforms serial transmission. This is mainly because the HeNB operates as a cache-aided full-duplex relay such that schemes utilizing serial transmission are special cases of the more general parallel transmission.    
\end{remark}
\begin{remark}[Cache-Only F-RAN and Cloud-Only F-RAN] When establishing converse and achievability for both serial and parallel transmission, we differentiate between \emph{cache-only F-RAN} and \emph{cloud-only F-RAN}. The former refers to the case when the fronthaul link is absent during the delivery phase ($C_{F}=0$). The latter, on the other hand, encompasses a C-RAN system, i.e., a radio network without edge caching capabilities ($\mu=0$). For the network unter study, the special case when $\mu=0$, $C_{F}=0$ reduces to a (Gaussian) broadcast channel \cite{Cover_2006}. This is because there is  neither local information stored in the HeNB's cache nor can useful information on requested files be made available through the cloud-to-HeNB link. In one of the following lemmas to come (Lemma \ref{lemma:2}), we will specify the optimal DTB for this broadcast setting. 

\end{remark}
\subsection{Linear-Determinstic Model}
\label{subsec:sym_model_LDM}
To gain insight into the DTB for the Gaussian system model, we suggest to approximate \eqref{eq:Gaus_mod} by the linear-deterministic model (LDM) \cite{Avestimehr}. 
In the LDM, an input symbol at the (H)eNB is given by the binary input vector $\mathbf{x}_k\in\mathbb{F}_2^{q}$ where $q=\max\{n_{d1},n_{d2},n_{d3}\}$. The integers $n_{dm}\in\mathbb{N}_{0}^{+}$, $m\in\{1,2,3\}$, given by  
\begin{equation}\label{eq:nd}
n_{dm}=\ceil{\log\big(P|h_m|^{2}\big)}
\end{equation} approximate the number of bits which can be communicated over each link reliably. Similarly to the parameters $n_{dm}$, we define $n_{F}=\ceil{C_{F}}$ in the LDM. The channel output symbols $\mathbf{y}_k^{T}$ received in $T\in\{T_{E},T_{P}\}$ channel uses (depending on whether the system operates under serial or parallel transmission) at $U_k$ is given by a deterministic function of the inputs; that is
\begin{equation}\label{eq:LDM_mod}
\mathbf{y}_k^{T}=\begin{cases}
\mathbf{S}^{q-n_{d1}}\mathbf{x}_{1}^{T}\oplus\mathbf{S}^{q-n_{d2}}\mathbf{x}_{2}^{T}&\text{ for }k=1\\\mathbf{S}^{q-n_{d3}}\mathbf{x}_{2}^{T}&\text{ for } k=2
\end{cases}, 
\end{equation} where $\mathbf{S}\in\mathbb{F}_2^{q\times q}$ is a down-shift matrix defined by \begin{equation}\label{eq:shift_mat}
\mathbf{S}=\begin{pmatrix}
\mathbf{0}_{q-1}^{T} & 0 \\ \mathbf{I}_{q-1} & \mathbf{0}_{q-1}
\end{pmatrix}.
\end{equation} The input-output equation \eqref{eq:LDM_mod} approximates the input-output equation of the Gaussian channel given in \eqref{eq:Gaus_mod} in the high SNR regime. A graphical representation of the transmitted and received binary vectors $\mathbf{x}_k[t]$ and $\mathbf{y}_k[t]$, $k\in\{1,2\}$, in the $t-$th channel use is shown in Fig. 3(a). Each circle in the figure represents a signal level which holds a binary digit for transmission. For any link between HeNB/eNB to $U_1$ or $U_2$, only the most $\mathbf{n}=(n_{d1},n_{d2},n_{d3})^{T}$ significant bits are received at the destinations while less significant bits are not. In analogy to \eqref{eq:DTB}, we denote the DTB for the LDM by $\Delta_{\text{det}}(\mu,n_{F},\mathbf{n})$. The remainder of this paper focuses on characterizing the DTB on the basis of the LDM. 

We show next that irrespective of the type of fronthaul-edge transmission the DTB is convex in the fractional cache size $\mu$ for any given value of fronthaul capacity $n_{F}$ and wireless channel parameters $\mathbf{n}$. This is proven through a memory-sharing argument that is based on splitting a file into two distinct fractions and applying different fronthaul-edge policies to each of it. Due to the additivity in delivery time, the overall DTB becomes the sum of individual DTB's obtained on each file fraction.  
\vspace{.5em}
\begin{lemma}[Convexity of Minimum DTB]\label{lemma:1}
The minimum DTB $\Delta^{*}_{\text{det}}$ is a convex function of $\mu\in[0,1]$ for any given value $n_{F}\geq 0$ and $\mathbf{n}\in\big(\mathbb{N}_{0}^{+}\big)^{3}$.
\end{lemma}
\noindent\begin{IEEEproof}
For any given $n_{F}\geq 0$ and $\mathbf{n}\in\big(\mathbb{N}_{0}^{+}\big)^{3}$, consider two \emph{feasible} policies that require fractional cache sizes $\mu_1$ and $\mu_2$ that achieve a minimum (optimal) DTB of 
$\Delta_{\text{det}}^{*}(\mu_1,n_{F},\mathbf{n})$ and $\Delta_{\text{det}}^{*}(\mu_2,n_{F},\mathbf{n})$, respectively. 
At a fractional cache size $\mu=\alpha\mu_1+(1-\alpha)\mu_2$ for any $\alpha\in[0,1]$, the system can apply \emph{file splitting} into subfiles; the first subfile being of size $\alpha L$ and the second being of size $(1-\alpha)L$. Note that this strategy is in agreement with the cache constraints. Using the first policy on the first subfile and the second policy on the second subfile consecutively through \emph{time sharing} achieves a DTB equal to the convex 
combination $\alpha\Delta_{\text{det}}^{*}(\mu_1,n_{F},\mathbf{n})+(1-\alpha)\Delta_{\text{det}}^{*}(\mu_2,n_{F},\mathbf{n})$. This achievable DTB is at best as low as the minimal DTB $\Delta^{*}_{\text{det}}(\mu,n_{F},\mathbf{n})$ at fractional cache size $\mu$. Thus, the convex combination $\alpha\Delta_{\text{det}}^{*}(\mu_1,n_{F},\mathbf{n})+(1-\alpha)\Delta_{\text{det}}^{*}(\mu_2,n_{F},\mathbf{n})$ functions as an upper bound on $\Delta^{*}_{\text{det}}(\mu,n_{F},\mathbf{n})$.
\end{IEEEproof}
\vspace{.5em}
\begin{lemma}[Optimal DTB of the Broadcast Channel]\label{lemma:2}
For the LDM-based cloud and cache-aided HetNet in Fig. \ref{fig:HetNet} with $n_{F}=0$ and $\mu=0$, the optimal DTB is given by \begin{equation}\label{eq:DTB_lemma}
\Delta_0(\mathbf{n})\triangleq\Delta_{\text{det}}^{*}(\mu=0,n_F=0,\mathbf{n})=\max\bigg\{\frac{2}{\max\{n_{d2},n_{d3}\}},\frac{1}{n_{d2}},\frac{1}{n_{d3}}\bigg\}.
\end{equation}
\end{lemma}
\begin{IEEEproof} 
For $\mu=0$ and $n_{F}=0$, the HeNB has no relevant information on the requested files $W_{d1}$ and $W_{d2}$. Thus, the eNB is involved in broadcasting files $W_{d1}$ and $W_{d2}$ to $U_1$ and $U_2$ while the HeNB remains silent. In the LDM, this is equivalent to $n_{d1}=0$. For a feasible scheme, either user can reliably decode $W_{dk}$ ($W_{d1}$ and $W_{d2}$) if it is aware of $\mathbf{y}_{k}^{T}$ ($\mathbf{y}_{1}^{T}$ and $\mathbf{y}_{2}^{T}$) for $T\in\{T_E,T_P\}$. These observations can be used to generate lower bounds on the DTB $\Delta_{\text{det}}^{*}(\mu=0)$ that correspond to \eqref{eq:DTB_lemma}. Since the requested files are of the same size, an optimal scheme that minimizes the latency would try to split the transmission load equally to the two users. Depending on the channel conditions, the load balancing is done as follows. For instance, when $n_{d2}\geq n_{d3}$, $\nicefrac{n_{d2}}{2}$ bits can be send in one channel use from the eNB to both $U_1$ and $U_2$, if the weaker channel $n_{d3}$ is stronger than $\nicefrac{n_{d2}}{2}$ ($\nicefrac{n_{d2}}{2}\leq n_{d3}$). On the other hand, if $n_{d2}\geq n_{d3}$ and $\nicefrac{n_{d2}}{2}\geq n_{d3}$, the latency is governed by the weaker channel and $n_{d3}$ bits are transmitted in one channel use to each user. In summary, for $n_{d2}\geq n_{d3}$ we can reliably convey $\min\{\nicefrac{n_{d2}}{2},n_{d3}\}$ bits per channel use to each user, or in other words, $\Delta^{*}_{\text{det}}(\mu=0)=\max\{\nicefrac{2}{n_{d2}},\nicefrac{1}{n_{d3}}\}$ channel uses are needed to provide each user with one bit. Due to symmetry, a similar observation holds for $n_{d2}\leq n_{d3}$. This concludes the proof.\end{IEEEproof}
\begin{remark}[Broadcast Conditions]\label{remark_bc_conditions} One can verify from Eq. \ref{eq:DTB_lemma} of Lemma \ref{lemma:2} that the optimal DTB corresponds to 
\begin{equation}
\Delta_{0}(\mathbf{n})=\begin{cases}\frac{2}{n_{d2}}&\text{ if }\mathbf{n}\in\mathcal{I}_0\triangleq\{2n_{d3}\geq n_{d2}\geq n_{d3}\}\\\frac{1}{n_{d3}}&\text{ if }\mathbf{n}\in\mathcal{I}_0^{C}\triangleq\{n_{d2}\geq 2n_{d3}\}\\\frac{2}{n_{d3}}&\text{ if }\mathbf{n}\in\mathcal{I}_1\triangleq\{2n_{d2}\geq n_{d3}\geq n_{d2}\}\\\frac{1}{n_{d2}}&\text{ if }\mathbf{n}\in\mathcal{I}_1^{C}\triangleq\{n_{d3}\geq 2n_{d2}\}\end{cases}.
\end{equation}
Hereby, $\mathcal{I}_0$, $\mathcal{I}_0^{C}$, $\mathcal{I}_1$ and $\mathcal{I}_1^{C}$ are mutually exclusive broadcast channel regimes, or broadcast conditions, for which the optimal DTB are distinct. Throughout this paper, we refer to $\Delta_{0}(\mathbf{n})$ as the \emph{broadcast DTB}.  
\end{remark}

%% file: content/serial.tex
\section{Serial Transmission -- Main Result}
\label{sec:ser_trans}
\newcommand\NoIndent[1]{%
  \par\vbox{\parbox[t]{\linewidth}{#1}}%
}
In  this  section, we state our main result on the minimum DTB for the F-RAN in Fig. \ref{fig:HetNet} for a serial fronthaul-edge transmission. Hereby, our result is a complete DTB characterization stated in the following theorem.
\begin{theorem}\label{th:1}
The DTB of the LDM-based cloud and cache-aided HetNet in Fig. \ref{fig:HetNet} for any $n_{F}\geq 0$ and $\mu\in[0,1]$ is given by 
\begin{equation}\label{eq:th_1}
\Delta_{\text{det}}^{*}(\mu,n_{F},\mathbf{n})=\begin{cases}\max\Big\{\frac{1-\mu}{n_{d2}},\frac{2-\mu}{\max\{n_{d2},n_{d3}\}},\Delta^{\prime}_{\text{LB}}(\mathbf{n})\Big\}\qquad&\text{for }n_{F}\leq n_{d2}\\\max\Big\{\frac{1-\mu}{n_F}+\big(1-\frac{n_{d2}}{n_F}\big)\Delta_{\text{LB}}^{\prime\prime}(\mathbf{n}),\frac{2-\mu}{\max\{n_{d2},n_{d3}\}},\Delta^{\prime}_{\text{LB}}(\mathbf{n})\Big\}\qquad&\text{for }n_{d2}\leq n_{F}\leq \max\{n_{d2},n_{d3}\}\\\max\begin{Bmatrix}\frac{2-\mu}{n_F}+\Big(1-\frac{\max\{n_{d2},n_{d3}\}}{n_{F}}\Big)\Delta_{\text{LB}}^{\prime}(\mathbf{n}),\\ \frac{1-\mu}{n_{F}}+\Big(1-\frac{n_{d2}}{n_{F}}\Big)\Delta_{\text{LB}}^{\prime}(\mathbf{n}),\\\Delta_{\text{LB}}^{\prime}(\mathbf{n})\end{Bmatrix}\qquad&\text{for }n_{F}\geq \max\{n_{d2},n_{d3}\}\end{cases},
\end{equation} where
\begin{equation}\label{eq:dlb_main}  
\Delta_{\text{LB}}^{\prime}(\mathbf{n})=\max\Bigg\{\frac{1}{n_{d3}},\frac{1}{\max\{n_{d1},n_{d2}\}}, \frac{2}{\max\{n_{d1}+n_{d3},n_{d2}\}}\Bigg\}
\end{equation} and 
\begin{equation}\label{eq:dlb_main_prime_prime}  
  \Delta_{\text{LB}}^{\prime\prime}(\mathbf{n})=\max\Bigg\{\frac{1}{n_{d3}-n_{d2}},\Delta_{\text{LB}}^{\prime}(\mathbf{n})\Bigg\}.
\end{equation}
\end{theorem}
\noindent \begin{IEEEproof} (Theorem \ref{th:1}) Lower (converse) and upper bounds (achievability) on the DTB are derived for the case of active and inactive ($n_{F}=0$) fronthaul links. Specifically, for the case of inactive fronthauling, we provide lower and upper bounds on the DTB in sub-sections \ref{subsec:ser_low_bound} and \ref{subsec:ser_upp_bound} through \ref{sec:optimizing_ser}, respectively, whereas we relegate the reader to sub-sections \ref{subsec:ser_low_bound_nF_greater_0} and \ref{subsec:ser_upp_bound_nF_greater_0}, respectively, for lower and upper bounds in case of active fronthauling.  
\end{IEEEproof}
\vspace{.5em}
We extract from Theorem \ref{th:1} that the behavior in DTB in terms of the fronthaul capacity $n_F$ changes in the intervals $[0,n_{d2}]$, $[n_{d2},\max\{n_{d2},n_{d3}\}]$ and $[\max\{n_{d2},n_{d3}\},\infty)$. The change in behavior occurs at cut-off/threshold fronthaul capacities $n_{d2}$ and $\max\{n_{d2},n_{d3}\}$. As shown in Fig. \ref{fig:distinction_classes}, there are in total four mutually exclusive \emph{classes of channel regimes} for which the number and/or the magnitude of the threshold fronthaul capacities differ. We term these classes as Class \rom{1}, \rom{2}, \rom{3} and \rom{4} for which the optimal DTB of Theorem \ref{th:1} simplifies to:
\begin{itemize}
\item Class \rom{1} $\Big(\text{if }\mathbf{n}\in \Big\{\mathcal{I}_0\cap\{n_{d1}+n_{d3}\geq n_{d2}\}\Big\}\cup\mathcal{I}_1\Big)$:
\begin{equation}\label{eq:th_1_Class_A}
\Delta_{\text{det}}^{*}(\mu,n_{F},\mathbf{n})=\begin{cases}\max\Big\{\frac{2-\mu}{\max\{n_{d2},n_{d3}\}},\Delta^{\prime}_{\text{LB}}(\mathbf{n})\Big\}\qquad&\text{for }n_{F}\leq \max\{n_{d2},n_{d3}\}\\\max\Big\{\frac{2-\mu}{n_F}+\Big(1-\frac{\max\{n_{d2},n_{d3}\}}{n_{F}}\Big)\Delta_{\text{LB}}^{\prime}(\mathbf{n}),\Delta_{\text{LB}}^{\prime}(\mathbf{n})\Big\}\qquad&\text{for }n_{F}\geq \max\{n_{d2},n_{d3}\}\end{cases},
\end{equation}
\item Class \rom{2} $\Big(\text{if }\mathbf{n}\in \Big\{\mathcal{I}_1^{C}\cap\big\{\frac{1}{n_{d3}-n_{d2}}\geq\Delta_{\text{LB}}^{\prime}(\mathbf{n})\big\}\cap\big\{\frac{1}{n_{d2}}\geq\Delta_{\text{LB}}^{\prime}(\mathbf{n})\big\}\Big\}\Big)$:
\begin{equation}\label{eq:th_1_Class_B}
\Delta_{\text{det}}^{*}(\mu,n_{F},\mathbf{n})=\begin{cases}\max\Big\{\frac{1-\mu}{n_{d2}},\frac{2-\mu}{n_{d3}},\Delta^{\prime}_{\text{LB}}(\mathbf{n})\Big\}\qquad&\text{for }n_{F}\leq n_{d2}\\\max\Big\{\frac{1-\mu}{n_F}+\big(1-\frac{n_{d2}}{n_F}\big)\Delta_{\text{LB}}^{\prime\prime}(\mathbf{n}),\frac{2-\mu}{n_{d3}},\Delta^{\prime}_{\text{LB}}(\mathbf{n})\Big\}\qquad&\text{for }n_{d2}\leq n_{F}\leq n_{d3}\\\max\Big\{\frac{2-\mu}{n_F}+\Big(1-\frac{n_{d3}}{n_{F}}\Big)\Delta_{\text{LB}}^{\prime}(\mathbf{n}),\Delta_{\text{LB}}^{\prime}(\mathbf{n})\Big\}\qquad&\text{for }n_{F}\geq n_{d3}\end{cases},
\end{equation}
\item Class \rom{3} $\Big(\text{if }\mathbf{n}\in \Big\{\mathcal{I}_1^{C}\cap\big\{\frac{1}{n_{d3}-n_{d2}}\leq\Delta_{\text{LB}}^{\prime}(\mathbf{n})\big\}\cap\big\{\frac{1}{n_{d2}}\geq\Delta_{\text{LB}}^{\prime}(\mathbf{n})\big\}\Big\}\Big)$:
\begin{equation}\label{eq:th_1_Class_C}
\Delta_{\text{det}}^{*}(\mu,n_{F},\mathbf{n})=\begin{cases}\max\Big\{\frac{1-\mu}{n_{d2}},\Delta^{\prime}_{\text{LB}}(\mathbf{n})\Big\}\qquad&\text{for }n_{F}\leq n_{d2}\\\max\Big\{\frac{1-\mu}{n_F}+\big(1-\frac{n_{d2}}{n_F}\big)\Delta_{\text{LB}}^{\prime}(\mathbf{n}),\Delta^{\prime}_{\text{LB}}(\mathbf{n})\Big\}\qquad&\text{for }n_{F}\geq n_{d2}\end{cases},
\end{equation}
\item Class \rom{4} $\Big(\text{if }\mathbf{n}\in \Big\{\mathcal{I}_0^{C}\cup\{n_{d1}+n_{d3}\leq n_{d2}\}\Big\}\cup\Big\{\mathcal{I}_1^{C}\cap\big\{\frac{1}{n_{d2}}\leq \Delta_{\text{LB}}^{\prime}(\mathbf{n})\big\}\Big\}\Big)$:
\begin{equation}\label{eq:th_1_Class_D}
\Delta_{\text{det}}^{*}(\mu,n_{F},\mathbf{n})=\Delta_{\text{LB}}^{\prime}(\mathbf{n})\qquad\forall n_{F}.
\end{equation}
\end{itemize}
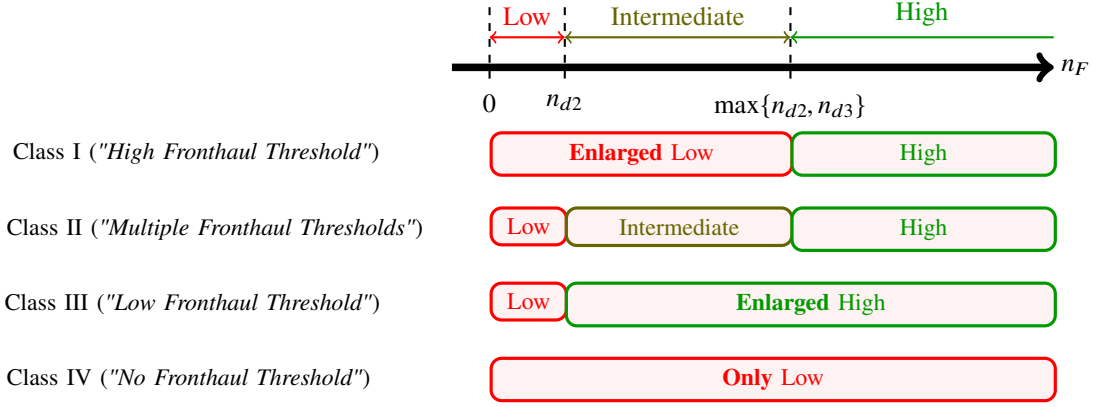
\begin{figure}
\centering
\begin{tikzpicture}[squarednode/.style={rectangle, rounded corners, draw=red!60, fill=red!5, very thick}]
\FronthaulRegime
\end{tikzpicture}
\caption{\small The figure illustrates the differences between the four classes of channel regimes in terms of their number and magnitude of fronthaul capacity thresholds. In the fronthaul regimes for classes \rom{1}, \rom{2}, \rom{3} and \rom{4} that are denoted as \textit{"Low"}, the DTB performance is as if the system operates with no fronthaul link in the delivery phase ($n_F=0$).} 
        \label{fig:distinction_classes}
\end{figure}

\begin{figure*}
        \centering
        \begin{subfigure}[b]{0.475\textwidth}
            \centering
            \begin{tikzpicture}[scale=1]
            \Resulttwo
            \end{tikzpicture}
            \caption{{\small Class \rom{1}}}    
            \label{fig:th_1}
        \end{subfigure}
        \hfill
        \begin{subfigure}[b]{0.475\textwidth}  
            \centering 
            \begin{tikzpicture}[scale=1]
            \Resultthree
            \end{tikzpicture}
            \caption{{\small Class \rom{2}}}       
            \label{fig:th_2}
        \end{subfigure}
        \vskip\baselineskip
        \begin{subfigure}[b]{0.475\textwidth}   
            \centering 
            \begin{tikzpicture}[scale=1]
            \Resultfour
            \end{tikzpicture}
            \caption{{\small Class \rom{3}}} 
            \label{fig:th_3}
        \end{subfigure}
        \quad
        \begin{subfigure}[b]{0.475\textwidth}   
            \centering 
            \begin{tikzpicture}[scale=1]
            \Resultfive
            \end{tikzpicture}
            \caption{{\small Class \rom{4}}}  
            \label{fig:th_4}
        \end{subfigure}
        \caption[DTB as a function of $\mu$ for four classes of channel regimes (Class \rom{1}, \rom{2}, \rom{3} and \rom{4}) at distinct fronthaul capacities]
        {\small DTB as a function of $\mu$ for four classes of channel regimes (Class \rom{1}, \rom{2}, \rom{3} and \rom{4}) at distinct fronthaul capacities. In Fig. \ref{fig:results}, depending on the operating channel regime, we use the notation $A_r$, $r\in\{1,\ldots,4\}$, $B_s$, $s\in\{1,2\}$, and $C_1$ to denote corner points in the convex latency-memory curves.} 
        \label{fig:results}
\end{figure*}
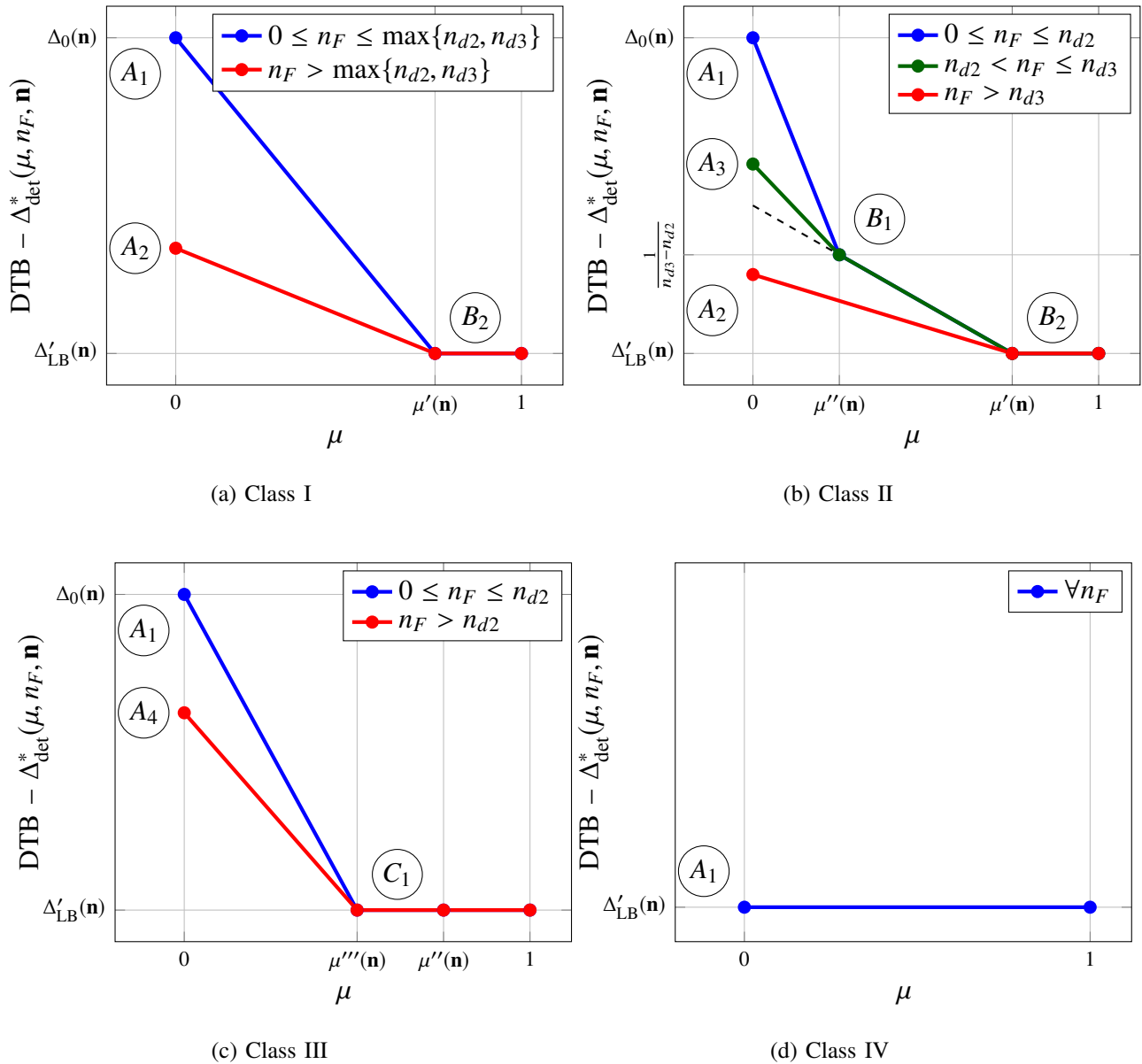

The DTB for each of these classes including their characteristic corner points $A_r$, $r\in\{1,\ldots,4\}$, $B_s$, $s\in\{1,2\}$ and $C_1$ are illustrated in Fig. \ref{fig:results}. Note that $\Delta_{\text{LB}}^{\prime}(\mathbf{n})$ is the \emph{lowest} attainable DTB as shown in the figure. It is independent of $\mu$ and $n_{F}$ such that neither an increase in the fractional cache size $\mu$ nor in the fronthaul capacity $n_{F}$ can lead to any further decrease in the DTB. Instead it only depends on the given wireless channel parameters $\mathbf{n}$ and as such characterizes the wireless bottleneck in the DTB. Thus, we term $\Delta_{\text{LB}}^{\prime}(\mathbf{n})$ as the \emph{wireless bottleneck DTB}. We now discuss our results through some remarks.   
\vspace{.5em}
\begin{remark}[Monotonously Decreasing DTB]
In all classes of channel regimes specified in Eqs. \eqref{eq:th_1_Class_A}, \eqref{eq:th_1_Class_B} and \eqref{eq:th_1_Class_C}, edge caching and fronthauling decrease the DTB
with increasing fractional cache size (cf. Fig. \ref{fig:th_1}--\ref{fig:th_3}). At the threshold cache size $\mu_{\text{th}}(\mathbf{n})$ (e.g., in channel regimes of Class \rom{1} and \rom{2}: $\mu_{\text{th}}(\mathbf{n})=\mu^{\prime}(\mathbf{n})$), the lowest overall DTB corresponds to the wireless bottleneck DTB $\Delta_{\text{LB}}^{\prime}(\mathbf{n})$. This incurred DTB is only governed by the wireless channel and thus cannot be further minimized by fronthaul-edge scheme adjustments\footnote{For instance, when $\mathbf{n}\in\{2n_{d3}\geq n_{d2}\geq n_{d1}\geq n_{d3}\}$, caching more than a fraction $\mu^{\prime}(\mathbf{n})=\nicefrac{(2n_{d3}-n_{d2})}{n_{d3}}$ of any file will not lead to any improvement of the optimal DTB of $\Delta_{\text{LB}}^{\prime}(\mathbf{n})=\nicefrac{1}{n_{d3}}$. This is because in this case the wireless link from eNB to $U_2$ is able to carry at most $n_{d3}$ bits per cannel use; thus, representing the bottleneck.}. In other words, at this cache size prefetched information at the HeNB is just enough so that the requested files are transmitted directly over the wireless channel from eNB/HeNB to the users without requiring any fronthauling transmissions ($T_{F}=0$). This observation is independent of the fronthaul capacity. When operating at any fractional cache size \emph{below} $\mu_{\text{th}}(\mathbf{n})$, however, edge caching along with fronthauling reduces the latency further than a system without cloud processing ($n_{F}=0$) either partially or entirely over the cache size range $\big[0,\mu_{\text{th}}(\mathbf{n})\big)$ as a function of the fronthaul capacity. For instance, in Class \rom{2} channel regimes (cf. Fig \ref{fig:th_2}) where $n_{d3}\geq n_{d2}$, optimal cloud-and cache-based schemes with fronthaul links
\begin{enumerate}[label=(\Alph*)]
\item in the \emph{medium-fronthaul capacity} range $(n_{d2},\max\{n_{d2},n_{d3}\}]$
\item and in the \emph{high-fronthaul capacity} range $(\max\{n_{d2},n_{d3}\},\infty)$
\end{enumerate} strictly outperform cache-only schemes 
\begin{enumerate}[label=(\Alph*)]
\item partially in the sub-interval $\Big[0,\mu^{\prime\prime}(\mathbf{n})\Big)$ of $\Big[0,\mu^{\prime}(\mathbf{n})\Big)$
\item and entirely over the interval $\Big[0,\mu^{\prime}(\mathbf{n})\Big)$.
\end{enumerate} As one would expect intuitively, we observe that for any fronthaul capacity the extent to which \emph{serial} fronthauling (in addition to edge caching) reduces the latency to a system without cloud capabilities decays with increasing cache size.
\end{remark}
\vspace{.5em}
\begin{remark}[Constant DTB]
In all channel regimes of Class \rom{4} \eqref{eq:th_1_Class_D}, edge caching and/or fronthauling is not beneficial. This means that neither a large HeNB cache nor high-fronthaul capacities from cloud to the HeNB can lead to any latency improvement in comparison to the case for $\mu=0,n_F=0$. Consequently, only the eNB is involved in the transmission while the HeNB remains silent. As a result, the optimal latency corresponds to the optimal DTB given in Lemma \ref{lemma:2} for the broadcast channel, i.e., when $\Delta_{\text{LB}}^{\prime}(\mathbf{n})=\Delta_{0}(\mathbf{n})$. This includes all channel regimes for which broadcast and wireless bottleneck DTB are identical.
\end{remark}
\vspace{.5em}
\begin{remark}[Threshold Fronthaul Capacity $n_{F,\text{th}}$]
We see from Theorem \ref{th:1} that for low fronthaul capacities below a threshold $n_{F,\text{th}}$ (e.g. $n_{F,\text{th}}=n_{d2}$ for channel regimes of Class \rom{3}), cache-only schemes remain optimal. 
This means that even though the fronthaul capacity is present/active, in DTB-sense the channel behaves as if the fronthaul link is absent in the delivery phase.
\end{remark}
\section{Serial Transmission -- Lower bound}
\label{sec:ser_trans_lb}

In this section, we develop lower bounds on the DTB for the cases of inactive (cf. \ref{subsec:ser_low_bound}) and active (cf. \ref{subsec:ser_low_bound_nF_greater_0}) fronthaul links. To this end, we specify two propositions applicable for either case. Details on the proofs can be found in the appendices \ref{app:lw_bound_ser} and \ref{app:lw_bound_ser_nF_greater_0}. First, we start with the case of inactive fronthaul links ($n_F=0$). 

\subsection{Lower Bound (Converse) on the Minimum DTB for $n_{F}=0$}
\label{subsec:ser_low_bound}

\begin{proposition}[Lower Bound on the Minimum DTB on Cache-Only F-RAN]\label{prop:1}
For the LDM-based cloud and cache-aided HetNet in Fig. \ref{fig:HetNet} with $n_F=0$, the optimal DTB $\Delta_{\text{det}}^{*}(\mu,\mathbf{n})$ is lower bounded as
\begin{equation}
\label{eq:ser_lb_cache_only}
\Delta_{\text{det}}^{*}(\mu,\mathbf{n})\geq\Delta_{\text{LB}}(\mu,\mathbf{n}),
\end{equation}
where \begin{equation}
\label{eq:lb_cache_only}
\Delta_{\text{LB}}(\mu,\mathbf{n})=\max\Bigg\{\frac{1}{n_{d3}},\frac{1}{\max\{n_{d1},n_{d2}\}}, \frac{2}{\max\{n_{d1}+n_{d3},n_{d2}\}},\frac{1-\mu}{n_{d2}},\frac{2-\mu}{\max\{n_{d2},n_{d3}\}}\Bigg\}.
\end{equation}
\end{proposition}
\noindent\begin{IEEEproof}
The proof of Proposition \ref{prop:1} is presented in Appendix \ref{app:lw_bound_ser}. 
Shortly, the first three bounds inside the outer $\max$-expression leverage the fact that reliable decoding of the user's requested files is feasible through the user's received signal(s) spanning $T_{E}$ channel uses. The remaining two bounds use two distinct combinations of cached and wireless information enabling reliable decoding of requested files to establish two lower bounds on the wireless delivery time $T_{E}$ and ultimately on the DTB as a function of $\mu$ when $n_{F}=0$. For instance, any receiver can decode the user's requested files as long as they are aware of the information subset $\{\mathbf{S}^{q-\max\{n_{d2},n_{d3}\}}\mathbf{x}_{2}^{T_{E}},S_{d_1},S_{d_2}\}$.
\end{IEEEproof}

\subsection{Lower Bound (Converse) on the Minimum DTB for $n_{F}\geq 0$}
\label{subsec:ser_low_bound_nF_greater_0}

We now state the DTB lower bound for the more general case of active fronthaul links ($n_F\geq 0$) through the following proposition.
\begin{proposition}[Lower Bound on the Minimum DTB]\label{prop:2}
For the LDM-based cloud and cache-aided HetNet in Fig. \ref{fig:HetNet} with $n_F\geq 0$, the optimal DTB $\Delta_{\text{det}}^{*}(\mu,n_F,\mathbf{n})$ is lower bounded as
\begin{equation}
\label{eq:ser_lb_cache_only_nF_greater_0}
\Delta_{\text{det}}^{*}(\mu,n_F,\mathbf{n})\geq\Delta_{\text{LB}}(\mu,n_F,\mathbf{n}),
\end{equation}
where $\Delta_{\text{LB}}(\mu,n_F,\mathbf{n})$ is the solution of the linear optimization problem
\begin{subequations}
\label{eq:lb_cache_only_nF_greater_0}
\begin{align}
& \underset{\Delta_{E},\Delta_{F}}{\text{minimize}}
& & \Delta_{E}+\Delta_{F} \label{eq:zielfunktion}\\
& \text{subject to}
& & \Delta_{E}+\Delta_{F}\frac{n_{F}}{\max\{n_{d2},n_{d3}\}} \geq \frac{2-\mu}{\max\{n_{d2},n_{d3}\}} \label{eq:constraint_1}\\
& & & \Delta_{E}+\Delta_{F}\frac{n_{F}}{n_{d2}} \geq \frac{1-\mu}{n_{d2}} \label{eq:constraint_2}\\
& & & \Delta_{E}\geq \max\Bigg\{\frac{1}{n_{d3}},\frac{1}{\max\{n_{d1},n_{d2}\}},\frac{2}{\max\{n_{d1}+n_{d3},n_{d2}\}}\Bigg\} \label{eq:constraint_3}\\
& & & \Delta_{F}\geq 0 \label{eq:constraint_4}
\end{align}
\end{subequations}
and $\Delta_{E}$ and $\Delta_{F}$ denoting the individual DTBs of wireless and fronthaul transmissions, respectively. 
\end{proposition}
\noindent\begin{IEEEproof}
The proof of Proposition \ref{prop:2} is presented in Appendix \ref{app:lw_bound_ser_nF_greater_0}.   
\end{IEEEproof}
We next present multiple corollaries that specialize the lower bound of Proposition \ref{prop:2} to different settings of fronthaul capacities $n_{F}$. 
\begin{corollary}[Lower Bound for Low-Fronthaul Capacity Regime]\label{corr:1} For the F-RAN under study in the low-fronthaul capacity regime with $n_{F}\leq n_{d2}$, the DTB is lower bounded as
\begin{equation}\label{eq:lw_bound_low_fronthaul_capacity_regime}
\Delta^{*}_{\text{det}}(\mu,n_{F},\mathbf{n})\geq \max\Bigg\{\frac{1}{n_{d3}},\frac{1}{\max\{n_{d1},n_{d2}\}},\frac{2}{\max\{n_{d1}+n_{d3},n_{d2}\}},\frac{1-\mu}{n_{d2}},\frac{2-\mu}{\max\{n_{d2},n_{d3}\}}\Bigg\}. 
\end{equation}
\end{corollary}
\noindent \begin{IEEEproof}
We note that the lower bound on the wireless DTB $\Delta_{E}$ (see \eqref{eq:constraint_3}) is also a valid bound on the DTB $\Delta_{E}+\Delta_{F}$ due to the non-negativity of the fronthaul DTB $\Delta_{F}$ (cf. \eqref{eq:constraint_4}). Furthermore, for $n_{F}\leq n_{d2}$ the left-hand side (LHS) of \eqref{eq:constraint_1} and \eqref{eq:constraint_2} are upper bounded by the overall DTB $\Delta_{E}+\Delta_{F}$. Thus, \eqref{eq:constraint_1}--\eqref{eq:constraint_3} are all active lower bounds on the DTB. This concludes the proof.   
\end{IEEEproof}
\begin{remark} We see that the lower bound on the DTB $\Delta_{E}+\Delta_{F}$ in \eqref{eq:lw_bound_low_fronthaul_capacity_regime} coincides with the lower bound on $\Delta_{E}$ for $n_{F}=0$ in Proposition \ref{prop:1}. This means in DTB sense that in the low-fronthaul capacity regime $n_F\leq n_{d2}$, the network behaves as if the fronthaul link is non-existent.
\end{remark}
\begin{corollary}[Lower Bound for High-Fronthaul Capacity Regime]\label{corr:2} For the F-RAN under study in the high-fronthaul capacity regime with $n_{F}\geq\max\{n_{d2},n_{d3}\}$, the DTB is lower bounded as
\begin{align}\label{eq:lw_bound_high_fronthaul_capacity_regime}
\Delta^{*}_{\text{det}}(\mu,n_{F},\mathbf{n})\geq \max\Bigg\{&\frac{2-\mu}{n_F}+\Bigg(1-\frac{\max\{n_{d2},n_{d3}\}}{n_{F}}\Bigg)\Delta_{\text{LB}}^{\prime}(\mathbf{n}),\nonumber\\& \frac{1-\mu}{n_{F}}+\Bigg(1-\frac{n_{d2}}{n_{F}}\Bigg)\Delta_{\text{LB}}^{\prime}(\mathbf{n}),\:\: \Delta_{\text{LB}}^{\prime}(\mathbf{n})\Bigg\}, 
\end{align} where $\Delta_{\text{LB}}^{\prime}(\mathbf{n})$ is defined in \eqref{eq:dlb_main}.
\end{corollary}
\noindent \begin{IEEEproof}
First, we note that $\Delta_{\text{LB}}^{\prime}(\mathbf{n})$ is identical to the right-hand side (RHS) of \eqref{eq:constraint_3}. This bound is a valid bound on the DTB $\Delta_{E}+\Delta_{F}$. For $n_{F}\geq\max\{n_{d2},n_{d3}\}$, we combine \eqref{eq:constraint_1} and \eqref{eq:constraint_2} with \eqref{eq:constraint_3}, which yields
\begin{subequations}
\label{eq:lb_reformulate_high_fronthaul}
\begin{align}
\Delta_{E}+\Delta_{F} & \geq \frac{2-\mu}{n_F}+\Bigg(1-\frac{\max\{n_{d2},n_{d3}\}}{n_{F}}\Bigg)\Delta_{\text{LB}}^{\prime}(\mathbf{n}) \label{eq:reformulate_constraint_1_high_fronthaul}\\
\Delta_{E}+\Delta_{F} & \geq \frac{1-\mu}{n_{F}}+\Bigg(1-\frac{n_{d2}}{n_{F}}\Bigg)\Delta_{\text{LB}}^{\prime}(\mathbf{n}). \label{eq:reformulate_constraint_2_high_fronthaul}
\end{align}
\end{subequations} The RHSs of inequalities \eqref{eq:reformulate_constraint_1_high_fronthaul}, \eqref{eq:reformulate_constraint_2_high_fronthaul} and \eqref{eq:constraint_3} result in \eqref{eq:lw_bound_high_fronthaul_capacity_regime} of Corollary \ref{corr:2}.    
\end{IEEEproof}
\begin{corollary}[Lower Bound for Medium-Fronthaul Capacity Regime in $\mathcal{I}_1$]\label{corr:3} For the F-RAN under study in the medium-fronthaul capacity regime $n_{F}\in[n_{d2},\max\{n_{d2},n_{d3}\}]$ and channel regime $\mathbf{n}\in\mathcal{I}_1$, the DTB is lower bounded as
\begin{align}\label{eq:lw_bound_medium_fronthaul_capacity_regime}
\Delta^{*}_{\text{det}}(\mu,n_{F},\mathbf{n})\geq \max\Bigg\{&\frac{2-\mu}{n_{d3}},\:\: \Delta_{\text{LB}}^{\prime}(\mathbf{n})\Bigg\}, 
\end{align} where $\Delta_{\text{LB}}^{\prime}(\mathbf{n})$ is defined in \eqref{eq:dlb_main}.
\end{corollary}
\noindent \begin{IEEEproof}
We observe that for $n_{F}\leq \max\{n_{d2},n_{d3}\}$, $\Delta_{E}+\Delta_{F}$ is an upper bound on the LHS of \eqref{eq:constraint_1}. For $\mathbf{n}\in\mathcal{I}_1$, it is easy to see that for any $\mu\in[0,1]$
\begin{equation}\label{eq:inequality_medium_fronthaul_I_1}
\Delta_{E}+\Delta_{F}\geq \frac{2-\mu}{n_{d3}}\geq\frac{1-\mu}{n_{d2}}
\end{equation} holds. Thus, only \eqref{eq:constraint_1} and \eqref{eq:constraint_3} are active lower bounds on the DTB. Combining these two bounds leads to \eqref{eq:lw_bound_medium_fronthaul_capacity_regime}.  
\end{IEEEproof}
\begin{corollary}[Lower Bound for Medium-Fronthaul Capacity Regime in $\mathcal{I}_1^{C}$]\label{corr:4} For the F-RAN under study in the medium-fronthaul capacity regime $n_{F}\in[n_{d2},\max\{n_{d2},n_{d3}\}]$ and channel regime $\mathbf{n}\in\mathcal{I}_1^{C}$, the DTB is lower bounded as
\begin{align}\label{eq:lw_bound_medium_fronthaul_capacity_regime_I1_C}
\Delta^{*}_{\text{det}}(\mu,n_{F},\mathbf{n})\geq\begin{cases}\frac{1-\mu}{n_F}+\big(1-\frac{n_{d2}}{n_F}\big)\Delta_{\text{LB}}^{\prime\prime}(\mathbf{n})&\text{ for }\mu\leq \mu^{\prime\prime}(\mathbf{n})\\\max\Big\{\frac{2-\mu}{n_{d3}},\:\: \Delta_{\text{LB}}^{\prime}(\mathbf{n})\Big\}&\text{ for }\mu\geq \mu^{\prime\prime}(\mathbf{n})\end{cases}, 
\end{align} where $\Delta_{\text{LB}}^{\prime}(\mathbf{n})$ is defined in \eqref{eq:dlb_main} and $\Delta_{\text{LB}}^{\prime\prime}(\mathbf{n})$ equals
\begin{equation}\label{eq:delta_LB_prime_prime}
\Delta_{\text{LB}}^{\prime\prime}(\mathbf{n})=\max\Bigg\{\frac{1}{n_{d3}-n_{d2}},\:\:\Delta_{\text{LB}}^{\prime}(\mathbf{n})\Bigg\}.
\end{equation}
\end{corollary}
\noindent \begin{IEEEproof}
We observe that for $n_{F}\leq n_{d3}$, $\Delta_{E}+\Delta_{F}$ is an upper bound on the LHS of \eqref{eq:constraint_1}. For $\mathbf{n}\in\mathcal{I}_1^{C}$, one can show that for any $\mu\in[\mu^{\prime\prime}(\mathbf{n}),1]$
\begin{equation}\label{eq:inequality_medium_fronthaul_I_1}
\Delta_{E}+\Delta_{F}\geq \frac{2-\mu}{n_{d3}}\geq\frac{1-\mu}{n_{d2}}
\end{equation} holds. We infer that for this case only \eqref{eq:constraint_1} and \eqref{eq:constraint_3} are active lower bounds on the DTB. Combining these two bounds leads to the second case of \eqref{eq:lw_bound_medium_fronthaul_capacity_regime_I1_C}. For $\mu\in[0,\mu^{\prime\prime}(\mathbf{n})]$, on the other hand, we bound \eqref{eq:constraint_1} as follows:
\begin{eqnarray}\label{eq:bounding_constraint_1}
\Delta_{E}+\Delta_{F}\geq \Delta_{E}+\Delta_{F}\frac{n_{F}}{n_{d3}}\geq \Delta_{E}\geq\frac{2-\mu}{n_{d3}}\geq \frac{2-\mu}{n_{d3}} \Bigg|_{\mu=\mu^{\prime\prime}(\mathbf{n})}=\frac{1}{n_{d3}-n_{d2}}
\end{eqnarray} 
Using \eqref{eq:bounding_constraint_1} and \eqref{eq:constraint_3} together gives us a new lower bound \eqref{eq:delta_LB_prime_prime} on $\Delta_{E}$. 
We obtain a bound on the DTB by combining \eqref{eq:delta_LB_prime_prime} and \eqref{eq:constraint_2}. This concludes the proof.     
\end{IEEEproof}

\section{Serial Transmission -- Upper bound}
\label{sec:ser_trans_ub}

In this section, we present the upper bounds on the DTB for the cases of active and inactive fronthaul links. In the respective sub-sections, we introduce transmission schemes which achieve the optimal DTB provided in Theorem \ref{th:1} (cf. Fig. \ref{fig:results}). To this end, various schemes are proposed to cover different operating channel regimes of the network under study. Initially, we will focus on the cache-only ($n_F=0$) case. This will turn to be useful when considering the achievability in the more general setting of non-negative fronthaul capacity links ($n_F\geq 0$) in sub-section \ref{subsec:ser_upp_bound_nF_greater_0}. Due to the convexity of the DTB (cf. Lemma \ref{lemma:1}), we establish the achievability of corner points in the latency-cache tradeoff curves. In Fig. \ref{fig:results}, depending on the operating channel regime, these corner points are denoted by $A_r$, $r\in\{1,\ldots,4\}$, $B_s$, $s\in\{1,2\}$, and $C_1$.  
In this regard, we will show that corner points 
\begin{itemize}
\item $A_1$, $B_1$, $B_2$ and $C_1$ involve only edge caching and do not require any fronthaul transmission ($n_{F}=0$); thus, representing cache-only transmission policies, whereas
\item $A_r$, $r\in\{2,3,4\}$, involve only fronthauling and do not require edge caching ($\mu=0$); thus, representing cloud-only transmission policies. 
\end{itemize} Generally speaking, the DTB at fractional cache size $\mu$ which lies between two neighboring corner points (say $D$ and $E$) cache sizes' is achieved through file splitting and time sharing between the policies at corner points $D$ and $E$. This strategy is only operational at a channel regime under which the transmission policy at $D$, $\mathcal{R}_{D}$, and the transmission policy at $E$, $\mathcal{R}_{E}$ are \emph{both} feasible. This is given by the \emph{non-empty} set $\mathcal{R}_{D}\cap\mathcal{R}_{E}$.


\subsection{Establishing Upper Bound (Achievability) for $n_{F}=0$ through Rate Maximization}
\label{subsec:ser_upp_bound}

We propose a transmission scheme which minimizes the delivery time per bit $\Delta_{\text{det}}$ for $n_F=0$ and $\mu> 0$ under various channel regimes. To this end, we determine the best possible achievability scheme by solving a \emph{per-user} rate maximization problem. In fact, minimizing the DTB is equivalent to \emph{maximizing} the number of \emph{desired} bits $\bar{L}$ that are conveyed to $U_1$ and $U_2$ in \emph{one} channel use. Essentially, our proposed transmission scheme seeks to determine the optimal vector of design variables\footnote{These design variables are in fact rate allocation parameters as we shall see later.} $\mathcal{\boldsymbol{r}}^{*}$ that solves the following two \emph{equivalent} optimization problems
  \begin{equation*}
	\begin{aligned}
	& \underset{\mathcal{\boldsymbol{r}}}{\min}
	& & \Delta_{\text{det}}(\mathcal{\boldsymbol{r}})\qquad\qquad\qquad\qquad\qquad & \underset{\mathcal{\boldsymbol{r}}}{\max} & & \bar{L}(\mathcal{\boldsymbol{r}}) 
	\end{aligned}
  \end{equation*}
for which the optimum becomes $\Delta_{\text{det}}^{*}=\nicefrac{1}{\bar{L}^{*}}$. The components of the design vector $\mathcal{\boldsymbol{r}}$ will be specified later after we present the general transmission scheme. The solution to above optimization problem reveals the DTB-achievability at corner points $B_1$, $B_2$ and $C_1$. The achievability of corner point $A_1$ readily follows from Lemma~\ref{lemma:2}. Remaining intermediate points in the tradeoff curves on the DTB follow from the argument of convexity (cf. Lemma \ref{lemma:1}). But before we describe the schemes for various channel regimes in detail, we suggest a general building block structure in the next section \ref{sec:building_blocks_ser} that all schemes have in common. This block structure is applicable to cache-only schemes ($n_{F}=0$) for $\mu>0$. 

\subsection{Building Blocks ($n_{F}=0$)}
\label{sec:building_blocks_ser}
Due to the partial wireless connectivity of the network under study, in which only the eNB is connected to both users while the HeNB is only connected to $U_1$, we propose that the eNB sends
\begin{itemize}
\item \underline{p}rivate information
\begin{itemize}
\item[--]  $\boldsymbol{w}_{1,p}[t]$, $w\in\{u,v\}$ intended for either $U_1$ if $w=u$ or for $U_2$ if $w=v$,
\end{itemize}
\item \underline{c}ommon information
\begin{itemize}
\item[--]  $\boldsymbol{u}_{c}[t]$ intended for $U_1$
\item[--]  $\boldsymbol{v}_{c}[t]$ intended for $U_2$,
\end{itemize}
\item and \underline{i}nterference \underline{n}eutralizing information
\begin{itemize}
\item[--]  $\boldsymbol{v}_{\text{IN}}[t]$ intended for $U_2$,
\end{itemize}
\end{itemize}
while the HeNB transmits
\begin{itemize}
\item additional \underline{p}rivate information $\boldsymbol{u}_{2,p}[t]$ intended for $U_2$
\item and XORed information $\boldsymbol{n}_{\text{IN}}[t]\triangleq\boldsymbol{v}_{\text{IN}}[t]\oplus \boldsymbol{u}_{\text{IN}}[t]$ for \underline{i}nterference \underline{n}eutralization
\end{itemize} 
(in the $t-$th channel use ($t=1,\ldots, T_{E}$)), where in all schemes $\boldsymbol{n}_{\text{IN}}[t]=\boldsymbol{v}_{\text{IN}}[t]\oplus \boldsymbol{u}_{\text{IN}}[t]$ is introduced such that the interference caused by the signal $\boldsymbol{v}_{\text{IN}}[t]$ is completely neutralized at $U_1$. Depending on the particular channel regime, certain signal levels are zero-padded to either avoid collision of private and common information at $U_1$ or to reduce the transmission power level of the HeNB. More details will follow as we ellaborate on the encoding (see section \ref{sec:encoding_Tx_ser}) and the decoding (see section \ref{sec:decoding_Rx_ser}) procedure. In the LDM, this is captured by null vectors $\boldsymbol{0}_{l}$ of length $l$. Since we are interested in maximizing $\bar{L}$, we focus on transmission schemes that utilize a single channel use $T_{E}=1$. Thus, in the sequel, we drop the time dependency in our notation. We denote the length of signal vectors $\boldsymbol{w}_{1,p}$, $\boldsymbol{u}_{2,p}$, $\boldsymbol{u}_{c}$, $\boldsymbol{v}_{c}$, $\boldsymbol{v}_{\text{IN}}$ and $\boldsymbol{n}_{\text{IN}}$ by $R^{w}_{1,p}$, $R^{u}_{2,p}$, $R^{u}_{c}$, $R^{v}_{c}$, $R^{v}_{\text{IN}}$ and $R^{n}_{\text{IN}}$, respectively. In the sequel, we will differentiate between DTB-optimal schemes for (a) strong cross-link (SCL) channel regimes and (b) weak cross-link regimes\footnote{SCL and WCL channel regimes include cases where either the cross-link is stronger or weaker than the eNB--$U_2$ link, i.e., $n_{d2}\geq n_{d3}$ or $n_{d2}\leq n_{d3}$, respectively.} (WCL). We now move to the description of the encoding at the transmitters.

%
\subsection{Encoding at the Transmitters ($n_F=0$)}
\label{sec:encoding_Tx_ser}
The transmission signal of the HeNB and eNB use the signal vectors described in section \ref{sec:building_blocks_ser} according to:
\begin{equation}\label{eq:ldm_stack}
\mathbf{x}_{1}=\begin{pmatrix}
\boldsymbol{0}_{l_4} \\ \boldsymbol{u}_{2,p} \\ \boldsymbol{0}_{l_3} \\ \boldsymbol{n}_{\text{IN}} \\ \boldsymbol{0}_{l_2}
\end{pmatrix},\qquad \boldsymbol{x}_{2}=\begin{pmatrix}
\boldsymbol{u}_{c} \\ \boldsymbol{v}_{c} \\ \boldsymbol{v}_{\text{IN}} \\ \boldsymbol{w}_{1,p} \\ \boldsymbol{0}_{l_1}
\end{pmatrix}\text{ for }w\in\{u,v\}.
\end{equation} 
As shown in Figs. \ref{fig:rx_sig} and \ref{fig:rx_sig_wcl} private information $\boldsymbol{w}_{1,p}$ corresponds to $\boldsymbol{u}_{1,p}$ in the SCL case ($n_{d2}\geq n_{d3}$) and to $\boldsymbol{v}_{1,p}$ in the WCL case ($n_{d2}\leq n_{d3}$). For $q=\max\{n_{d1},n_{d2},n_{d3}\}$, it is obvious that the eNB, on the one hand, can generate its signal vector $\mathbf{x}_{2}$ as long as
\begin{equation}\label{eq:tx_condition_eNB}
l_1+R_{1,p}^{w}+R_{\text{IN}}^{v}+R_{c}^{v}+R_{c}^{u}\leq q,
\end{equation} while the HeNB, on the other hand, can construct its signal vector $\mathbf{x}_{1}$ only if 
\begin{align}
\label{eq:tx_condition_HeNB_1}
l_2+l_3+l_4+R_{2,p}^{u}+R_{\text{IN}}^{n}&\leq q, \\
\label{eq:tx_condition_HeNB_2}
R_{2,p}^{u}+R_{\text{IN}}^{n}&\leq \mu \bar{L}.
\end{align} The condition \eqref{eq:tx_condition_HeNB_2} guarantees that the HeNB is endowed with a finite-size cache of fractional size $\mu$. Next, we outline the decoding strategy for both users.  

\subsection{Decoding at the Receivers ($n_F=0$)}
\label{sec:decoding_Rx_ser}

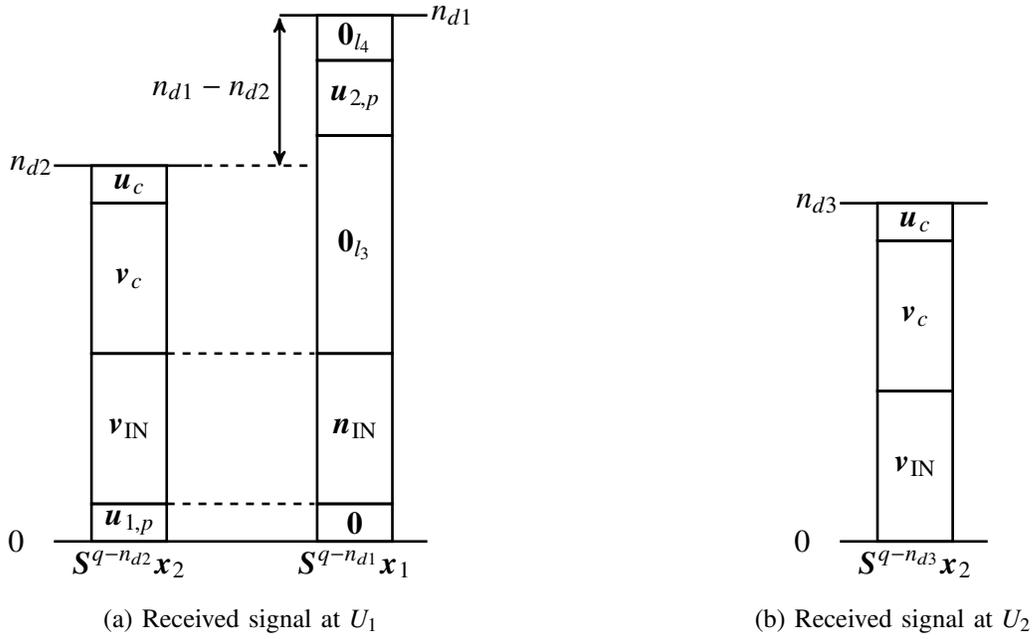
\begin{figure*}
        \centering
        \begin{subfigure}[b]{0.475\textwidth}
            \centering
			\begin{tikzpicture}[->,>=stealth',shorten >=1pt,auto,node 							distance=3cm,thick,scale=1]
				\SCHEMERxoneSCL
			\end{tikzpicture}  
            \caption{{\small Received signal at $U_1$}}    
            \label{fig:rx_u1}
        \end{subfigure}
        \hfill
        \begin{subfigure}[b]{0.475\textwidth}  
            \centering 
			\begin{tikzpicture}[->,>=stealth',shorten >=1pt,auto,node 							distance=3cm,thick,scale=1]
				\SCHEMERxtwoSCL
			\end{tikzpicture} 
            \caption{{\small Received signal at $U_2$}}       
            \label{fig:rx_u2}
        \end{subfigure}
                \caption{\small The received signal vector of (a) $U_1$ and (b) $U_2$ are shown for the SCL case $n_{d2}\geq n_{d3}$. In the SCL case, we fix $\boldsymbol{w}_{1,p}=\boldsymbol{u}_{1,p}$. Note that $U_1$ receives the superposition $\mathbf{S}^{q-n_{d1}}\mathbf{x}_1\oplus\mathbf{S}^{q-n_{d2}}\mathbf{x}_2$.} 
        \label{fig:rx_sig}
\end{figure*}

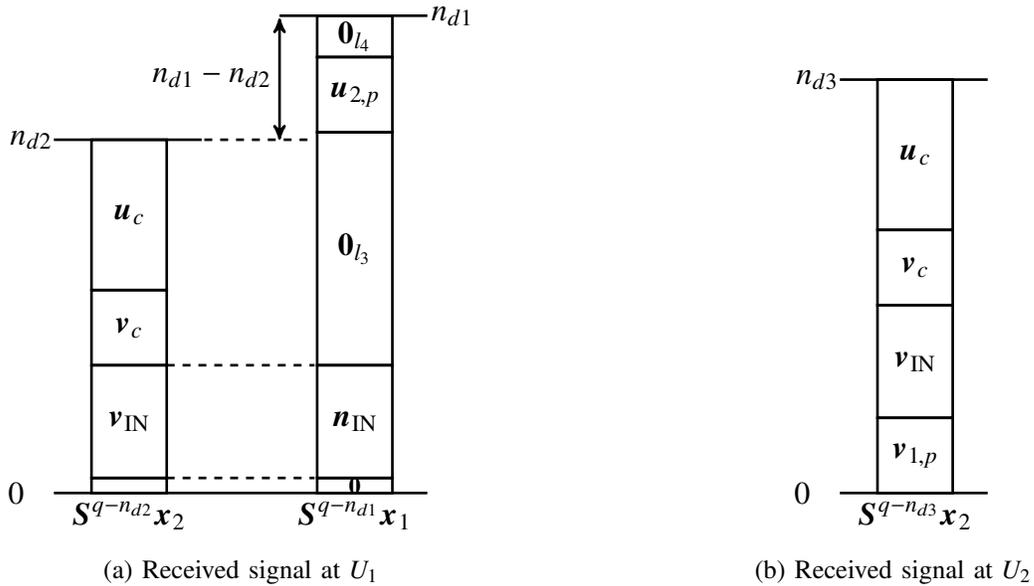
\begin{figure*}
        \centering
        \begin{subfigure}[b]{0.475\textwidth}
            \centering
			\begin{tikzpicture}[->,>=stealth',shorten >=1pt,auto,node 							distance=3cm,thick,scale=1]
				\SCHEMERxoneWCL
			\end{tikzpicture}  
            \caption{{\small Received signal at $U_1$}}    
            \label{fig:rx_u1_wcl}
        \end{subfigure}
        \hfill
        \begin{subfigure}[b]{0.475\textwidth}  
            \centering 
			\begin{tikzpicture}[->,>=stealth',shorten >=1pt,auto,node 							distance=3cm,thick,scale=1]
				\SCHEMERxtwoWCL
			\end{tikzpicture} 
            \caption{{\small Received signal at $U_2$}}       
            \label{fig:rx_u2_wcl}
        \end{subfigure}
                \caption{\small The received signal vector of (a) $U_1$ and (b) $U_2$ are shown for the WCL case $n_{d2}\leq n_{d3}$. In the WCL case, we fix $\boldsymbol{w}_{1,p}=\boldsymbol{v}_{1,p}$. Note that $U_1$ receives the superposition $\mathbf{S}^{q-n_{d1}}\mathbf{x}_1\oplus\mathbf{S}^{q-n_{d2}}\mathbf{x}_2$.} 
        \label{fig:rx_sig_wcl}
\end{figure*}

The transmission according to the block structure of Eq. \eqref{eq:ldm_stack} requires that at the respective receivers interference-decoding and treating interference as noise (TIN) is applied successively. The received signal vectors at $U_1$ and $U_2$ are shown in Figs. \ref{fig:rx_sig} and \ref{fig:rx_sig_wcl} for the SCL ($n_{d2}\geq n_{d3}$) and WCL ($n_{d2}\leq n_{d3}$) case, respectively. In what follows, we will specify the conditions under which both $U_1$ and $U_2$ can decode their desired signal components \emph{reliably}. 

First, let us consider $U_2$. This receiver is interested in retrieving $\boldsymbol{v}_{c}$, $\boldsymbol{v}_{\text{IN}}$ and $\boldsymbol{w}_{1,p}$ if $w=v$ from its received signal $\mathbf{S}^{q-n_{d3}}\mathbf{x}_2$. Successive decoding, however, starting from the top-most significant bit levels necessitates that prior to decoding these signals, the undesired common signal vector $\boldsymbol{u}_{c}$ has to be decoded and canceled from the received signal vector $\mathbf{S}^{q-n_{d3}}\mathbf{x}_2$. Thus, $U_2$ can only decode $\boldsymbol{v}_{c}$, $\boldsymbol{v}_{\text{IN}}$ and $\boldsymbol{w}_{1,p}$ if $w=v$ (cf. Figs. \ref{fig:rx_u2} and \ref{fig:rx_u2_wcl}), as long as 
\begin{equation}\label{eq:rx_condition_U2}
\begin{cases}
R_{c}^{u}+R_{c}^{v}+R_{\text{IN}}^{v}\leq n_{d3}&\text{ if }n_{d2}\geq n_{d3}\\R_{c}^{u}+R_{c}^{v}+R_{\text{IN}}^{v}+R_{1,p}^{w}\leq n_{d3}&\text{ if }n_{d2}\leq n_{d3}\end{cases}.
\end{equation}   

We now move to state the reliability conditions for $U_1$. Recall that $U_1$ receives the superposition $\mathbf{S}^{q-n_{d1}}\mathbf{x}_1\oplus\mathbf{S}^{q-n_{d2}}\mathbf{x}_2$ and aims at obtaining $\boldsymbol{u}_{2,p}$, $\boldsymbol{u}_{c}$ and $\boldsymbol{w}_{1,p}$ if $w=u$ directly as well as $\boldsymbol{u}_{\text{IN}}$ indirectly by superimposing $\boldsymbol{v}_{\text{IN}}$ and $\boldsymbol{n}_{\text{IN}}=\boldsymbol{v}_{\text{IN}}\oplus\boldsymbol{u}_{\text{IN}}$. Similarly to $U_2$'s condition \eqref{eq:rx_condition_U2}, we infer that the individual signals $\mathbf{S}^{q-n_{d1}}\mathbf{x}_1$ and $\mathbf{S}^{q-n_{d2}}\mathbf{x}_2$ have to be able to carry the set of signal components $\{\boldsymbol{n}_{\text{IN}},\boldsymbol{u}_{2,p}\}$ and $\{\boldsymbol{w}_{1,p},\boldsymbol{v}_{\text{IN}},\boldsymbol{v}_{c},\boldsymbol{u}_{c}\}$ if $w=u$ and only $\{\boldsymbol{v}_{\text{IN}},\boldsymbol{v}_{c},\boldsymbol{u}_{c}\}$ if $w=v$, respectively, i.e., 
\begin{align}
\label{eq:rx_condition_U1_1}
&\quad l_3+l_4+R_{2,p}^{u}+R_{\text{IN}}^{n}\leq n_{d1}, \\
\label{eq:rx_condition_U1_2}
&\begin{cases}R_{1,p}^{w}+R_{\text{IN}}^{v}+R_{c}^{v}+R_{c}^{u}\leq n_{d2}&\text{ if } n_{d2}\geq n_{d3}\\R_{\text{IN}}^{v}+R_{c}^{v}+R_{c}^{u}\leq n_{d2}&\text{ if } n_{d2}\leq n_{d3}\end{cases},
\end{align} such that the desired signals $\boldsymbol{u}_{2,p}$, $\boldsymbol{u}_{c}$, $\boldsymbol{u}_{\text{IN}}$ and $\boldsymbol{w}_{1,p}$ if $w=u$ are all received above noise level. The remaining conditions specify how overlaps between desired components are precluded. We avoid an overlap between $\boldsymbol{u}_{2,p}$ and $\boldsymbol{u}_{c}$ if
\begin{equation}\label{eq:rx_condition_U1_3}
l_4+R_{2,p}^{u}\leq (n_{d1}-n_{d2})^{+}
\end{equation} is satisfied. As shown in Figs. \ref{fig:rx_u1} and \ref{fig:rx_u1_wcl}, full interference neutralization
by superposing $\boldsymbol{v}_{\text{IN}}$ and $\boldsymbol{n}_{\text{IN}}$ is ensured if on the one hand the lowest bit levels of $\boldsymbol{v}_{\text{IN}}$ and $\boldsymbol{n}_{\text{IN}}$ are aligned, i.e.,
\begin{equation}\label{eq:rx_condition_U1_4}
R_{c}^{u}+R_{c}^{v}+R_{\text{IN}}^{v}+n_{d1}-n_{d2}=l_3+l_4+R_{\text{IN}}^{n}+R_{2,p}^{u},
\end{equation} and, on the other hand, the allocated number of bits for $\boldsymbol{v}_{\text{IN}}$ and $\boldsymbol{n}_{\text{IN}}$ are identical, i.e., 
\begin{equation}\label{eq:rx_condition_U1_5}
R_{\text{IN}}^{v}=R_{\text{IN}}^{n}.
\end{equation} We point out that through the alignment conditions \eqref{eq:rx_condition_U1_4} and \eqref{eq:rx_condition_U1_5}, overlaps between $\boldsymbol{w}_{1,p}$ and $\boldsymbol{n}_{\text{IN}}$ are prohibited. When all conditions \eqref{eq:tx_condition_eNB}--\eqref{eq:rx_condition_U1_5} are satisfied, the achievable rates of $U_1$ and $U_2$ become
\begin{subequations}
\begin{alignat}{2}
R_{U_1}&=\begin{cases}R_{2,p}^{u}+R_{c}^{u}+R_{\text{IN}}^{n}+R_{1,p}^{w}\:\:&\text{ if }n_{d2}\geq n_{d3}\\R_{2,p}^{u}+R_{c}^{u}+R_{\text{IN}}^{n}\:\:&\text{ if }n_{d2}\leq n_{d3}\end{cases},\label{eq:ach_rate_U1}\\
R_{U_2}&=\begin{cases}R_{c}^{v}+R_{\text{IN}}^{v}\qquad\quad\:&\text{ if }n_{d2}\geq n_{d3}\\R_{c}^{v}+R_{\text{IN}}^{v}+R_{1,p}^{w}\qquad\quad\:&\text{ if }n_{d2}\leq n_{d3}\end{cases},\label{eq:ach_rate_U2}
\end{alignat}   
\end{subequations} respectively. 

\subsection{Maximizing $\bar{L}$ ($n_F=0$)} 
\label{sec:optimizing_ser}

Through sections \ref{sec:encoding_Tx_ser}--\ref{sec:decoding_Rx_ser}, we were able to formulate conditions for feasible encoding and decoding at given fractional cache size $\mu$ for $n_F=0$. We now formulate the underlying optimization problem to establish the achievability. Recall that our scheme maximizes $\bar{L}$ by determining the optimal vector of design variables $\mathcal{\boldsymbol{r}}^{*}$. The design parameters of our scheme are the rate allocation parameters and the respective null vectors. To this end, all design parameters are combined to the vector
\begin{equation*}
\mathcal{\boldsymbol{r}}=\Big(R_{1,p}^{w},R_{2,p}^{u},R_{c}^{u},R_{\text{IN}}^{v},R_{c}^{v},R_{\text{IN}}^{n},l_1,l_2,l_3,l_4\Big)^{T}.
\end{equation*} We then solve the following linear optimization problem:
\begin{equation}\label{eq:lin_opt_problem}
\begin{aligned}
& \underset{\mathcal{\boldsymbol{r}}}{\max}
& & \bar{L}\triangleq\min\{R_{U_1},R_{U_2}\} \\
& \text{s.t.}
& & \eqref{eq:tx_condition_eNB}\text{--}\eqref{eq:rx_condition_U1_5}\text{ are satisfied}, \\
&&& \mathcal{\boldsymbol{r}} \geq \mathbf{0}.
\end{aligned}
\end{equation} 
In the sequel, we will solve the optimization problem for various SCL and WCL channel regimes and thus determine $\bar{L}^{*}$. Recall that this means that at most $\Delta_{\text{det}}(\mu,n_F=0,\mathbf{n})=\nicefrac{1}{\bar{L}^{*}}$ channel uses are required to provide each user with a single bit. Before we solve the problem \eqref{eq:lin_opt_problem} for SCL and WCL channel regimes, we will exemplify our scheme for a specific SCL and WCL channel setting.  

\begin{example}[Class \rom{1} SCL channel regime]\label{example_1}
In this example, we consider the channel $\mathbf{n}=(2,5,4)^{T}$ for 
$n_{F}=0$. This channel belongs to the Class \rom{1} SCL channel regime. Theorem \ref{th:1} gives us the broadcast corner point 
$A_1=\Big(\mu=0,\Delta_{\text{det}}^{*}(\mu=0,n_{F}=0,\mathbf{n})=\nicefrac{2}{5}\Big)$ and the wireless bottleneck corner point
$B_2=\Big(\mu^{\prime}=\nicefrac{1}{3},\Delta_{\text{det}}^{*}(\mu=\nicefrac{1}{3},n_{F}=0,\mathbf{n})=\nicefrac{1}{3}\Big)$.

\begin{figure*}[t!]
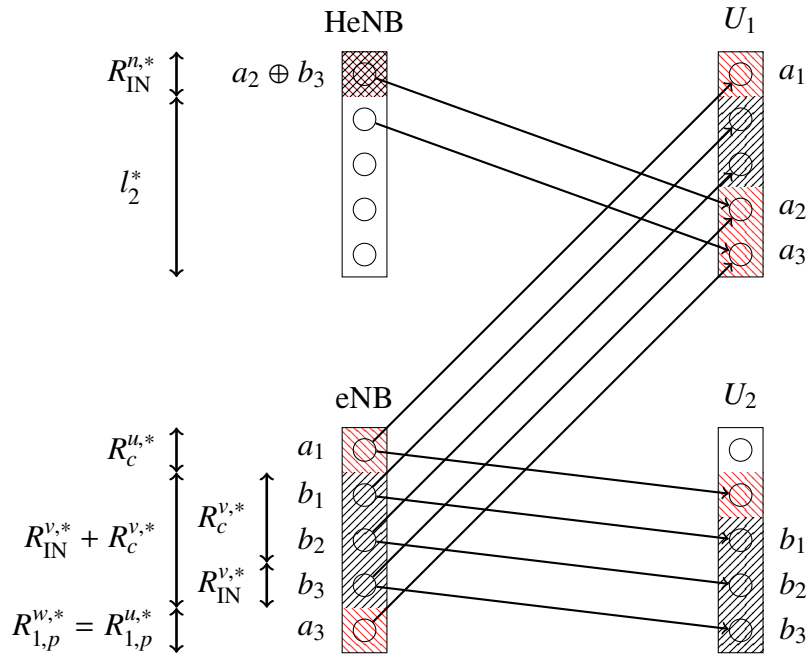

\centering
\LDMblockone
\caption{{\small Example of proposed scheme for $n_{d1}=2$, $n_{d2}=5$ and $n_{d3}=4$. The cached content of the HeNB on the requested files $W_{d_1}$ and $W_{d_2}$ are $S_{d_1}=(a_2)$ and $S_{d_2}=(b_3)$. This corresponds to a fractional cache size of $\mu^{\prime}(\mathbf{n})=\nicefrac{1}{3}$.}}
\label{fig:AcScheme}
\end{figure*} 

At extreme point $A_1$ neither fronthauling nor edge caching involving the HeNB is performed. Thus, the F-RAN under study reduces to a broadcast channel in which the eNB is responsible for transmitting the requested files while the HeNB remains silent. Thus, the only relevant channel realizations are the eNB-$U_1$ and eNB-$U_2$ links with $n_{d2}=5$ and $n_{d3}=4$. We observe that the stronger of those two channels is the eNB-$U_1$ link with $n_{d2}=5$ ($n_{d3}\leq n_{d2}$). The optimal scheme equally distributes the transmission load of $n_{d2}=5$ bits to both users. That is, by sending $\nicefrac{n_{d2}}{2}=\nicefrac{5}{2}$ desired information bits in a single channel use; or, alternatively by sending one desired bit in $\Delta_{\text{det}}^{*}(\mu=0,n_{F}=0,\mathbf{n})=\nicefrac{2}{5}$ channel uses, to $U_1$ and $U_2$, respectively. This scheme is feasible since the weaker channel $n_{d3}$ supports the transmission of $\nicefrac{n_{d2}}{2}$ bits ($\nicefrac{n_{d2}}{2}\leq n_{d3}$). These conditions are equivalent with the broadcast condition $\mathbf{n}\in\mathcal{I}_0$. We refer to the proof of Lemma \ref{lemma:2} and Remark \ref{remark_bc_conditions} for further details on this scheme.   


The extreme point $B_2$ is achievable through a cache-only transmission scheme. It requires one channel use and achieves the wireless bottleneck DTB of $\nicefrac{2}{(n_{d1}+n_{d3})}=\nicefrac{1}{3}$ at fractional cache size $\mu^{\prime}(\mathbf{n})=\nicefrac{(2n_{d1}+2n_{d3}-2n_{d2})}{(n_{d1}+n_{d3})}=\nicefrac{1}{3}$. Fig. \ref{fig:AcScheme} illustrates the scheme for the case when $U_1$ and $U_2$ request $W_{d_1}=(a_1,a_2,a_3)$ and $W_{d_2}=(b_1,b_2,b_3)$, respectively. In this case where the channel of the cross link $n_{d2}$ is the strongest, it is advantageous that the eNB transmits not only common information to $U_2$ but also private and common information to $U_1$. This implies that the rates $R_{1,w}^{u}$ and $R_{c}^{u}$ are chosen to be non-zero whereas $l_1^{*}=0$. When we fix $R_{\text{IN}}^{v,*}+R_{c}^{v,*}=\nicefrac{(n_{d1}+n_{d3})}{2}=3$ all bits ($b_1,b_2$ and $b_3$) of $W_{d_2}$ are conveyed to $U_2$. Note that the eNB-$U_2$ link is capable of reliably carrying $R_{\text{IN}}^{v,*}+R_{c}^{v,*}=3$ bits to $U_2$ because $R_{\text{IN}}^{v,*}+R_{c}^{v,*}=3<n_{d3}=4$ holds. The unused signal levels at the eNB of rate $R_{c}^{u,*}=\nicefrac{(n_{d3}-n_{d1})}{2}=1$ and $R_{1,p}^{w,*}=n_{d2}-n_{d3}=1$ are allocated to send common information $a_1$ and private information $a_3$ to $U_1$, respectively. To retrieve the remaining desired bit of $U_1$ ($a_2$), the interfering bit $b_3$ of $\boldsymbol{v}_{\text{IN}}$ is neutralized by sending the single XOR combination ($R_{\text{IN}}^{n,*}=n_{d1}+n_{d3}-n_{d2}=1$) $a_2\oplus b_3$ at the \emph{highest} signal level of the HeNB. This is feasible since the HeNB retains $a_2$ and $b_3$ separately in its cache during the placement phase. This constitutes a fractional cache size $\mu^{\prime}(\mathbf{n})=\nicefrac{1}{3}$. With this strategy, $U_1$ receives $R_{1,p}^{w,*}+R_{c}^{u,*}=\nicefrac{(2n_{d2}-n_{d1}-n_{d3})}{2}=2$ bits ($a_1$ and $a_3$) of $W_{d_1}$ from the eNB while the remaining $R_{\text{IN}}^{n,*}=n_{d1}+n_{d3}-n_{d2}=1$ bits ($a_2$) are received through network coding which involves both HeNB and eNB. Since the HeNB's fractional cache size is $\nicefrac{1}{3}$, the HeNB cannot send any additional private information. For that reason, $R_{2,p}^{u,*}=l_3^{*}=l_4^{*}=0$ and $l_2^{*}=2n_{d2}-n_{d1}-n_{d3}=4$. Interestingly, as shown in Fig. \ref{fig:AcScheme} increasing the fractional cache size by some $\epsilon>0$ to $\mu^{\prime}(\mathbf{n})+\epsilon$ cannot decrease the DTB any further. This is due to the fact that additional $\epsilon$ information on the requested files at the HeNB cannot be constructively used for extra interference neutralization. E.g., using the second most significant bit level at the HeNB in Fig. \ref{fig:AcScheme} for transmission can only distort the desired bit $a_3$ which $U_1$ already receives. This observation justifies why the DTB remains constant for $\mu>\mu^{\prime}(\mathbf{n})$ as illustrated in Fig. \ref{fig:th_1}.      

Any point between the two corner points $A_1$ and $B_2$ is achieved through file splitting and time sharing between the policies at those two corner points. Such scheme is often termed \emph{memory sharing}. 
\end{example}
\begin{example}[Class \rom{1} WCL channel regime]\label{example_2}
In this example, we consider the channel $\mathbf{n}=(2,5,6)^{T}$ for $n_{F}=0$. This channel belongs to the Class \rom{1} WCL channel regime. Theorem \ref{th:1} gives us the corner points $A_1=\Big(\mu=0,\Delta_{\text{det}}^{*}(\mu=0,n_{F}=0,\mathbf{n})=\nicefrac{1}{3}\Big)$ and 
$B_2=\Big(\mu^{\prime}=\nicefrac{1}{2},\Delta_{\text{det}}^{*}(\mu=\nicefrac{1}{2},n_{F}=0,\mathbf{n})=\nicefrac{1}{4}\Big)$.
\begin{figure*}[t!]
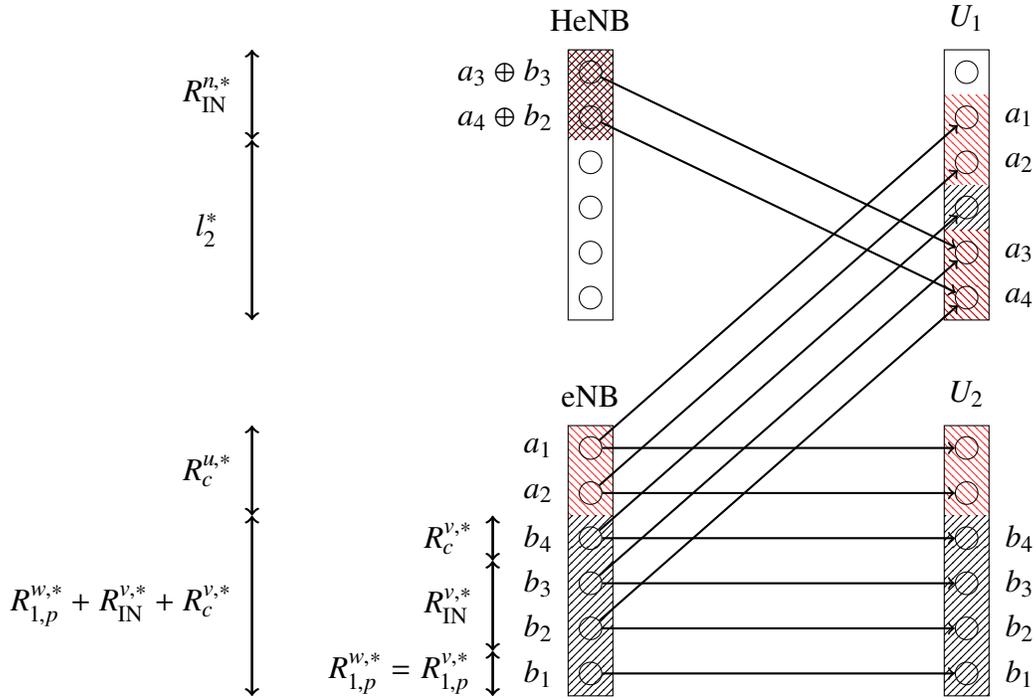

\centering
\LDMblockthree
\caption{{\small Example of proposed scheme for $n_{d1}=2$, $n_{d2}=5$ and $n_{d3}=6$. The cached content of the HeNB on the requested files $W_{d_1}$ and $W_{d_2}$ are $S_{d_1}=(a_3,a_4)$ and $S_{d_2}=(b_2,b_3)$. This constitutes a fractional cache size of $\mu^{\prime}(\mathbf{n})=\nicefrac{1}{2}$.}}
\label{fig:AcScheme2}
\end{figure*} 
At extreme point $A_1$, we apply a similar broadcasting scheme as in example \ref{example_1} for $\mu=0$. Recall that for $\mu=0,n_F=0$, the F-RAN under study simplifies to a broadcast channel in which the eNB is solely responsible for transmitting the requested files. In comparison to example \ref{example_1}, however, the difference is that now out of the two channel links -- eNB-$U_1$ and eNB-$U_2$ -- the stronger channel is the eNB-$U_2$ link with $n_{d3}=6$ ($n_{d3}\geq n_{d2}$). Again, we apply load balancing by equally distributing the transmission load of $n_{d3}=6$ bits to both users. Thus, we send $\nicefrac{n_{d3}}{2}=3$ desired bits in a single channel use to each user such that the DTB becomes $\Delta_{\text{det}}^{*}(\mu=0,n_{F}=0,\mathbf{n})=\nicefrac{1}{3}$. This scheme is feasible since both the weaker and the stronger channel allow the reliable transmission of $\nicefrac{n_{d3}}{2}=3$ bits since $\nicefrac{n_{d3}}{2}\leq n_{d2}$ and $\nicefrac{n_{d3}}{2}\leq n_{d3}$ holds true. Recall that these two conditions are equivalent to the broadcast condition $\mathbf{n}\in\mathcal{I}_1$.  

At the other extreme point $B_2$, a cache-only transmission scheme at $\mu^{\prime}(\mathbf{n})=\nicefrac{2n_{d1}}{(n_{d1}+n_{d3})}=\nicefrac{1}{2}$ is required to attain the wireless bottleneck DTB of $\nicefrac{2}{(n_{d1}+n_{d3})}=\nicefrac{1}{4}$. Fig. \ref{fig:AcScheme2} illustrates the scheme for the case when $U_1$ and $U_2$ request the files $W_{d_1}=(a_1,a_2,a_3,a_4)$ and $W_{d_2}=(b_1,b_2,b_3,b_4)$, respectively. In this example, the strongest link in the network is the eNB-$U_2$ link capable of conveying at most $n_{d3}=6$ bits in total to the receivers reliably. Since the cross-link is weaker than the eNB-$U_2$ link ($n_{d2}\leq n_{d3}$) and the HeNB-$U_1$ link is significantly weaker than the cross-link ($n_{d1}\leq \nicefrac{n_{d2}}{2}$), we allocate parts of the requested file $W_{d_1}$ of $U_1$  to the top-most bit levels of the eNB. As $U_2$ is only connected to the eNB, the remaining less significant bit levels of the eNB have to be able to carry and convey \emph{all} $R_{U_2}^{*}=R_{1,p}^{w,*}+R_{\text{IN}}^{v,*}+R_{c}^{v,*}=\nicefrac{(n_{d1}+n_{d3})}{2}=4$ bits of the requested file $W_{d2}$. Thus, we are able to utilize $R_{c}^{u,*}=n_{d3}-\nicefrac{(n_{d1}+n_{d3})}{2}=\nicefrac{(n_{d3}-n_{d1})}{2}=2$ most significant bit levels of the eNB for transmitting desired information $a_1$ and $a_2$ to $U_1$. Note that this is feasible since $R_{c}^{u,*}=\nicefrac{(n_{d3}-n_{d1})}{2}=2\leq n_{d2}=5$. At fractional cache size $\mu^{\prime}(\mathbf{n})=\nicefrac{1}{2}$, the HeNB is aware of half of each file content. For instance, with respect to $W_{d1}$ and $W_{d2}$ we assume in Fig. \ref{fig:AcScheme2} that the cache content is $S_{d1}=(a_3,a_4)$ and $S_{d2}=(b_2,b_3)$, respectively. With $S_{d1}$ and $S_{d2}$, the HeNB is capable of forming exactly $R_{\text{IN}}^{n,*}=n_{d1}=2$ XOR-combinations $a_3\oplus b_3$ and $a_4\oplus b_2$ to neutralize the received interference of $R_{\text{IN}}^{v,*}=n_{d1}=2$ bits $b_2$ and $b_3$ at $U_1$. Thus, in this case $U_1$ receives $R_{U_1}^{*}=R_{\text{IN}}^{n,*}+R_{c}^{u,*}=n_{d1}+\nicefrac{(n_{d3}-n_{d1})}{2}=4$ desired bits and $R_{c}^{v,*}=n_{d2}-R_{\text{IN}}^{n,*}-R_{c}^{u,*}=\nicefrac{(2n_{d2}-n_{d1}-n_{d3})}{2}=1$ undesired bit $b_4$. Note that $U_1$ does not receive $R_{1,p}^{w,*}=n_{d3}-n_{d2}=1$ bit ($b_1$) at all. It is easy to see that the remaining parameters are set to $R_{2,p}^{u,*}=l_1^{*}=l_3^{*}=l_4^{*}=0$ and $l_2^{*}=n_{d3}-n_{d1}=4$. Interestingly, similarly to example \ref{example_1}, we can see in Fig. \ref{fig:AcScheme2} that increasing the fractional cache size by some $\epsilon>0$ to $\mu^{\prime}(\mathbf{n})+\epsilon$ cannot help the HeNB to decrease the DTB any further. This is due to the fact that additional $\epsilon$ information on the requested files at the HeNB cannot be constructively used at all since all bit levels at the HeNB are already occupied with XOR combinations $a_3\oplus b_3$ and $a_4\oplus b_2$. This explains why the DTB remains constant for $\mu>\mu^{\prime}(\mathbf{n})$.       
\end{example}
\vspace{0.5em}
In the remainder of sub-section \ref{subsec:ser_upp_bound}, we will show the achievability of all relevant corner points that are based on cache-only transmission policies ($n_{F}=0$). To this end, we specify the optimal rate allocation parameters introduced as part of the generalized block structure that optimizes the linear program \eqref{eq:lin_opt_problem}. We start with Class \rom{1} channel regimes and end with Class \rom{4} channel regimes. 
\vspace{0.5em}    

\subsubsection*{Class \rom{1}}
Now, we present the DTB-optimal scheme for all Class \rom{1} channel regimes. We distinguish between Class \rom{1} SCL channel regimes for which $n_{d2}\geq n_{d3}$ holds and Class \rom{1} WCL channel regimes where $n_{d2}\leq n_{d3}$ applies. The achievable DTB for both cases is given by the following proposition. 
\begin{proposition}\label{prop_class_a}
The achievable DTB for Class \rom{1} channel regimes of the network under study for $n_{F}=0$ and $\mu\in[0,1]$ is given for $n_{d2}\geq n_{d3}$ by 
\begin{equation}
\label{eq:prop_class_A_SCL_ser}
\Delta_{\text{det}}(\mu,n_{F}=0,\mathbf{n})=\begin{cases}\max\Big\{\frac{2-\mu}{n_{d2}},\frac{2}{n_{d1}+n_{d3}}\Big\}&\text{ if }n_{d1}+n_{d3}\geq n_{d2}\geq n_{d3}\geq n_{d1}\\\max\Big\{\frac{2-\mu}{n_{d2}},\frac{1}{n_{d3}}\Big\}&\text{ if }2n_{d3}\geq n_{d2}\geq n_{d1}\geq n_{d3}\\\max\Big\{\frac{2-\mu}{n_{d2}},\frac{1}{n_{d3}}\Big\}&\text{ if }n_{d1}\geq n_{d2}\geq n_{d3},2n_{d3}\geq n_{d2}\end{cases},
\end{equation} and for $n_{d2}\leq n_{d3}$ by 
\begin{equation}
\label{eq:prop_class_A_WCL}
\Delta_{\text{det}}(\mu,n_{F}=0,\mathbf{n})=\begin{cases}\max\Big\{\frac{2-\mu}{n_{d3}},\frac{1}{n_{d3}}\Big\}&\text{ if }n_{d1}\geq n_{d3}\geq n_{d2},2n_{d2}\geq n_{d3}\\\max\Big\{\frac{2-\mu}{n_{d3}},\frac{1}{n_{d2}}\Big\}&\text{ if }n_{d1}+n_{d3}\geq 2n_{d2}\geq n_{d3}\geq n_{d2}\geq n_{d1}\\\max\Big\{\frac{2-\mu}{n_{d3}},\frac{2}{n_{d1}+n_{d3}}\Big\}&\text{ if }2n_{d2}\geq n_{d1}+n_{d3}\geq n_{d3}\geq n_{d2}\geq n_{d1}\\\max\Big\{\frac{2-\mu}{n_{d3}},\frac{1}{n_{d1}}\Big\}&\text{ if }2n_{d2}\geq n_{d3}\geq n_{d1}\geq n_{d2}\end{cases}.
\end{equation}
\end{proposition}
\vspace{0.5em}
In what follows, we present the scheme for all Class \rom{1} channel regimes specified in proposition \ref{prop_class_a} in detail. To this end, we establish the achievability at corner points $A_1$ and $B_2$ and apply arguments of convexity for achievability at intermediary points.  

The achievability for corner point $A_1$ readily follows from Lemma \ref{lemma:2} and Eq. \eqref{eq:DTB_lemma} (cf. Example \ref{example_1} and \ref{example_2}) for the SCL and WCL cases. At this point, the optimal DTB corresponds to
\begin{subnumcases}{\Delta_0(\mathbf{n})=}
\frac{2}{n_{d2}}\text{ if }2n_{d3}\geq n_{d2}\geq n_{d3}\\\frac{2}{n_{d3}}\text{ if }2n_{d2}\geq n_{d3}\geq n_{d2}.\end{subnumcases}
We observe that the achievable DTB is in accordance with proposition \ref{prop_class_a} for $\mu=0$ for all Class \rom{1} channel regimes. Next, we consider corner point $B_2$. At this point the fractional cache size corresponds to:
\begin{equation}\label{eq:given_mu_prime_SCL_WCL}
\mu^{\prime}(\mathbf{n})=2-\max\{n_{d2},n_{d3}\}\Delta_{\text{LB}}^{\prime}(\mathbf{n})=\begin{cases}2-n_{d2}\Delta_{\text{LB}}^{\prime}(\mathbf{n})&\text{ if }n_{d2}\geq n_{d3}\\2-n_{d3}\Delta_{\text{LB}}^{\prime}(\mathbf{n})&\text{ if }n_{d2}\leq n_{d3}\end{cases}.
\end{equation} 
For given $\mu^{\prime}(\mathbf{n})$ according to Eq. \eqref{eq:given_mu_prime_SCL_WCL}, we solve the linear optimization problem \eqref{eq:lin_opt_problem}. The solution for the channel regimes to which the channel realizations of Example \ref{example_1} and Example \ref{example_2} belong to are specified in Table \ref{tab:rate_alloc_param_ex_1_and_2}. Concretely, using the rate allocation specified in columns two and three of Table \ref{tab:rate_alloc_param_ex_1_and_2} for $\mathbf{n}=(2,5,4)^{T}$ and $\mathbf{n}=(2,5,6)^{T}$, respectively, reduces the generalized scheme to the schemes described in Examples \ref{example_1} and \ref{example_2}. The solution for all Class \rom{1} SCL and WCL channel regimes are provided in Table \ref{tab:rate_alloc_param} and Table \ref{tab:rate_alloc_param_WCL} of Appendix \ref{app:up_bound_ser_ach}. Under this rate allocation the achievable DTB is given in Eqs. \eqref{eq:prop_class_A_SCL_ser} and \eqref{eq:prop_class_A_WCL} for Class \rom{1} channel regimes with $n_{d2}\geq n_{d3}$ and $n_{d2}\leq n_{d3}$ at $\mu^{\prime}(\mathbf{n})$, respectively. Achievable points between corner points $A_1$ and $B_2$ are attainable if memory sharing is applied. 
Recall that memory sharing schemes are typically composed of two schemes of neighboring corner points, e.g., $A_1$ and $B_2$. Thus, for Class \rom{1}, memory sharing is only feasible in intersecting channel regimes $\mathcal{R}_{A_1}\cap\mathcal{R}_{B_2}$, where \emph{both} schemes, i.e., for $A_1$ and $B_2$, are feasible.  
\begin{table}[!htbp]
\centering
\begin{tabular}{|c||c|c|}
\hline
Regimes $\mathcal{R}_{B_2}$ & $n_{d1}+n_{d3}\geq n_{d2}\geq n_{d3}\geq n_{d1}$ & $2n_{d2}\geq n_{d1}+n_{d3}\geq n_{d3}\geq n_{d2}\geq n_{d1}$
\\ \hline 
$R_{1,p}^{w,*}=R_{1,p}^{u,*}$ & $n_{d2}-n_{d3}$ & $n_{d3}-n_{d2}$
\\ \hline
$R_{c}^{u,*}$ & $\frac{n_{d3}-n_{d1}}{2}$ & $\frac{n_{d3}-n_{d1}}{2}$
\\ \hline
$R_{2,p}^{u,*}$ & $0$ & $0$ 
\\ \hline 
$R_{\text{IN}}^{v,*}$  & $n_{d1}+n_{d3}-n_{d2}$ & $n_{d1}$ 
\\ \hline
$R_{c}^{v,*}$  & $\frac{2n_{d2}-n_{d1}-n_{d3}}{2}$ & $\frac{2n_{d2}-n_{d1}-n_{d3}}{2}$
\\ \hline
$R_{\text{IN}}^{n,*}$  & $n_{d1}+n_{d3}-n_{d2}$ & $n_{d1}$
\\ \hline
$l_1^{*}$ & $0$ & $0$
\\ \hline
$l_2^{*}$ & $2n_{d2}-n_{d1}-n_{d3}$ & $n_{d3}-n_{d1}$
\\ \hline
$l_3^{*}$ & $0$ & $0$
\\ \hline
$l_4^{*}$ & $0$ & $0$
\\ \hline\hline
$\bar{L}^{*}$  & $\frac{n_{d1}+n_{d3}}{2}$ & $\frac{n_{d1}+n_{d3}}{2}$
\\ \hline
$\Delta_{\text{LB}}^{\prime}(\mathbf{n})$  & $\frac{2}{n_{d1}+n_{d3}}$ & $\frac{2}{n_{d1}+n_{d3}}$ 
\\ \hline
$\mu^{\prime}(\mathbf{n})$  & $2-\frac{2n_{d2}}{n_{d1}+n_{d3}}$ & $2-\frac{2n_{d3}}{n_{d1}+n_{d3}}$ 
\\ \hline
\end{tabular}
\caption{\small Rate allocation parameters for corner point $B_2$ at fractional cache size $\mu^{\prime}(\mathbf{n})$ for channel regimes of Example \ref{example_1} (second column) and \ref{example_2} (third column).}
\label{tab:rate_alloc_param_ex_1_and_2}
\end{table}   
\subsubsection*{Class \rom{2}}
Now, we present the DTB-optimal scheme for all Class \rom{2} channel regimes. Hereby, the achievable DTB for this class is given by the following proposition. 
\begin{proposition}\label{prop_class_b}
The achievable DTB for Class \rom{2} channel regimes of the network under study for $n_{F}=0$ and $\mu\in[0,1]$ equals 
\begin{equation}
\label{eq:prop_class_B_SCL}
\Delta_{\text{det}}(\mu,n_{F}=0,\mathbf{n})=\begin{cases}\max\Big\{\frac{1-\mu}{n_{d2}},\frac{2-\mu}{n_{d3}},\frac{1}{n_{d3}}\Big\}&\text{ if }n_{d1}\geq n_{d3}\geq 2n_{d2}\\\max\Big\{\frac{1-\mu}{n_{d2}},\frac{2-\mu}{n_{d3}},\frac{1}{n_{d1}}\Big\}&\text{ if }n_{d1}+n_{d2}\geq n_{d3}\geq n_{d1}\geq n_{d2},n_{d3}\geq 2n_{d2}\end{cases}.
\end{equation}
\end{proposition}
\vspace{0.5em}
The achievability for corner point $A_1$ readily follows from Lemma \ref{lemma:2} and Eq. \eqref{eq:DTB_lemma}. At this point, the optimal DTB for this extreme point corresponds to
\begin{equation}\Delta_0(\mathbf{n})=
\frac{1}{n_{d2}}\text{ if }n_{d3}\geq 2n_{d3}.
\end{equation}
The achievability of the neighboring extreme point $B_1$ is derived as the solution of the optimization problem \eqref{eq:lin_opt_problem} for
\begin{equation}
\mu^{\prime\prime}(\mathbf{n})=\frac{n_{d3}-2n_{d2}}{n_{d3}-n_{d2}}.
\end{equation} 
The optimal rate allocation parameters of this optimization problem are provided in Table \ref{tab:rate_alloc_B_1} of Appendix \ref{app:up_bound_ser_ach}. The achievability at the extreme point $B_2$ applicable for Class \rom{2} channel regimes is listed in columns two and five of Table \ref{tab:rate_alloc_param_WCL} included in Appendix \ref{app:up_bound_ser_ach}. All intermediary points are achievable through memory sharing in the channel regime $\mathcal{R}_{A_1}\cap\mathcal{R}_{B_1}\cap\mathcal{R}_{B_2}$. This establishes the results of proposition \ref{prop_class_b}. 
\subsubsection*{Class \rom{3}}
The DTB-optimal scheme for the Class \rom{3} channel regime is presented. For Class \rom{3}, the following proposition quantifies the achievable DTB. 
\begin{proposition}\label{prop_class_c}
The achievable DTB for Class \rom{3} channel regime of the network under study for $n_{F}=0$ and $\mu\in[0,1]$ corresponds to
\begin{equation}
\label{eq:prop_class_C_SCL}
\Delta_{\text{det}}(\mu,n_{F}=0,\mathbf{n})=\max\Bigg\{\frac{1-\mu}{n_{d2}},\frac{1}{n_{d1}}\Bigg\}\text{ if }n_{d3}\geq n_{d1}+n_{d2}\geq n_{d1}\geq n_{d2}.
\end{equation}
\end{proposition}
\vspace{0.5em}
Lemma \ref{lemma:2} establishes the achievability of extreme point $A_1$ for Class \rom{3}. At this point, the optimal DTB for this extreme point is given by
\begin{equation}\Delta_0(\mathbf{n})=
\frac{1}{n_{d2}}\text{ if }n_{d3}\geq 2n_{d3}.
\end{equation} At corner point $C_1$, we solve the optimization problem of \eqref{eq:lin_opt_problem} at given fractional cache size
\begin{equation}
\mu^{\prime\prime\prime}(\mathbf{n})=\frac{n_{d1}-n_{d2}}{n_{d1}}.
\end{equation} to show the DTB achievability of $\nicefrac{1}{n_{d1}}$ at $\mu^{\prime\prime\prime}(\mathbf{n})$. The resulting decision variables of this linear program are listed in Table \ref{tab:rate_alloc_C_1} of Appendix \ref{app:up_bound_ser_ach}. For $\mu<\mu^{\prime\prime\prime}(\mathbf{n})$, memory sharing establishes the achievability for intermediary points as long as the operating channel regime is given by $\mathcal{R}_{A_1}\cap \mathcal{R}_{C_1}$. 
\subsubsection*{Class \rom{4}}
The DTB for Class \rom{4} channel regimes remains constant. The following proposition states its optimal value. 
\begin{proposition}\label{prop_class_D}
The achievable DTB for Class \rom{4} channel regime of the network under study remains constant for any $\mu\in[0,1]$ and corresponds to
\begin{equation}
\label{eq:prop_class_D}
\Delta_{\text{det}}(\mu,n_{F}=0,\mathbf{n})=\begin{cases}\frac{1}{n_{d3}}&\text{ if }n_{d1}+n_{d3}\geq n_{d2}\geq n_{d1}\geq n_{d3},n_{d2}\geq 2n_{d3}\\\frac{1}{n_{d3}}&\text{ if }n_{d1}\geq n_{d2}\geq 2n_{d3}\\\frac{1}{n_{d2}}&\text{ if }n_{d3}\geq n_{d2}\geq n_{d1},n_{d3}\geq 2n_{d2}\\\max\Big\{\frac{1}{n_{d3}},\frac{2}{n_{d2}}\Big\}&\text{ if }n_{d2}\geq n_{d1}+n_{d3}\end{cases}.
\end{equation}
\end{proposition}
\vspace{0.5em}
Proposition \ref{prop_class_D} readily follows from Lemma \ref{lemma:2}. 

\subsection{Upper Bound (Achievability) for $n_{F}\geq 0$}
\label{subsec:ser_upp_bound_nF_greater_0}

In this sub-section, we propose schemes to cover different operating channel regimes of the network under study for the cloud-only ($\mu=0$) case. As previously explained, we will only focus on the achievability at corner points $A_r$, $r\in\{2,3,4\}$. Remaining intermediate points in the tradeoff curves on the DTB follow from the the convexity of the DTB (cf. Lemma \ref{lemma:1}). To show the achievability at extreme point
\begin{enumerate}[label=(\Alph*)]
\item $A_2$,
\item $A_3$,
\item and $A_4$,
\end{enumerate} we borrow results on the achievability of the \emph{cache-only case} of their respective, neighboring extreme point
\begin{enumerate}[label=(\Alph*)]
\item $B_2$
\item $B_1$
\item and $C_1$
\end{enumerate} at cache sizes $\mu^{\prime}(\mathbf{n}), \mu^{\prime\prime}(\mathbf{n})$ and $\mu^{\prime\prime\prime}(\mathbf{n})$, respectively. In the sequel, we use the notation of the building block structure of sub-section \ref{sec:building_blocks_ser}. We denote the requested file of $U_k$, $k=\{1,2\}$, by $W_{dk}$. 

The general idea of our proposed scheme is to apply a fronthaul transmission policy in such a way that the optimal \emph{wireless} DTB $\Delta_{E}^{*}(\mu_{G}(\mathbf{n}),n_{F}=0,\mathbf{n})$ of the neighboring extreme points $B_2$, $B_1$ and $C_1$ achieved at cache sizes $\mu_{G}(\mathbf{n})\in\{\mu^{\prime}(\mathbf{n}), \mu^{\prime\prime}(\mathbf{n}), \mu^{\prime\prime\prime}(\mathbf{n})\}$ become feasible at $\mu=0$ at the cost of \emph{additional} fronthaul latency. Recall that $\Delta_{E}^{*}$ is the DTB which incurs when at least $\bar{L}^{*}$ bits are conveyed to each user through the wireless channel without invoking the fronthaul link. Recall that the achievability of $\Delta_{E}^{*}(\mu_{G}(\mathbf{n}),n_{F}=0,\mathbf{n})$ was discussed in sub-section \ref{subsec:ser_upp_bound}. The optimal rate allocation parameters $\mathbf{r}^{*}$ are available in Tables \ref{tab:rate_alloc_param} through \ref{tab:rate_alloc_C_1} in Appendix \ref{app:up_bound_ser_ach}. To this end, the cloud server forms 
\begin{enumerate}[label=(\alph*)]
\item $R_{2,p}^{u,*}$ bits of $U_1$'s requested file $W_{d1}$
\item and $R_{\text{IN}}^{n,*}$ XORed bits of both users requested files $W_{d1}$ and $W_{d2}$
\end{enumerate} in the same manner as in the achievability scheme of $B_2$, $B_1$ and $C_1$, respectively. These bits are transmitted from the cloud server through the fronthaul link to the HeNB. Recall from Eq. \eqref{eq:tx_condition_HeNB_2} and Tables \ref{tab:rate_alloc_param} through \ref{tab:rate_alloc_C_1} that at the optimum solution $\mu_{G}(\mathbf{n})\bar{L}^{*}=R_{2,p}^{u,*}+R_{\text{IN}}^{n,*}$. Transmitting these $\mu_{G}(\mathbf{n})\bar{L}^{*}$ bits induces a fronthaul DTB of $\Delta_{F}=\nicefrac{\mu_{G}(\mathbf{n})}{n_F}$. Thus, the overall DTB becomes $\Delta_{E}+\Delta_{F}$, that is, for extreme points
\begin{enumerate}[label=(\Alph*)]
\item $A_2$: \begin{equation}
\Delta_{\text{LB}}^{\prime}(\mathbf{n})+\frac{\mu^{\prime}(\mathbf{n})}{n_F}, \end{equation}
\item $A_3$: \begin{equation}
\Delta_{\text{LB}}^{\prime\prime}(\mathbf{n})+\frac{\mu^{\prime\prime}(\mathbf{n})}{n_F}, \end{equation}
\item and $A_4$: \begin{equation}
\Delta_{\text{LB}}^{\prime}(\mathbf{n})+\frac{\mu^{\prime\prime\prime}(\mathbf{n})}{n_F}, \end{equation}
\end{enumerate} which is identical to $\Delta^{*}_{\text{det}}(\mu=0,n_F,\mathbf{n})$ of Theorem \ref{th:1} for the underlying classes of channel regimes.

%% file: content/parallel.tex
\section{Parallel Transmission -- Main Result}
\label{sec:par_trans}

In  this  section, we outline and discuss our main result on the minimum DTB for the F-RAN in Fig. \ref{fig:HetNet} for a parallel fronthaul-edge transmission. Furthermore, we will contrast parallel transmission from serial transmission. The main result is stated in the following theorem.
\vspace{.5em}
\begin{theorem}\label{th:2}
The DTB of the LDM-based cloud and cache-aided HetNet in Fig. \ref{fig:HetNet} under parallel fronthaul-edge transmissions for any $n_{F}\geq 0$ and $\mu\in[0,1]$ is given by
\begin{equation}\label{eq:DTB_theorem_th_2}
\Delta_{\text{det}}^{*}(\mu,n_F,\mathbf{n})=\max\Bigg\{\frac{1-\mu}{n_{F}+n_{d2}},\frac{2-\mu}{n_{F}+\max\{n_{d2},n_{d3}\}},\Delta_{\text{LB}}^{\prime}(\mathbf{n})\Bigg\}.
\end{equation}
\end{theorem}
\noindent \begin{IEEEproof} (Theorem \ref{th:2}) Lower bounds on the DTB are provided in section \ref{sec:par_low_bound} while upper bounds can be found in section \ref{sec:par_upp_bound}. 
\end{IEEEproof}
Identically to Theorem \ref{th:1}, in the \emph{same} four classes of channel regimes -- Class \rom{1}, \rom{2}, \rom{3} and \rom{4} -- the DTB of Theorem \ref{th:2} simplifies to:
\begin{itemize}
\item Class \rom{1}:
\begin{equation}\label{eq:th_1_Class_A_par}
\Delta_{\text{det}}^{*}(\mu,n_{F},\mathbf{n})=\begin{cases}\max\Big\{\frac{2-\mu}{n_{F}+\max\{n_{d2},n_{d3}\}},\Delta^{\prime}_{\text{LB}}(\mathbf{n})\Big\}\qquad&\text{for }n_{F}\leq n_{F,\max}(\mathbf{n})\\\Delta_{\text{LB}}^{\prime}(\mathbf{n})\qquad&\text{for }n_{F}\geq n_{F,\max}(\mathbf{n})\end{cases},
\end{equation}
\item Class \rom{2}:
\begin{equation}\label{eq:th_1_Class_B_par}
\Delta_{\text{det}}^{*}(\mu,n_{F},\mathbf{n})=\begin{cases}\max\Big\{\frac{1-\mu}{n_{F}+n_{d2}},\frac{2-\mu}{n_{F}+n_{d3}},\Delta^{\prime}_{\text{LB}}(\mathbf{n})\Big\}\qquad&\text{for }n_{F}\leq n_{F,\text{IM}}(\mathbf{n})\\\max\Big\{\frac{2-\mu}{n_{F}+n_{d3}},\Delta^{\prime}_{\text{LB}}(\mathbf{n})\Big\}\qquad&\text{for }n_{F,\text{IM}}(\mathbf{n})\leq n_{F}\leq n_{F,\max}(\mathbf{n})\\\Delta_{\text{LB}}^{\prime}(\mathbf{n})\qquad&\text{for }n_{F}\geq n_{F,\max}(\mathbf{n})\end{cases},
\end{equation}
\item Class \rom{3}:
\begin{equation}\label{eq:th_1_Class_C_par}
\Delta_{\text{det}}^{*}(\mu,n_{F},\mathbf{n})=\begin{cases}\max\Big\{\frac{1-\mu}{n_{F}+n_{d2}},\Delta^{\prime}_{\text{LB}}(\mathbf{n})\Big\}\qquad&\text{for }n_{F}\leq n_{F,\max}(\mathbf{n})\\\Delta^{\prime}_{\text{LB}}(\mathbf{n})\qquad&\text{for }n_{F}\geq n_{F,\max}(\mathbf{n})\end{cases},
\end{equation}
\item Class \rom{4}:
\begin{equation}\label{eq:th_1_Class_D_par}
\Delta_{\text{det}}^{*}(\mu,n_{F},\mathbf{n})=\Delta_{\text{LB}}^{\prime}(\mathbf{n})\qquad\forall n_{F}.
\end{equation}
Hereby, the intermediary and maximum fronthaul capacities $n_{F,\text{IM}}(\mathbf{n})$ and $n_{F,\max}(\mathbf{n})$ correspond to:
\begin{subequations}
\begin{equation}
n_{F,\text{IM}}(\mathbf{n})=n_{d3}-2n_{d2}
\end{equation}
\begin{equation}\label{eq:nF_max}
n_{F,\max}(\mathbf{n})=\max\Bigg\{\frac{2}{\Delta^{\prime}_{\text{LB}}(\mathbf{n})}-\max\{n_{d2},n_{d3}\},\frac{1}{\Delta^{\prime}_{\text{LB}}(\mathbf{n})}-n_{d2}\Bigg\}
\end{equation}
\end{subequations}
\end{itemize}   
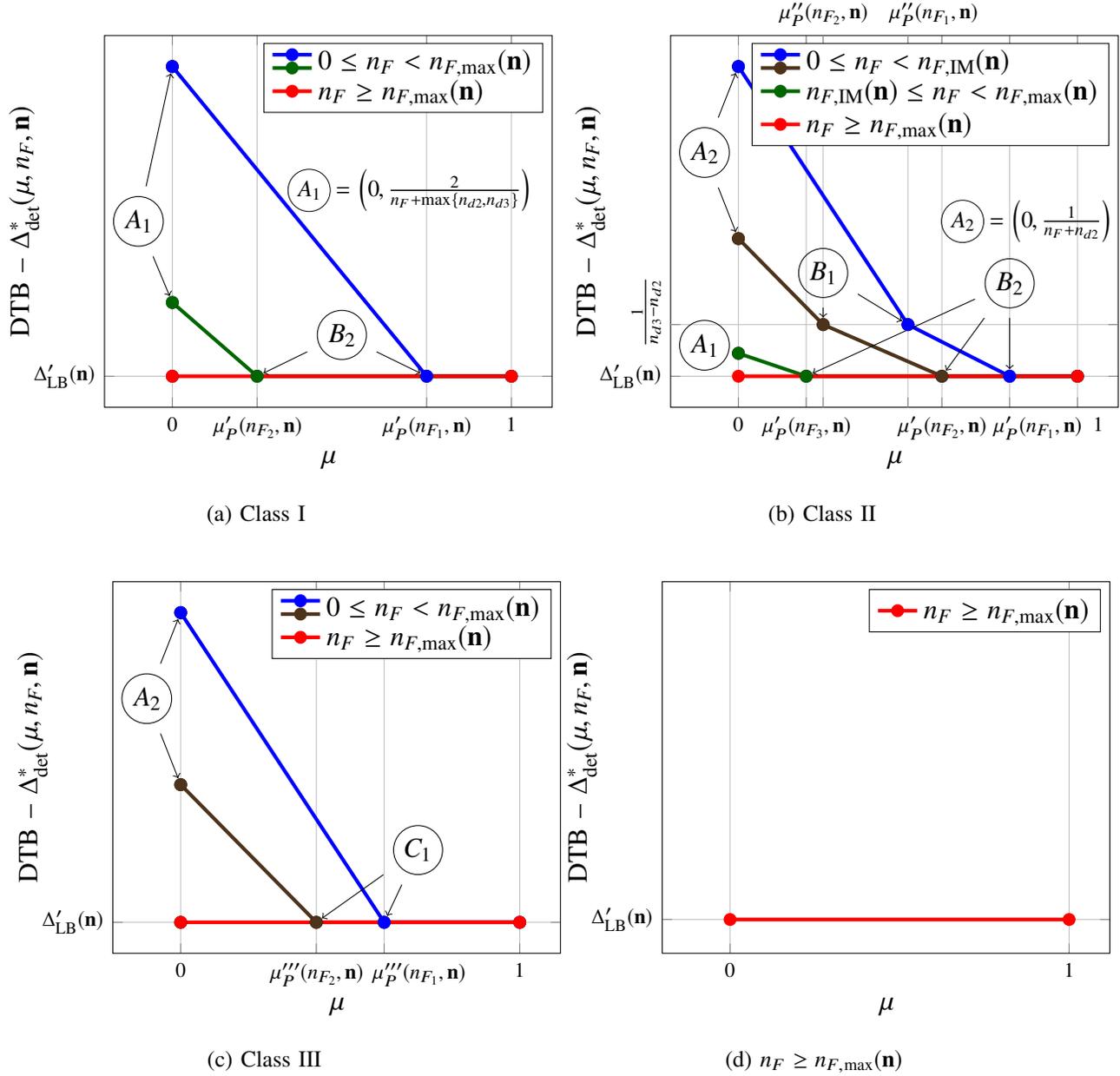
\begin{figure*}
        \centering
        \begin{subfigure}[b]{0.475\textwidth}
            \centering
            \begin{tikzpicture}[scale=1]
            \Resultonepar
            \end{tikzpicture}
            \caption{{\small Class \rom{1}}}    
            \label{fig:th_5}
        \end{subfigure}
        \hfill
        \begin{subfigure}[b]{0.475\textwidth}  
            \centering 
            \begin{tikzpicture}[scale=1]
            \Resulttwopar
            \end{tikzpicture}
            \caption{{\small Class \rom{2}}}     
            \label{fig:th_6}
        \end{subfigure}
        \vskip\baselineskip
        \begin{subfigure}[b]{0.475\textwidth}   
            \centering 
            \begin{tikzpicture}[scale=1]
            \Resulthreepar
            \end{tikzpicture}
            \caption{{\small Class \rom{3}}}    
            \label{fig:th_7}
        \end{subfigure}
        \quad
        \begin{subfigure}[b]{0.475\textwidth}   
            \centering 
            \begin{tikzpicture}[scale=1]
            \Resultfourpar
            \end{tikzpicture}
            \caption{{\small $n_{F}\geq n_{F,\max}(\mathbf{n})$}}  
            \label{fig:th_8}
        \end{subfigure}
        \caption[DTB as a function of $\mu$ for various channel regimes and distinct fronthaul capacities]
        {\small DTB as a function of $\mu$ for three classes of channel regimes (Class \rom{1}, \rom{2}, \rom{3}) at distinct fronthaul capacities $n_{F_1}$, $n_{F_2}$ and $n_{F_3}$. Specifically, we assume that $0\leq n_{F_1}\leq n_{F_2}\leq n_{F_3}\leq n_{F,\max}(\mathbf{n})$. As the fronthaul capacity increases, corner points $B_1$, $B_2$ and $B_3$ shift to the left along the $\mu$-axis until for $n_{F}\geq n_{F,\max}(\mathbf{n})$ the DTB does not change anymore as shown in Fig. \ref{fig:th_8}.} 
        \label{fig:results_par}
\end{figure*}
\vspace{.5em}
\begin{remark}[Serial vs. Parallel Tranmission Schemes]
According to Theorem \ref{th:2}, we observe similar performance curves at channel regimes of classes \rom{1}, \rom{2} and \rom{3} for both serial and parallel fronthaul-edge transmission policies (cf. Figs. \ref{fig:results} and \ref{fig:results_par}). Two main differences are: 
\begin{enumerate} \item First, only for the latter, fractional cache sizes of corner points $B_1$, $B_2$ and $C_1$ explicitly depend on the fronthaul capacity $n_{F}$. This dependency causes that increasing $n_{F}$ by some positive $\delta$ will lead to a linear proportional decrease by $\delta$ in these points' fractional cache sizes. This behavior is shown by a left shift of $B_1$, $B_2$ and $C_1$ in Figs. \ref{fig:th_5}--\ref{fig:th_7} as we increase the fronthaul capacities from $n_{F_1}$ to $n_{F_2}$ (and $n_{F_3}$ exclusively shown in Fig. \ref{fig:th_6}).
\item Second, only for the latter we require a finite fronthaul capacity $n_{F}=n_{F,\max}(\mathbf{n})$ to attain the lowest possible DTB $\Delta_{\text{LB}}^{\prime}(\mathbf{n})$ (see Fig. \ref{fig:th_8}). In contrast, serial fronthaul-edge schemes needed an \emph{infinite} fronthaul capacity ($n_{F}=\infty$) to achieve the same DTB. In other words, fronthaul capacities in the range $n_{F}\in(n_{F,\max}(\mathbf{n}),\infty)$ have a DTB-reducing 
effect only under serial fronthaul-edge schemes. The intuition behind this result is that in parallel transmission policies cloud resources are effectively utilized to overcome partial caching while incurring less fronthaul latency than serial transmission schemes. As a consequence, a finite $n_{F}$ suffices to obtain the DTB $\Delta_{\text{LB}}^{\prime}(\mathbf{n})$.     
\end{enumerate}  
\end{remark}
\vspace{.5em}
\begin{remark}[Maximum Fronthaul Capacity $n_{F,\max}(\mathbf{n})$]
One can verify from Eq. \eqref{eq:nF_max} that $n_{F,\max}(\mathbf{n})\leq n_{d1}$. This suggests that the full-duplex HeNB is capable to forward a fronthaul message of a single time instant, say $t-1$, through the wireless HeNB-to-$U_1$ link in the consecutive channel use $t$. Our achievability schemes (see section \ref{sec:par_upp_bound}) make implicit use of this observation.     
\end{remark}

\section{Parallel Transmission -- Lower Bound}
\label{sec:par_low_bound}

The lower bounds on the DTB for the case of parallel fronthaul transmission are presented in this section through the following proposition.

\begin{proposition}[Lower Bound on the Minimum DTB for F-RAN with Parallel Fronthaul-Edge Transmission]\label{prop:3}
For the LDM-based cloud and cache-aided HetNet in Fig. \ref{fig:HetNet} under a parallel fronthaul-edge transmission setting, the optimal DTB $\Delta_{\text{det}}^{*}(\mu,n_F,\mathbf{n})$ is lower bounded as
\begin{equation}
\label{eq:par_lb}
\Delta_{\text{det}}^{*}(\mu,n_F,\mathbf{n})\geq\Delta_{\text{LB},P}(\mu,\mathbf{n}),
\end{equation}
where \begin{equation}
\label{eq:lb_par}
\Delta_{\text{LB},P}(\mu,n_F,\mathbf{n})=\max\Bigg\{\frac{1}{n_{d3}},\frac{1}{\max\{n_{d1},n_{d2}\}}, \frac{2}{\max\{n_{d1}+n_{d3},n_{d2}\}},\frac{1-\mu}{n_F+n_{d2}},\frac{2-\mu}{n_F+\max\{n_{d2},n_{d3}\}}\Bigg\}.
\end{equation}
\end{proposition}
\noindent\begin{IEEEproof}
The proof of Proposition \ref{prop:3} is presented in Appendix \ref{app:lw_bound_par}. In comparison to the established bounds for the case of serial transmission with $n_F=0$ (cf. Proposition \ref{prop:1}), we ought to account for the received fronthaul message $\mathbf{S}^{q-n_F}\mathbf{x}_{F}^{T_{P}}$ at the HeNB when establishing the DTB lower bounds. This fact becomes noticeable in the last two terms inside the outer $\max$-expression of Eq. \eqref{eq:lb_par}.   
\end{IEEEproof}

\section{Parallel Transmission -- Upper Bound}
\label{sec:par_upp_bound}

In this section, we will focus on establishing the achievability (upper bound) on the DTB for the case of parallel fronthaul-edge transmissions. To this end, first, we introduce the main ingredients of our transmission schemes, that includes the large block number assumption (see section \ref{sec:lg_block_number_par}) and the generalized signal structure that all schemes, irrespective of the channel regime, have in common. Subsequently, we describe the transmission schemes for all channel regimes that achieve the optimal DTB provided in Theorem \ref{th:2} (cf. Fig. \ref{fig:results_par}). Due to the convexity of the DTB, we only establish the achievability of corner points in the latency-cache tradeoff curves. 
In Fig. \ref{fig:results_par}, depending on the operating channel regime, the corner points are denoted by $A_r$, $r\in\{1,2\}$, $B_s$, $s\in\{1,2\}$, and $C_1$. In this regard, we will show that corner points 
\begin{itemize}
\item $A_r$, $r\in\{1,2\}$, involve only fronthauling and do not require edge caching ($\mu=0$); thus, representing cloud-only transmission policies, whereas 
\item $B_1$, $B_2$ and $C_1$ involve both edge caching and fronthauling; thus, representing fronthaul-edge transmission policies.
\end{itemize} 
In general, the DTB at fractional cache size $\mu$ which lies between two neighboring corner points (say $D$ and $E$) cache sizes' is achieved through memory sharing between the policies at corner point $D$ and $E$. For parallel fronthaul edge transmissions, corner points $D$ and $E$ are operational at channel regimes $\mathcal{R}_{D}$ and $\mathcal{R}_{E}$ for non-negative fronthaul capacities below $n_{F,\max}^{D}$ and $n_{F,\max}^{E}$, respectively. Thus, any intermediate point between $D$ and $E$ is achievable through memory sharing when we restrict the channel regime to be 
$\mathcal{R}_{D}\cap\mathcal{R}_{E}$ for fronthaul capacities less than $\min\Big\{n_{F,\max}^{D},n_{F,\max}^{E}\Big\}$.  

\subsection{Large Block Number}
\label{sec:lg_block_number_par}
\begin{figure*}[t]
  \centering
	\begin{tikzpicture}[scale=1]
			\ParBlockMod
	\end{tikzpicture}
\caption{\small The figure depicts the relationship between fronthaul and wireless transmission for a parallel fronthaul-edge transmission scheme. In this scheme, the fronthaul transmission starts and terminates prior to the wireless transmission by a finite time offset of $T_O$ channel uses. This offset enables the HeNB to receive a-priori information which is used to initiate the wireless transmission of consecutive transmission blocks.}
\label{fig:par_operation}
\end{figure*}
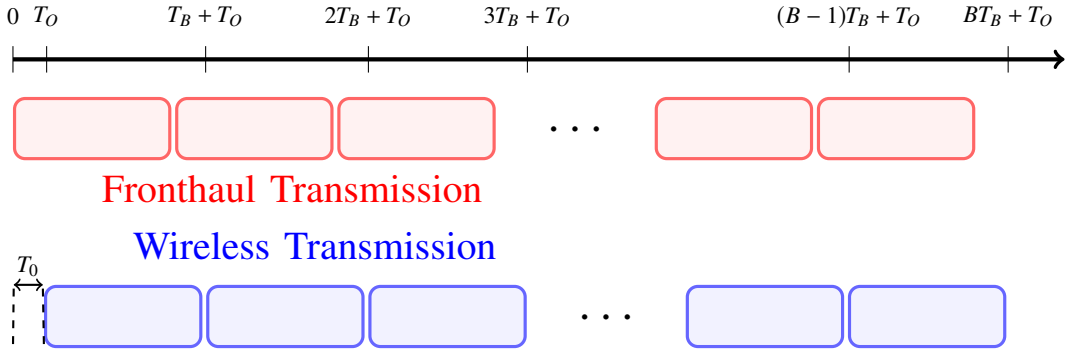  
We recall that in the case of parallel fronthaul-edge transmission, the HeNB operates as a causal \emph{full-duplex} relay endowed with additional cache capabilities. At any time instant $t$, $t=1,2,\ldots,T_P$, the HeNB is thus able to \emph{simultaneously} receive useful \emph{uncached} information through fronthauling and to transmit local information as a function of its cached content and \emph{past} fronthaul messages from time instants $1,2,\ldots,t-1$. Our transmission scheme operates over $B$ blocks for which we use block-Markov coding \cite{Cover79,Willems82}. To this end, we split each file in the library into $B$ blocks, so that each block is of finite size $\tilde{L}=\nicefrac{L}{B}$ bits. We design our scheme in such a way that the wireless delivery time of each block is $T_B$ channel uses; or in other words, the achievable per-block DTB equals $\Delta_{\text{det},B}(\mu,n_F,\mathbf{n})=\nicefrac{T_B}{\tilde{L}}$. This DTB becomes feasible if the fronthaul transmission initiates and terminates by an offset of $T_O$ ($1\leq T_O\leq T_B$) channel uses prior to the wireless transmission (see Fig. \ref{fig:par_operation}). Thus, prior to any per-block wireless transmission just enough a-priori information on each block is made available through fronthauling. The total delivery time to provide both users with their requested files adds up to $T_P=BT_B+T_O$. Hence, the total delivery time per bit for finite $T_O$ that does not scale with $B$ is given by
\begin{equation}\label{eq:tot_DTB_par}
\Delta_{\text{det}}(\mu,n_F,\mathbf{n})=\limsup_{L\rightarrow\infty}\frac{T_P}{L}=\Delta_{\text{det},B}(\mu,n_F,\mathbf{n})+\lim_{B\rightarrow\infty}\frac{T_O}{B\tilde{L}}=\Delta_{\text{det},B}(\mu,n_F,\mathbf{n}).
\end{equation}
One can infer that for arbitrarily large number of blocks $B$, the offset duration $T_O$ becomes negligible and thus the overall DTB converges to the per-block DTB. In the remainder of this sub-section, we will make use of the large block observation and therefore present optimal achievability schemes for a single block that require a finite offset $T_O=1$. For that purpose, we now summarize the main transmit signal vector structure of eNB and HeNB that all block transmissions have in common.
\subsection{Building Blocks}
\label{sec:building_blocks_par}       

Similar to sub-section \ref{sec:building_blocks_ser}, we propose a general signal vector structure that all schemes have in common. Due to the large block number assumption, we may restrict our focus to a single transmission block and \emph{minimize} the per-block DTB $\Delta_{\text{det},B}$ or \emph{maximize} the number of at least $\tilde{L}$ desired bits that are sent to both $U_1$ and $U_2$ in $T_{B}$ channel uses, i.e., we solve one of the two \emph{equivalent} optimization problems 
  \begin{equation*}
	\begin{aligned}
	& \underset{\mathcal{\boldsymbol{\tilde{r}}}}{\min}
	& & \Delta_{\text{det},B}(\mathcal{\boldsymbol{\tilde{r}}})\qquad\qquad\qquad\qquad\qquad & \underset{\mathcal{\boldsymbol{\tilde{r}}}}{\max} & & \tilde{L}(\mathcal{\boldsymbol{\tilde{r}}}) 
	\end{aligned}.
  \end{equation*}
We denote the optimal solution and the optimal decision variables of above optimization problem on the right by $\tilde{L}^{*}$ and $\boldsymbol{\tilde{r}}^{*}$, respectively. Specifically, in our proposed scheme, we will use $T_B=2$ channel uses to transmit a single block of the requested file requiring an offset of $T_O=1$ channel uses between fronthaul and wireless transmission. When $T_B=2$, the optimal per-block DTB becomes $\Delta_{\text{det},B}^{*}=\nicefrac{2}{\tilde{L}^{*}}$. Recall that in the limiting case of large blocks, i.e., $B\rightarrow\infty$, the overall DTB $\Delta_{\text{det}}^{*}$ converges to the per-block DTB specified above. Thus, in what follows, we will outline the \emph{per-block} signal vector structure of the $b$-th block, $b=1,2,\ldots,B$ which encompasses consecutive channel uses $t_1=2b$ and $t_2=2b+1$ if $T_O=1$. 
For the $b$-th block the eNB reserves fronthaul 
\begin{enumerate}[label=(\alph*)]
\item capacity-\emph{dependent}
\item and capacity-\emph{independent}
\end{enumerate} resources. All resources are transmitted in $T_B=2$ channel instants $t\in[t_1:t_2]$. 
First type of resource includes 
\begin{itemize}
\item \underline{p}rivate information
\begin{itemize}
\item[--]  $\boldsymbol{w}_{1,p}[t]$, $w\in\{u,v\}$ intended for either $U_1$ if $w=u$ or for $U_2$ if $w=v$,
\end{itemize}
\item and \underline{c}ommon information
\begin{itemize}
\item[--]  $\boldsymbol{u}_{2,c}[t]$ intended for $U_1$
\item[--]  $\boldsymbol{v}_{2,c}[t]$ intended for $U_2$.
\end{itemize}
\end{itemize} Latter resource type, which we will henceforth call fronthaul-edge block, comprises of 
\begin{itemize}
\item \underline{c}ommon information 
\begin{itemize}
\item[--] $\boldsymbol{u}_{1,c}[t]$ intended for $U_1$
\item[--] $\boldsymbol{v}_{1,c}[t]$ intended for $U_2$
\end{itemize}
\item and \underline{i}nterference \underline{n}eutralizing information
\begin{itemize}
\item[--]  $\boldsymbol{v}_{\text{IN}}[t]$ intended for $U_2$.
\end{itemize}
\end{itemize} The transmission structure of the HeNB, on the other hand, remains identical to the one presented in section \ref{sec:ser_trans_ub}. The HeNB uses sub-signal $\boldsymbol{u}_{2,p}[t]$ to sent additional private information to $U_1$ and sub-signal $\boldsymbol{n}_{\text{IN}}[t]=\boldsymbol{v}_{\text{IN}}[t]\oplus \boldsymbol{u}_{\text{IN}}[t]$ for complete interference neutralization of $\boldsymbol{v}_{\text{IN}}[t]$ at $U_1$. We use the notation of sub-section \ref{sec:building_blocks_ser} to denote the length of all signal vectors. As we shall see, at the eNB side only the rates $R_{\text{IN}}^{v}[t],R_{1,c}^{u}[t]$ and $R_{1,c}^{v}[t]$ will be dependent on \emph{both} the time $t$ and fronthaul capacity $n_F$. The rate of transmission signal $\boldsymbol{w}_{1,p}[t]$, on the other hand, will be both time-independent and fronthaul capacity-independent ($R_{1,w}^{p}[t]=R_{1,w}^{p}$). All remaining rates of the eNB signals will only be independent of the 
fronthaul capacity $n_F$. As before, we will differentiate between SCL and WCL channel regimes. Next, we describe the encoding at the transmitters. 

\subsection{Encoding at the Transmitters}
\label{sec:encoding_Tx_par}
The transmission signals of the HeNB and eNB are chosen according to:
\begin{subequations}\label{eq:ldm_stack_par}
\begin{equation}\label{eq:ldm_stack_par_HeNB}
\mathbf{x}_{1}[t]=\begin{pmatrix}
\boldsymbol{0}_{l_4[t]} \\ \boldsymbol{u}_{2,p}[t] \\ \boldsymbol{0}_{l_3[t]} \\ \boldsymbol{n}_{\text{IN}}[t] \\ \boldsymbol{0}_{l_2[t]}
\end{pmatrix},
\end{equation} 
\begin{equation}\label{eq:ldm_stack_par_eNB}
\qquad\qquad\qquad\mathbf{x}_{2}[t]=\begin{pmatrix}
\boldsymbol{u}_{2,c}[t] \\ \boldsymbol{v}_{2,c}[t] \\ \boldsymbol{u}_{1,c}[t] \\ \boldsymbol{v}_{1,c}[t] \\ \boldsymbol{v}_{\text{IN}}[t] \\ \boldsymbol{w}_{1,p}[t] \\ \boldsymbol{0}_{l_1}
\end{pmatrix}\text{ for }w\in\{u,v\}.
\end{equation}
\end{subequations}
Hereby, $\boldsymbol{w}_{\text{IN}}[t]=\boldsymbol{v}_{\text{IN}}[t]\oplus\boldsymbol{u}_{\text{IN}}[t]$ completely neutralizes the interfering signal $\boldsymbol{v}_{\text{IN}}[t]$ at $U_1$. Furthermore, note that private information $\boldsymbol{w}_{1,p}[t]$ corresponds to $\boldsymbol{u}_{1,p}[t]$ in the SCL case ($n_{d2}\geq n_{d3}$) and to $\boldsymbol{v}_{1,p}[t]$ in the WCL case ($n_{d2}\leq n_{d3}$). For $q=\max\{n_{d1},n_{d2},n_{d3}\}$, it is obvious that the eNB, on the one hand, can generate its signal vector $\mathbf{x}_{2}[t]$ 
as long as
\begin{equation}\label{eq:tx_condition_eNB_par}
l_1+R_{1,p}^{w}+R_{\text{IN}}^{v}[t]+R_{1,c}^{v}[t]+R_{1,c}^{u}[t]+R_{2,c}^{v}[t]+R_{2,c}^{u}[t]\leq q,\qquad\forall t\in[t_1:t_2],
\end{equation} while the HeNB, on the other hand, can construct its signal vector $\mathbf{x}_{1}[t]$ only if 
\begin{align}
\label{eq:tx_condition_HeNB_1_par}
l_2[t]+l_3[t]+l_4[t]+R_{2,p}^{u}[t]+R_{\text{IN}}^{n}[t]&\leq q,&\qquad\forall t\in[t_1:t_2],\\
\label{eq:tx_condition_HeNB_2_par}
R_{2,p}^{u}[t]+R_{\text{IN}}^{n}[t]&\leq \mu\frac{\tilde{L}}{2}+\min\{n_{F},n_{F,\max}(\mathbf{n})\},&\qquad\forall t\in[t_1:t_2].
\end{align} The condition \eqref{eq:tx_condition_HeNB_2_par} holds since the HeNB is endowed with a finite-size cache of fractional size $\mu$ and a fronthaul link of finite capacity $n_{F}$ bits per channel use. Specifically, in our scheme, at every time instant $t$, the HeNB uses its cache content and its received fronthaul signal $\mathbf{x}_{F}[t-1]$ at time instant $t-1$ to encode signals $\boldsymbol{u}_{2,p}[t]$ and $\boldsymbol{n}_{\text{IN}}[t]$. To this end, for $T_B=2$ at most $\nicefrac{\mu\tilde{L}}{T_B}=\nicefrac{\mu\tilde{L}}{2}$ and $\min\{n_{F},n_{F,\max}(\mathbf{n})$ bits emanate from the HeNB's local cache and the fronthaul signal $\mathbf{x}_{F}[t-1]$, respectively. Following, we will outline how the users decode their desired signals.  
    
\subsection{Decoding at the Receivers}
\label{sec:decoding_Rx_par}

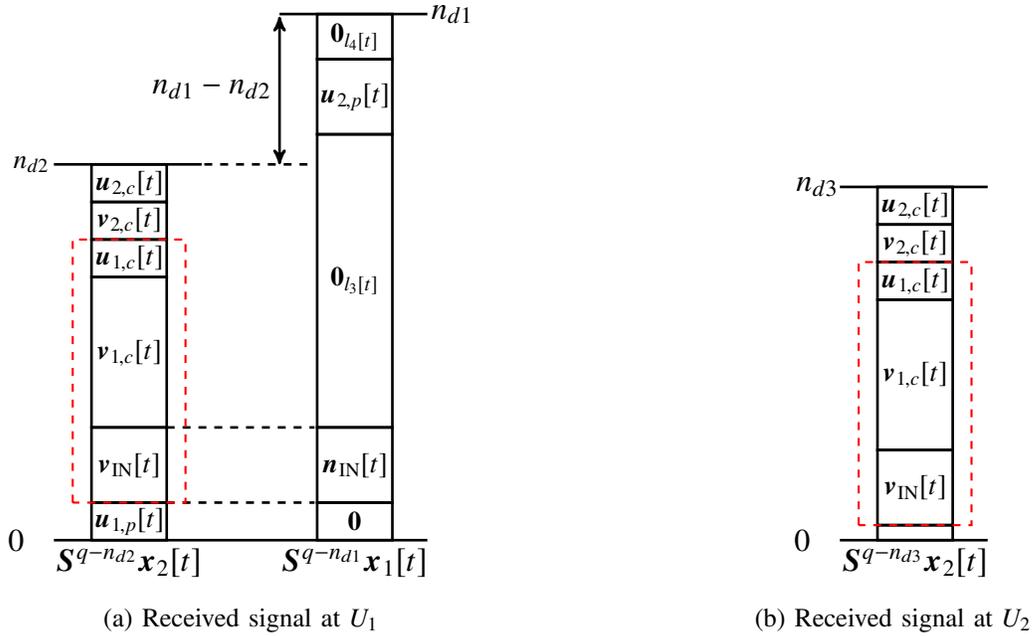
\begin{figure*}
        \centering
        \begin{subfigure}[b]{0.475\textwidth}
            \centering
			\begin{tikzpicture}[->,>=stealth',shorten >=1pt,auto,node 							distance=3cm,thick,scale=1]
				\SCHEMERxoneSCLPar
			\end{tikzpicture}  
            \caption{{\small Received signal at $U_1$}}    
            \label{fig:rx_u1_par}
        \end{subfigure}
        \hfill
        \begin{subfigure}[b]{0.475\textwidth}  
            \centering 
			\begin{tikzpicture}[->,>=stealth',shorten >=1pt,auto,node 							distance=3cm,thick,scale=1]
				\SCHEMERxtwoSCLPar
			\end{tikzpicture} 
            \caption{{\small Received signal at $U_2$}}       
            \label{fig:rx_u2_par}
        \end{subfigure}
                \caption{\small The received signal vector of (a) $U_1$ and (b) $U_2$ are shown for the SCL case $n_{d2}\geq n_{d3}$. In the SCL case, we fix $\boldsymbol{w}_{1,p}[t]=\boldsymbol{u}_{1,p}[t]$. Note that at time instant $t$, $U_1$ receives the superposition $\mathbf{S}^{q-n_{d1}}\mathbf{x}_1[t]\oplus\mathbf{S}^{q-n_{d2}}\mathbf{x}_2[t]$. We highlight the fronthaul-edge block through a red dashed line.} 
        \label{fig:rx_sig_par}
\end{figure*}

\begin{figure*}
        \centering
        \begin{subfigure}[b]{0.475\textwidth}
            \centering
			\begin{tikzpicture}[->,>=stealth',shorten >=1pt,auto,node 							distance=3cm,thick,scale=1]
				\SCHEMERxoneWCLPar
			\end{tikzpicture}  
            \caption{{\small Received signal at $U_1$}}    
            \label{fig:rx_u1_wcl_par}
        \end{subfigure}
        \hfill
        \begin{subfigure}[b]{0.475\textwidth}  
            \centering 
			\begin{tikzpicture}[->,>=stealth',shorten >=1pt,auto,node 							distance=3cm,thick,scale=1]
				\SCHEMERxtwoWCLPar
			\end{tikzpicture} 
            \caption{{\small Received signal at $U_2$}}       
            \label{fig:rx_u2_wcl_par}
        \end{subfigure}
                \caption{\small The received signal vector of (a) $U_1$ and (b) $U_2$ are shown for the WCL case $n_{d2}\leq n_{d3}$. In the WCL case, we fix $\boldsymbol{w}_{1,p}[t]=\boldsymbol{v}_{1,p}[t]$. Note that at time instant $t$, $U_1$ receives the superposition $\mathbf{S}^{q-n_{d1}}\mathbf{x}_1[t]\oplus\mathbf{S}^{q-n_{d2}}\mathbf{x}_2[t]$.} 
        \label{fig:rx_sig_wcl_par}
\end{figure*}
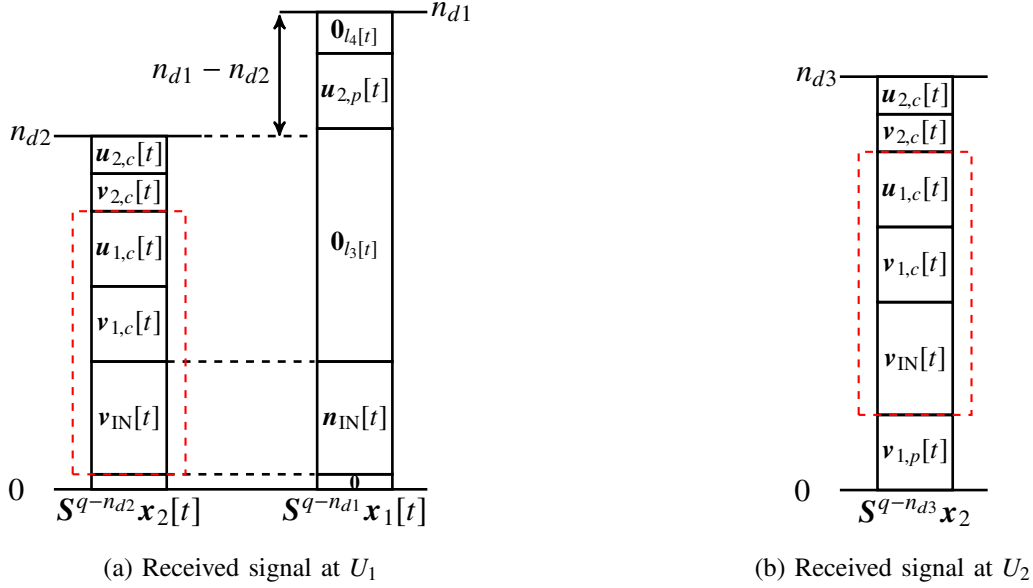

The transmission according to the block structure of Eq. \eqref{eq:ldm_stack_par} forces that at the respective receivers successive decoding is applied. The received signal vector at $U_1$ and $U_2$ of the $b$-th block spanning consecutive time instants $t_1$ and $t_2$ are shown in Figs. \ref{fig:rx_sig_par} and \ref{fig:rx_sig_wcl_par} for the SCL ($n_{d2}\geq n_{d3}$) and WCL ($n_{d2}\leq n_{d3}$) case, respectively. On this basis, we will specify the conditions under which both $U_1$ and $U_2$ can decode their desired signal components \emph{reliably}. 

We first consider the per-block decoding operation of $U_2$ at time instants $t\in[t_1:t_2]$. $U_2$ is interested in retrieving $\boldsymbol{v}_{1,c}[t]$, $\boldsymbol{v}_{2,c}[t]$, $\boldsymbol{v}_{\text{IN}}[t]$ and $\boldsymbol{w}_{1,p}[t]$ if $w=v$ from its received signal $\mathbf{S}^{q-n_{d3}}\mathbf{x}_2[t]$. In successive decoding, a top-down decoding approach is necessary, i.e., $U_2$ starts decoding from the most significant bit levels and moves down to the least significant bits of $\boldsymbol{v}_{\text{IN}}[t]$ if $w=u$ or $\boldsymbol{w}_{1,p}[t]$ if $w=v$. Thus, $U_2$ can only decode all of its desired signals (cf. Figs. \ref{fig:rx_u2_par} and \ref{fig:rx_u2_wcl_par}), as long as \begin{equation}\label{eq:rx_condition_U2_par}
\begin{cases}
R_{1,c}^{u}[t]+R_{2,c}^{u}[t]+R_{1,c}^{v}[t]+R_{2,c}^{v}[t]+R_{\text{IN}}^{v}[t]\leq n_{d3}&\text{ if }n_{d2}\geq n_{d3}\\R_{1,c}^{u}[t]+R_{2,c}^{u}[t]+R_{1,c}^{v}[t]+R_{2,c}^{v}[t]+R_{\text{IN}}^{v}[t]+R_{1,p}^{w}\leq n_{d3}&\text{ if }n_{d2}\leq n_{d3}\end{cases}.
\end{equation}   

Now let us move to state the reliability conditions for $U_1$. Recall that at channel use $t\in[t_1:t_2]$, $U_1$ receives the superposition $\mathbf{S}^{q-n_{d1}}\mathbf{x}_1[t]\oplus\mathbf{S}^{q-n_{d2}}\mathbf{x}_2[t]$ and aims at retrieving $\boldsymbol{u}_{2,p}[t]$, $\boldsymbol{u}_{2,c}[t]$, $\boldsymbol{u}_{1,c}[t]$ and $\boldsymbol{w}_{1,p}[t]$ if $w=u$ directly as well as $\boldsymbol{u}_{\text{IN}}[t]$ through interference neutralization by overlaying $\boldsymbol{v}_{\text{IN}}[t]$ and $\boldsymbol{n}_{\text{IN}}[t]=\boldsymbol{v}_{\text{IN}}[t]\oplus\boldsymbol{u}_{\text{IN}}[t]$. Thus, we infer on the one hand that the individual received signals $\mathbf{S}^{q-n_{d1}}\mathbf{x}_1[t]$ and $\mathbf{S}^{q-n_{d2}}\mathbf{x}_2[t]$ have to contain the set of signal components $\{\boldsymbol{n}_{\text{IN}}[t],\boldsymbol{u}_{2,p}[t]\}$ and $\big\{\boldsymbol{w}_{1,p}[t],\boldsymbol{v}_{\text{IN}}[t],\allowbreak\boldsymbol{v}_{1,c}[t],\boldsymbol{u}_{1,c}[t],\allowbreak\boldsymbol{v}_{2,c}[t],\boldsymbol{u}_{2,c}[t]\big\}$ if $w=u$ and only $\big\{\boldsymbol{v}_{\text{IN}}[t],\boldsymbol{v}_{1,c}[t],\allowbreak\boldsymbol{u}_{1,c}[t],\boldsymbol{v}_{2,c}[t],\allowbreak\boldsymbol{u}_{2,c}[t]\big\}$ if $w=v$, respectively, i.e., 
\begin{align}
\label{eq:rx_condition_U1_1_par}
&\quad l_3[t]+l_4[t]+R_{2,p}^{u}[t]+R_{\text{IN}}^{n}[t]\leq n_{d1}, \\
\label{eq:rx_condition_U1_2_par}
&\begin{cases}R_{1,p}^{w}+R_{\text{IN}}^{v}[t]+R_{1,c}^{u}[t]+R_{2,c}^{u}[t]+R_{1,c}^{v}[t]+R_{2,c}^{v}[t]\leq n_{d2}&\text{ if } n_{d2}\geq n_{d3}\\R_{\text{IN}}^{v}[t]+R_{1,c}^{u}[t]+R_{2,c}^{u}[t]+R_{1,c}^{v}[t]+R_{2,c}^{v}[t]\leq n_{d2}&\text{ if } n_{d2}\leq n_{d3}\end{cases}.
\end{align} On the other hand, at $U_1$ we avoid overlaps of desired signals if the conditions
\begin{equation}\label{eq:rx_condition_U1_3_par}
l_4[t]+R_{2,p}^{u}[t]\leq (n_{d1}-n_{d2})^{+},
\end{equation} 
\begin{equation}\label{eq:rx_condition_U1_4_par}
R_{1,c}^{u}[t]+R_{2,c}^{u}[t]+R_{1,c}^{v}[t]+R_{2,c}^{v}[t]+R_{\text{IN}}^{v}[t]+n_{d1}-n_{d2}=l_3[t]+l_4[t]+R_{\text{IN}}^{n}[t]+R_{2,p}^{u}[t],
\end{equation}
\begin{equation}\label{eq:rx_condition_U1_5_par}
R_{\text{IN}}^{v}[t]=R_{\text{IN}}^{n}[t],
\end{equation}
are satisfied. If all conditions \eqref{eq:tx_condition_eNB_par}--\eqref{eq:rx_condition_U1_5_par} are met for $t\in[t_1:t_2]$, the achievable \emph{per-block} rates of $U_1$ and $U_2$ become
\begin{subequations}
\begin{alignat}{2}
R_{U_1}&=\begin{cases}2R_{1,p}^{w}+\sum_{t\in[t_1:t_2]}R_{1,c}^{u}[t]+R_{2,c}^{u}[t]+R_{\text{IN}}^{n}[t]+R_{2,p}^{u}[t]\:\:&\text{ if }n_{d2}\geq n_{d3}\\\sum_{t\in[t_1:t_2]}R_{1,c}^{u}[t]+R_{2,c}^{u}[t]+R_{\text{IN}}^{n}[t]+R_{2,p}^{u}[t]\:\:&\text{ if }n_{d2}\leq n_{d3}\end{cases},\label{eq:ach_rate_U1_par}\\
R_{U_2}&=\begin{cases}\sum_{t\in[t_1:t_2]}R_{\text{IN}}^{v}[t]+R_{1,c}^{v}[t]+R_{2,c}^{v}[t]\qquad\quad\:&\text{ if }n_{d2}\geq n_{d3}\\2R_{1,p}^{w}+\sum_{t\in[t_1:t_2]}R_{\text{IN}}^{v}[t]+R_{1,c}^{v}[t]+R_{2,c}^{v}[t]\qquad\quad\:&\text{ if }n_{d2}\leq n_{d3}\end{cases},\label{eq:ach_rate_U2_par}
\end{alignat}   
\end{subequations} respectively. 

\subsection{Maximizing $\tilde{L}$} 
\label{sec:optimizing_par}

Recall that our scheme maximizes $\tilde{L}$ by determining the optimal vector of design variables $\mathcal{\boldsymbol{\tilde{r}}}^{*}$. The design parameters of our scheme are the rate allocation parameters and the respective null vectors for time instants $t_1$ and $t_2$. We combine all design parameters to the vector
\begin{align*}
\mathcal{\boldsymbol{\tilde{r}}}=\Big(&R_{1,p}^{w},R_{1,c}^{u}[t_1:t_2],R_{2,c}^{u}[t_1:t_2],R_{2,p}^{u}[t_1:t_2],R_{\text{IN}}^{v}[t_1:t_2],R_{1,c}^{v}[t_1:t_2],R_{2,c}^{v}[t_1:t_2],\\&R_{\text{IN}}^{n}[t_1:t_2],l_1,l_2[t_1:t_2],l_3[t_1:t_2],l_4[t_1:t_2]\Big)^{T}.
\end{align*} The per-block user rate maximization problem can then be cast by the following linear optimization program:
\begin{equation}\label{eq:lin_opt_problem_par}
\begin{aligned}
& \underset{\mathcal{\boldsymbol{\tilde{r}}}}{\max}
& & \tilde{L}\triangleq\min\{R_{U_1},R_{U_2}\} \\
& \text{s.t.}
& & \eqref{eq:tx_condition_eNB_par}\text{--}\eqref{eq:rx_condition_U1_5_par}\text{ are satisfied } \forall t\in[t_1:t_2], \\
&&& \mathcal{\boldsymbol{\tilde{r}}} \geq \mathbf{0}.
\end{aligned}
\end{equation} 
For various SCL and WCL channel regimes in all four classes of channel regimes, we will determine $\tilde{L}^{*}$ and $\mathcal{\boldsymbol{\tilde{r}}}^{*}$. Recall that this means under the large block number assumption that at most $\Delta_{\text{det}}(\mu,n_F=0,\mathbf{n})=\nicefrac{2}{\tilde{L}^{*}}$ channel uses are required to provide each user with a single bit. In the remainder of this sub-section, we will show the achievability of corner points $A_1$, $A_2$, $B_1$, $B_2$ and $C_1$. To this end, we specify the optimal rate allocation parameters $\mathcal{\boldsymbol{\tilde{r}}}^{*}$.
\subsubsection*{Class \rom{1}}
\label{subsec_class_a_par}
Now, we present the DTB-optimal scheme for all Class \rom{1} channel regimes. Again, we distinguish between Class \rom{1} SCL channel regimes for which $n_{d2}\geq n_{d3}$ holds and Class \rom{1} WCL channel regimes where $n_{d2}\leq n_{d3}$ applies. The achievable DTB for both cases is given by the following proposition. 
\begin{proposition}\label{prop_class_a_par}
The achievable DTB for Class \rom{1} channel regimes of the network under study for $n_{F}\geq 0$ and $\mu\in[0,1]$ under a parallel-fronthaul edge transmission is given for $n_{d2}\geq n_{d3}$ by 
\begin{equation}
\label{eq:prop_class_A_SCL}
\Delta_{\text{det}}(\mu,n_{F},\mathbf{n})=\begin{cases}\max\Big\{\frac{2-\mu}{n_F+n_{d2}},\frac{2}{n_{d1}+n_{d3}}\Big\}&\text{ if }n_{d1}+n_{d3}\geq n_{d2}\geq n_{d3}\geq n_{d1}\\\max\Big\{\frac{2-\mu}{n_F+n_{d2}},\frac{1}{n_{d3}}\Big\}&\text{ if }2n_{d3}\geq n_{d2}\geq n_{d1}\geq n_{d3}\\\max\Big\{\frac{2-\mu}{n_F+n_{d2}},\frac{1}{n_{d3}}\Big\}&\text{ if }n_{d1}\geq n_{d2}\geq n_{d3},2n_{d3}\geq n_{d2}\end{cases},
\end{equation} and for $n_{d2}\leq n_{d3}$ by 
\begin{equation}
\label{eq:prop_class_A_WCL_par}
\Delta_{\text{det}}(\mu,n_{F},\mathbf{n})=\begin{cases}\max\Big\{\frac{2-\mu}{n_F+n_{d3}},\frac{1}{n_{d3}}\Big\}&\text{ if }n_{d1}\geq n_{d3}\geq n_{d2},2n_{d2}\geq n_{d3}\\\max\Big\{\frac{2-\mu}{n_F+n_{d3}},\frac{1}{n_{d2}}\Big\}&\text{ if }n_{d1}+n_{d3}\geq 2n_{d2}\geq n_{d3}\geq n_{d2}\geq n_{d1}\\\max\Big\{\frac{2-\mu}{n_F+n_{d3}},\frac{2}{n_{d1}+n_{d3}}\Big\}&\text{ if }2n_{d2}\geq n_{d1}+n_{d3}\geq n_{d3}\geq n_{d2}\geq n_{d1}\\\max\Big\{\frac{2-\mu}{n_F+n_{d3}},\frac{1}{n_{d1}}\Big\}&\text{ if }2n_{d2}\geq n_{d3}\geq n_{d1}\geq n_{d2}\end{cases}.
\end{equation}
\end{proposition}
\vspace{0.5em}
In what follows, we present the scheme for all Class \rom{1} channel regimes specified in proposition \ref{prop_class_a_par} in detail. To this end, we establish the achievability at corner points $A_1$ and $B_2$ and apply arguments of convexity for achievability at intermediary points.  

First, we consider corner point $A_1$ for which $\mu=0$. At this fractional cache size, we solve the optimization problem \eqref{eq:lin_opt_problem_par}. Table \ref{tab:rate_alloc_I_a_R1_R3} and Tables \ref{tab:rate_alloc_I_a_R51_R52} and \ref{tab:rate_alloc_I_a_R4_R62} of Appendix \ref{app:up_bound_par_ach} specify the optimal rate allocation parameters $\mathcal{\boldsymbol{\tilde{r}}}^{*}$ for Class \rom{1} SCL and WCL channel regimes, respectively, that attain a DTB of $\nicefrac{2}{(n_F+\max\{n_{d2},n_{d3}\})}$ at $\mu=0$. We point out that Table \ref{tab:rate_alloc_I_a_R4_R62} in general is applicable for $n_{F}\geq (n_{d3}-2n_{d2})^{+}$. For Class \rom{1} WCL channel regimes, however, $(n_{d3}-2n_{d2})^{+}=0$.   

Next, we show the achievability of corner point $B_2$ at fractional cache size $\mu_{P}^{\prime}(n_F,\mathbf{n})$. We will show that the achievability at this corner point is directly deducible from the scheme for the corner point $B_2$ under serial fronthaul-edge transmission (see sub-section \ref{subsec:ser_upp_bound}). The same observation applies to corner points $B_1$ and $C_1$ as we shall see later. We use the indices $(S)$ and $(P)$ to distinguish corner points of serial transmissions from those for parallel transmissions. Through careful comparison of Theorems \ref{th:1} and \ref{th:2}, one can verify that 
\begin{equation}\label{eq:mu_par_vs_mu_ser_1}
\mu_{S}^{\prime}(\mathbf{n})\bar{L}^{*}=\mu_{P}^{\prime}(n_F,\mathbf{n})\bar{L}^{*}+n_{F}.
\end{equation} Recall that in the delivery phase of serial transmission, $\mu_{S}^{\prime}(\mathbf{n})\bar{L}^{*}=R_{2,p}^{u,S,*}+R_{\text{IN}}^{v,S,*}$ bits are required in total to attain the DTB of corner point $B_{2}^{S}$. Hereby, $R_{\text{IN}}^{v,S,*}$ bits originate as XOR combinations of the files $W_{d_1}$ and $W_{d_2}$ requested by $U_1$ and $U_2$ while $R_{2,p}^{u,S,*}$ bits are a raw data subset of file $W_{d_1}$. In this regard, Eq. \eqref{eq:mu_par_vs_mu_ser_1} suggests that the HeNB's conveyed content (during the delivery phase) consisting of $\mu_{S}^{\prime}(\mathbf{n})\bar{L}^{*}$ bits for corner point $B_{2}^{S}$ is separable into two distinct components attributed to edge caching and fronthauling for the case of parallel transmission (similar observation for $B_{1}^{S}$ and $C_{1}^{S}$). In the following, we explicitly account for the \emph{separability in delivery content} suggested in Eq. \eqref{eq:mu_par_vs_mu_ser_1} and propose how to map the achievability scheme of serial to parallel fronthaul-edge transmission.  
The mapping is done as follows:

We keep the signal structure for both parallel and serial transmissions identical by adjusting the rate allocation parameters of parallel transmissions according to
\begin{subequations}
\label{eq:decomposing_signals}
\begin{alignat}{6}
R_{1,p}^{w,P,*}[t]&=R_{1,p}^{w,S,*},\label{eq:decomposing_signals_1}\\ 
R_{1,c}^{u,P,*}[t]=R_{1,c}^{v,P,*}[t]&=0,\label{eq:decomposing_signals_2}\\
R_{2,c}^{u,P,*}[t]&=R_{c}^{u,S,*},\label{eq:decomposing_signals_3}\\
R_{2,c}^{v,P,*}[t]&=R_{c}^{v,S,*},\label{eq:decomposing_signals_4}\\
R_{2,p}^{u,P,*}[t]&=R_{2,p}^{u,S,*},\label{eq:decomposing_signals_5}\\
R_{\text{IN}}^{n,P,*}[t]=R_{\text{IN}}^{v,P,*}[t]&=R_{\text{IN}}^{n,S,*}=R_{\text{IN}}^{v,S,*},\label{eq:decomposing_signals_6} \\ l_i^{P,*}[t]&=l_i^{S,*},\qquad i=1,\ldots,4\label{eq:decomposing_signals_7}
\end{alignat}
\end{subequations} for $t=[t_1:t_2]$. As previously suggested, at the HeNB we account for the separability in delivery content. To this end, we decompose signals $\boldsymbol{u}_{2,p}[t]$ and $\boldsymbol{n}_{\text{IN}}[t]$ for the parallel transmission case into sub-signals attributed to edge caching and fronthauling (superscripts $(C)$ and $(F)$) in agreement with 
\begin{subequations}
\label{eq:decomposing}
\begin{alignat}{2}
\boldsymbol{u}_{2,p}[t]&=\Big(\boldsymbol{u}_{2,p}^{C}[t]),\boldsymbol{u}_{2,p}^{F}[t]\Big)^{T},\label{eq:decomposing_1}\\ \boldsymbol{n}_{\text{IN}}[t]&=\Big(\boldsymbol{n}_{\text{IN}}^{C}[t],\boldsymbol{n}_{\text{IN}}^{F}[t]\Big)^{T},\label{eq:decomposing_2}
\end{alignat}
\end{subequations} with rate allocation 
\begin{subequations}
\label{eq:decomposing_rate}
\begin{alignat}{2}
R_{2,p}^{F,*}[t]+R_{\text{IN}}^{n,F,*}[t]&=n_{F},
\label{eq:decomposing_rate_1}\\ 
R_{2,p}^{C,*}[t]+R_{\text{IN}}^{n,C,*}[t]&=\mu_{S}^{\prime}(\mathbf{n})\bar{L}^{*}-n_F\quad&\text{ at corner point $B_2$},\label{eq:decomposing_rate_2}
\end{alignat}
\end{subequations} for $t=[t_1:t_2]$. Recall that Tables \ref{tab:rate_alloc_param} and \ref{tab:rate_alloc_param_WCL} of Appendix \ref{app:up_bound_ser_ach} specify the optimal rate allocation parameters at corner point $B_2^{S}$ (of serial transmissions) for Class \rom{1} channel regimes. These tables can be readily utilized to construct the achievability scheme of parallel transmissions at corner point $B_2^{P}$ in accordance with Eq. \eqref{eq:decomposing_signals}. This concludes the achievability at corner point $B_2^{P}$ for the case of parallel transmission.  

\subsubsection*{Class \rom{2}}
Now, we establish the achievability for all Class \rom{2} channel regimes for the low ($n_{F}\leq n_{d3}-2n_{d2}$) and high ($n_{F}\geq n_{d3}-2n_{d2}$) fronthaul capacity regimes. The following proposition specifies the achievable DTB.   
\begin{proposition}\label{prop_class_b_par}
The achievable DTB for Class \rom{2} channel regimes of the network under study for $\mu\in[0,1]$ under a parallel-fronthaul edge transmission is given for $n_{F}\leq n_{d3}-2n_{d2}$ by
\begin{equation}
\label{eq:prop_class_B_SCL_par_low_fronthaul}
\Delta_{\text{det}}(\mu,n_{F},\mathbf{n})=\begin{cases}\max\Big\{\frac{1-\mu}{n_F+n_{d2}},\frac{2-\mu}{n_F+n_{d3}},\frac{1}{n_{d3}}\Big\}&\text{ if } n_{d1}\geq n_{d3}\geq 2n_{d2}\\\max\Big\{\frac{1-\mu}{n_F+n_{d2}},\frac{2-\mu}{n_F+n_{d3}},\frac{1}{n_{d1}}\Big\}&\text{ if }n_{d1}+n_{d2}\geq n_{d3}\geq n_{d1}\geq n_{d2},n_{d3}\geq 2n_{d2}\end{cases},
\end{equation} and for $n_{F}\geq n_{d3}-2n_{d2}$ by 
\begin{equation}
\label{eq:prop_class_B_SCL_par_high_fronthaul}
\Delta_{\text{det}}(\mu,n_{F},\mathbf{n})=\begin{cases}\max\Big\{\frac{2-\mu}{n_F+n_{d3}},\frac{1}{n_{d3}}\Big\}&\text{ if } n_{d1}\geq n_{d3}\geq 2n_{d2}\\\max\Big\{\frac{2-\mu}{n_F+n_{d3}},\frac{1}{n_{d1}}\Big\}&\text{ if }n_{d1}+n_{d2}\geq n_{d3}\geq n_{d1}\geq n_{d2},n_{d3}\geq 2n_{d2}\end{cases},
\end{equation}
\end{proposition}
\vspace{0.5em}
Next, we present the scheme for all Class \rom{2} channel regimes specified in proposition \ref{prop_class_a_par} in detail. To this end, we establish the achievability at corner points $A_2$, $B_1$ and $B_2$ for low fronthaul capacities $n_{F}\leq n_{d3}-2n_{d2}$ as well as $A_1$ and $B_2$ for high fronthaul capacities $n_{F}\geq n_{d3}-2n_{d2}$. Points between these corner points are achievable due to arguments of DTB-convexity. 

To establish the achievability of $A_1$ and $A_2$, we solve the linear optimization problem at $\mu=0$. Hereby, by choosing the rate allocation parameters according to Table \ref{tab:rate_alloc_I_a_R4_R62} included in Appendix \ref{app:up_bound_par_ach} for $n_{F}\geq n_{d3}-2n_{d2}\geq 0$ the DTB $\nicefrac{2}{(n_F+n_{d3})}$ of corner point $A_1$ is attainable. On the other hand, fixing $\mathcal{\boldsymbol{\tilde{r}}}$ in accordance with Table \ref{tab:rate_alloc_I_b_R4_R61_prime_low_fronthaul} of Appendix \ref{app:up_bound_par_ach} for $n_{F}\leq n_{d3}-2n_{d2}$ accomplishes the achievable DTB $\nicefrac{1}{(n_F+n_{d2})}$ of corner point $A_2$. 

We now move to corner points $B_1$ and $B_2$ for Class \rom{2} channel regimes. For both corner points, the mapping of achievability schemes from serial to parallel transmission as given in Eq. \eqref{eq:decomposing_signals} applies. In fact, as far as $B_2$ for Class \rom{2} channel regimes are concerned, the separability in the HeNB's delivery content as stated in Eq. \eqref{eq:mu_par_vs_mu_ser_1} is identical. Thus, columns two and five of Table \ref{tab:rate_alloc_param_WCL} of Appendix \ref{app:up_bound_ser_ach} allow for the direct mapping of serial and parallel transmission of Class \rom{2} channel regimes. In terms of $B_1$, however, the HeNB's delivery content separability in Eq. \eqref{eq:mu_par_vs_mu_ser_1} changes to 
\begin{equation}\label{eq:mu_par_vs_mu_ser_2}
\mu_{S}^{\prime\prime}(\mathbf{n})\bar{L}^{*}=\mu_{P}^{\prime\prime}(n_F,\mathbf{n})\bar{L}^{*}+n_{F}. 
\end{equation} such that the RHS of Eq. \eqref{eq:decomposing_rate_2} is replaced by \eqref{eq:mu_par_vs_mu_ser_2}. Hereby with Table \ref{tab:rate_alloc_B_1}, the mapping can be readily established. This completes the achievability of corner points $A_1,A_2,B_1$ and $B_2$ for Class \rom{2} channel regimes.   
\vspace{0.5em}
\subsubsection*{Class \rom{3}}
Now, we present the DTB-optimal scheme for the Class \rom{3} channel regime under a parallel fronthaul-edge transmission for any $n_{F}\geq 0$. The following proposition quantifies the achievable DTB.   
\begin{proposition}\label{prop_class_c_par}
The achievable DTB for Class \rom{3} channel regime of the network under study for $\mu\in[0,1]$ and $n_{F}\geq 0$ under a parallel-fronthaul edge transmission corresponds to
\begin{equation}
\label{eq:prop_class_C_par}
\Delta_{\text{det}}(\mu,n_{F},\mathbf{n})=\max\Bigg\{\frac{1-\mu}{n_F+n_{d2}},\frac{1}{n_{d1}}\Bigg\}\text{ if } n_{d3}\geq n_{d1}+n_{d2}\geq n_{d1}\geq n_{d2}.
\end{equation}
\end{proposition} For corner point $A_1$, we solve the optimization problem \eqref{eq:lin_opt_problem_par} at Class \rom{3} channel regime for $\mu=0$. This results in Table \ref{tab:rate_alloc_I_b_C_62} of Appendix \ref{app:up_bound_par_ach}. 

The mapping of achievability schemes from serial to parallel fronthaul transmissions given in Eq. \ref{eq:decomposing_signals} also applies to corner point $C_1$. The only difference is that the HeNB's delivery content is separable in consonance with \begin{equation}\label{eq:mu_par_vs_mu_ser_3}
\mu_{S}^{\prime\prime\prime}(\mathbf{n})\bar{L}^{*}=\mu_{P}^{\prime\prime\prime}(n_F,\mathbf{n})\bar{L}^{*}+n_{F}. 
\end{equation} For this particular corner point, Table \ref{tab:rate_alloc_C_1} ought to be used to construct the mapping. This concludes the achievability of the Class \rom{3} channel regime. 

%% file: content/conclusion.tex
\section{Conclusion}
\label{sec:conclusion}

In this paper, we have studied the fundamental limit on the delivery time of a cloud and cache-aided HetNet in the downlink. We utilize the delivery time per bit (DTB) as the performance metric that captures the worst-case per-bit delivery latency of requested files. This metric inherently captures aspects of achievable per-user rate fairness. The linear deterministic model is used to completely characterize the optimal tradeoff between storage and delivery time for a HetNet with two transmitters and two receivers operating under parallel or serial fronthaul-edge transmissions at various channel regimes. A combination of private, common and interference-neutralizing information subject to optimized rate allocation is shown to be DTB-optimal. We identify channel regimes, where edge caching and fronthauling provide synergestic and non-synergestic benefits. Interestingly, for the case of serial transmission we determine three distinct classes of channel regimes, Class \rom{1}, \rom{2} and \rom{3}, for which operating below a threshold fronthaul capacity is  
as if the fronthaul link is absent. On the contrary, in case of parallel transmission where the HeNB operates in full-duplex mode such threshold fronthaul capacity does not exist. Instead any positive fronthaul capacity below a maximum fronthaul capacity has a DTB-reducing effect. This is due to the fact that fronthaul and wireless transmission can occur simultaneously enabling the exploitation of cloud and caching resources in the most efficient manner. 

%% file: content/appendixa.tex
\section{Lower Bound for Serial Transmission for $n_{F}=0$}
\label{app:lw_bound_ser}
In this section, we develop lower bounds on the DTB $\Delta_{\text{det}}^{*}(\mu,n_F,\mathbf{n})$ for the cache-only F-RAN ($n_F=0$) to settle the optimality of our proposed achievability scheme for various regimes of channel parameters $n_{d1},n_{d2}$ and $n_{d3}$. Based on the definition of the DTB in \eqref{eq:DTB} for serial transmission, we need to first consider the worst-case demand pattern $\mathbf{d}=(d_1, d_2)^{T}$; that is $U_1$ and $U_2$ request \emph{distinct} files $W_{d_1}$ and $W_{d_2}$ ($d_1\neq d_2$). Without loss of generality, we assume that $\mathbf{d}=(1,2)^{T}$, i.e., $W_{d_1}=W_1$ and $W_{d_2}=W_2$. Given channel realization $\mathbf{n}=(n_{d1},n_{d2},n_{d3})^{T}$, we establish lower bounds on the delivery time $T_E$ as the converse on $\Delta_{\text{det}}^{*}(\mu,n_F=0,\mathbf{n})$. This suffices because in the cache-only setting, the cloud-to-HeNB is inactive during the delivery phase such that $T_{F}=0$. The first lower bound is based on the idea that for any feasible scheme, reliable decoding of $W_1$ and $W_2$ is possible, when any arbitrary decoder is provided with side information containing $\mathbf{S}^{q-\max\{n_{d2},n_{d3}\}}\mathbf{x}_{2}^{T_{E}}$, as well as cached contents $S_1$ and $S_2$. This is due to the fact that through this side information, the decoder can recover $\mathbf{y}_1^{T_E}$ and hence $W_1$ as well as $\mathbf{y}_2^{T_E}$ and thus $W_2$.
We obtain the lower bound as follows:   
\begin{align}\label{eq:conv_1} 
2L\quad=&\quad H\big(W_1,W_2\big)\nonumber \\ =&\quad I\Big(W_1,W_2;\mathbf{S}^{q-\max\{n_{d2},n_{d3}\}}\mathbf{x}_{2}^{T_{E}},S_1,S_2\Big)
 +H\big(W_1,W_2|\mathbf{S}^{q-\max\{n_{d2},n_{d3}\}}\mathbf{x}_{2}^{T_{E}},S_1,S_2\big)\nonumber\\=&\quad I\Big(W_1,W_2;\mathbf{S}^{q-\max\{n_{d2},n_{d3}\}}\mathbf{x}_{2}^{T_{E}},S_1,S_2\Big)+H\big(W_1|\mathbf{S}^{q-\max\{n_{d2},n_{d3}\}}\mathbf{x}_{2}^{T_{E}},S_1,S_2\big)\nonumber\\ &\quad+H\big(W_2|\mathbf{S}^{q-\max\{n_{d2},n_{d3}\}}\mathbf{x}_{2}^{T_{E}},S_1,S_2,W_1\big)\nonumber\\
\stackrel{(a)}\leq&\quad I\Big(W_1,W_2;\mathbf{S}^{q-\max\{n_{d2},n_{d3}\}}\mathbf{x}_{2}^{T_{E}},S_1,S_2\Big)+H\big(W_1|\mathbf{S}^{q-\max\{n_{d2},n_{d3}\}}\mathbf{x}_{2}^{T_{E}},S_1,S_2,\mathbf{y}_{1}^{T_{E}}\big)\nonumber\\ &\quad+H\big(W_2|\mathbf{y}_{2}^{T_{E}},S_1,S_2,W_1,\mathbf{y}_{1}^{T_{E}}\big)\nonumber \\ \stackrel{(b)}\leq &\quad I\Big(W_1,W_2;\mathbf{S}^{q-\max\{n_{d2},n_{d3}\}}\mathbf{x}_{2}^{T_{E}},S_1,S_2\Big)+L\epsilon_L\nonumber \\ =&\quad H\Big(\mathbf{S}^{q-\max\{n_{d2},n_{d3}\}}\mathbf{x}_{2}^{T_{E}},S_1,S_2\Big)+L\epsilon_L\nonumber \\\leq &\quad H\Big(\mathbf{y}_{2}^{T_{E}},\mathbf{S}^{q-\max\{n_{d2},n_{d3}\}}\mathbf{x}_{2}^{T_{E}},S_1,S_2\Big)+L\epsilon_L\nonumber\\ \stackrel{(c)} \leq&\quad H\Big(\mathbf{S}^{q-\max\{n_{d2},n_{d3}\}}\mathbf{x}_{2}^{T_{E}}\Big)+H\big(S_1\big)
+H\Big(\mathbf{y}_{2}^{T_{E}}|\mathbf{S}^{q-\max\{n_{d2},n_{d3}\}}\mathbf{x}_{2}^{T_{E}},S_1\Big)
\nonumber\\&\quad+H\Big(S_2|\mathbf{S}^{q-\max\{n_{d2},n_{d3}\}}\mathbf{x}_{2}^{T_{E}},S_1,\mathbf{y}_{2}^{T_{E}}\Big)+L\epsilon_L\nonumber\\ 
\stackrel{(d)}\leq&\quad H\Big(\mathbf{S}^{q-\max\{n_{d2},n_{d3}\}}\mathbf{x}_{2}^{T_{E}}\Big)+H\big(S_1\big)+H\Big(W_2,S_2|\mathbf{S}^{q-\max\{n_{d2},n_{d3}\}}\mathbf{x}_{2}^{T_{E}},S_1,\mathbf{y}_{2}^{T_{E}}\Big)+L\epsilon_L\nonumber\\ 
\stackrel{(e)}\leq&\quad H\Big(\mathbf{S}^{q-\max\{n_{d2},n_{d3}\}}\mathbf{x}_{2}^{T_{E}}\Big)+H\big(S_1\big)+L\epsilon_L\nonumber\\ 
\leq&\quad T_{E}\max\{n_{d2},n_{d3}\}+\mu L+L\epsilon_L,
\end{align} 
where (a) is because $\mathbf{y}_{1}^{T_E}$ is a function of $\mathbf{S}^{q-\max\{n_{d2},n_{d3}\}}\mathbf{x}_{2}^{T_{E}}$, $S_1$, $S_2$ and it is because $\mathbf{S}^{q-\max\{n_{d2},n_{d3}\}}\mathbf{x}_{2}^{T_{E}}$ contains all information on $\mathbf{y}_2^{T_{E}}$, (b) follows from Fano's inequality with $\epsilon_L$ being a term that vanishes as $L\rightarrow\infty$, (c) follows since conditioning does not increase entropy, (d) is since $H\Big(\mathbf{y}_{2}^{T_{E}}|\mathbf{S}^{q-\max\{n_{d2},n_{d3}\}}\mathbf{x}_{2}^{T_{E}}\Big)=0$ and the non-negativity of the discrete entropy and (e) is due to Fano's inequality $H\Big(W_2|\mathbf{y}_{2}^{T_{E}}\Big)\leq L\epsilon_L$ and $H(S_2|W_2)=0$. 

Since the requested files $W_1$ and $W_2$ can also be retrieved from the received signals $\mathbf{y}_{1}^{T_{E}}$ and $\mathbf{y}_{2}^{T_{E}}$, another lower bound is obtained as follows:
\begin{eqnarray}\label{eq:conv_2} 
2L&=&H\big(W_1,W_2\big)\nonumber \\&=&I\Big(W_1,W_2;\mathbf{y}_{1}^{T_{E}},\mathbf{y}_{2}^{T_{E}}\Big)+H\Big(W_1,W_2|\mathbf{y}_{1}^{T_{E}},\mathbf{y}_{2}^{T_{E}}\Big)\nonumber \\&\stackrel{(a)}\leq & I\Big(W_1,W_2;\mathbf{y}_{1}^{T_{E}},\mathbf{y}_{2}^{T_{E}}\Big)+L\epsilon_L\nonumber\\ 
&=&H\big(\mathbf{y}_{2}^{T_{E}}\big)+H\big(\mathbf{y}_{1}^{T_{E}}|\mathbf{y}_{2}^{T_{E}}\big)+L\epsilon_L\nonumber\\&=&\sum_{t=1}^{T_E}\Big[H\big(\mathbf{y}_2[t]|\mathbf{y}_2^{t-1}\big)+H\big(\mathbf{y}_1[t]|\mathbf{y}_1^{t-1},\mathbf{y}_2^{T_E}\big)\Big]+L\epsilon_L\nonumber\\&\stackrel{(b)}\leq & \sum_{t=1}^{T_E}\Big[H\big(\mathbf{y}_2[t]\big)+H\big(\mathbf{y}_1[t]|\mathbf{y}_2[t]\big)\Big]+L\epsilon_L\nonumber\\&\stackrel{(c)}\leq & T_{E}n_{d3}+\sum_{t=1}^{T_E}\Big[H\big(\mathbf{S}^{q-n_{d1}}\mathbf{x}_{1}[t]\oplus\mathbf{S}^{q-n_{d2}}\mathbf{x}_{2}[t]|\mathbf{S}^{q-n_{d3}}\mathbf{x}_{2}[t]\big)\Big]+L\epsilon_L\nonumber\\&\stackrel{(d)}\leq &L\epsilon_L+T_{E}\cdot\begin{cases} n_{d1}+n_{d3}&\text{ for } n_{d3}\geq n_{d2} \\ \max\{n_{d1}+n_{d3},n_{d2}\}&\text{ for }n_{d3}\leq n_{d2}
\end{cases}\nonumber\\&=&T_{E}\max\{n_{d1}+n_{d3},n_{d2}\}+L\epsilon_L,\end{eqnarray} 
where (a) follows from Fano's inequality, (b) follows since conditioning does not increase the entropy and the independence of $\mathbf{y}_{2}[t]$ from $\mathbf{y}_{2}^{t-1}$, (c) is because the $\text{Bern}(\nicefrac{1}{2})$ distribution maximizes the binary entropy of each component of all $n_{d3}$ random elements of $\mathbf{y}_{2}[t]$, (d) follows from the fact that conditioning does not increase entropy and that the randomness in $\mathbf{S}^{q-n_{d2}}\mathbf{x}_{2}[t]$ is fully or partially contained in $\mathbf{S}^{q-n_{d3}}\mathbf{x}_{2}[t]$ depending on whether $n_{d3}\geq n_{d2}$ or $n_{d3}\leq n_{d2}$. 

Also, file $W_k$, $k\in\{1,2\}$, must be decodable if $U_{k}$ is aware of $\mathbf{y}_{k}^{T_{E}}$, yielding the lower bound on $T_{E}$ 
\begin{eqnarray}\label{eq:conv_3} 
L&=&H\big(W_k\big)\nonumber \\& = & I\big(W_k;\mathbf{y}_{k}^{T_{E}}\big)+H\big(W_k|\mathbf{y}_{k}^{T_{E}}\big)\nonumber \\&\stackrel{(a)}\leq& I\big(W_k;\mathbf{y}_{k}^{T_{E}}\big)+L\epsilon_L\nonumber\\&\leq& H\big(\mathbf{y}_{k}^{T_{E}}\big)+L\epsilon_{L}\nonumber\\&\leq&L\epsilon_L+T_{E}\cdot\begin{cases}
\max\{n_{d1},n_{d2}\}&\text{ for }k=1\\n_{d3}&\text{ for }k=2\end{cases}, \end{eqnarray} where (a) is due to Fano's inequality.
 
Finally, file $W_1$, is also decodable if $U_{1}$ is aware of the information subset $\big\{S_1,S_2,\mathbf{S}^{q-n_{d2}}\mathbf{x}_{2}^{T_{E}}\big\}$. Thus, we find the lower on $T_{E}$ as follows: 
\begin{align}\label{eq:conv_4} 
L\quad=&\quad H\big(W_1\big)\stackrel{(a)}=H\big(W_1|W_2\big)\nonumber \\ =&\quad I\Big(W_1;S_1,S_2,\mathbf{S}^{q-n_{d2}}\mathbf{x}_{2}^{T_{E}}|W_2\Big)
 +H\Big(W_1|S_1,S_2,\mathbf{S}^{q-n_{d2}}\mathbf{x}_{2}^{T_{E}},W_2\Big)\nonumber\\ \stackrel{(b)}\leq &\quad I\Big(W_1;S_1,S_2,\mathbf{S}^{q-n_{d2}}\mathbf{x}_{2}^{T_{E}}|W_2\Big)+L\epsilon_L\nonumber \\ =&\quad H\Big(S_1,S_2,\mathbf{S}^{q-n_{d2}}\mathbf{x}_{2}^{T_{E}}|W_2\Big)+L\epsilon_L\nonumber \\ \stackrel{(c)} \leq&\quad H\Big(\mathbf{S}^{q-n_{d2}}\mathbf{x}_{2}^{T_{E}}\Big)+H\big(S_1\big)
+L\epsilon_L\nonumber\\ \leq&\quad T_{E}n_{d2}+\mu L+L\epsilon_L.
\end{align} 
Step (a) follows from the independence of files $W_1$ and $W_2$, (b) follows by applying Fano's inequality to the conditional entropy and (c) is due to $H(S_2|W_2)=0$ and conditioning does not increase entropy. 

Rearranging the expressions \eqref{eq:conv_1}, \eqref{eq:conv_2}, \eqref{eq:conv_3} and \eqref{eq:conv_4} and letting $L\rightarrow\infty$ such that $\epsilon_L\rightarrow 0$, yields the desired lower bound on the DTB.

%% file: content/appendixb.tex
\section{Lower Bound for Serial Transmission for $n_{F}\geq 0$}
\label{app:lw_bound_ser_nF_greater_0}
In this section, we develop lower bounds on the DTB $\Delta_{\text{det}}^{*}(\mu,n_F,\mathbf{n})$ for the F-RAN with non-negative fronthauling capacity ($n_F\geq 0$). Through this setting, we change the definition of $q$ in the LDM to $q=\max\{n_F,n_{d1},n_{d2},n_{d3}\}$. We use the same request pattern as in Appendix \ref{app:lw_bound_ser} and develop two lower bounds on weighted linear combinations of wireless and fronthaul delivery time $T_{E}$ and $T_{F}$. For this purpose, we extend the bounds \eqref{eq:conv_1} and \eqref{eq:conv_4} by incorporating the fronthaul transmission through the message $\mathbf{S}^{q-n_{F}}\mathbf{x}_{F}^{T_{F}}$. The bounds \eqref{eq:conv_2} and \eqref{eq:conv_3} on the wireless delivery time $T_{E}$ remain valid. 

We modify bound \eqref{eq:conv_4} as follows:      
\begin{align}\label{eq:conv_1_nF} 
2L\quad=&\quad H\big(W_1,W_2\big)\nonumber \\ =&\quad I\Big(W_1,W_2;\mathbf{S}^{q-n_{F}}\mathbf{x}_{F}^{T_{F}},\mathbf{S}^{q-\max\{n_{d2},n_{d3}\}}\mathbf{x}_{2}^{T_{E}},S_1,S_2\Big)\nonumber
 \\&\quad+H\big(W_1,W_2|\mathbf{S}^{q-n_{F}}\mathbf{x}_{F}^{T_{F}},\mathbf{S}^{q-\max\{n_{d2},n_{d3}\}}\mathbf{x}_{2}^{T_{E}},S_1,S_2\big)\nonumber \\\stackrel{(a)}\leq&\quad T_{E}\max\{n_{d2},n_{d3}\}+T_{F}n_{F}+\mu L+L\epsilon_L.
\end{align} 
Similarly to \eqref{eq:conv_4}, we obtain the bound
\begin{align}\label{eq:conv_4_nF} 
L\quad=&\quad H\big(W_1\big)=H\big(W_1|W_2\big)\nonumber \\ =&\quad I\Big(W_1;S_1,S_2,\mathbf{S}^{q-n_{F}}\mathbf{x}_{F}^{T_{F}},\mathbf{S}^{q-n_{d2}}\mathbf{x}_{2}^{T_{E}}|W_2\Big)
 +H\Big(W_1|S_1,S_2,\mathbf{S}^{q-n_{F}}\mathbf{x}_{F}^{T_{F}},\mathbf{S}^{q-n_{d2}}\mathbf{x}_{2}^{T_{E}},W_2\Big)\nonumber\\ \stackrel{(a)}\leq &\quad T_{E}n_{d2}+T_{F}n_{F}+\mu L+L\epsilon_L,
\end{align} 
where (a) follows by applying the exact same bounding technique as in \eqref{eq:conv_1} and \eqref{eq:conv_4}. Rearranging the expressions \eqref{eq:conv_1_nF} and \eqref{eq:conv_4_nF} and letting $L\rightarrow\infty$ such that $\epsilon_L\rightarrow 0$, yields the following two lower bounds on linear combinations of $T_{E}$ and $T_{F}$.
\begin{subequations}
\begin{alignat}{2}
\frac{T_E}{L}+\frac{T_F}{L}\frac{n_F}{\max\{n_{d2},n_{d3}\}}&\geq\frac{2-\mu}{\max\{n_{d2},n_{d3}\}},\label{eq:lb_lin_comb1}\\
\frac{T_E}{L}+\frac{T_F}{L}\frac{n_F}{n_{d2}}&\geq \frac{1-\mu}{n_{d2}}.\label{eq:lb_lin_comb2}
\end{alignat}   
\end{subequations} 
We can easily see from \eqref{eq:lb_lin_comb1} and \eqref{eq:lb_lin_comb2} that for $n_{F}\leq\max\{n_{d2},n_{d3}\}$ and $n_{F}\leq n_{d2}$, the RHS of both inequalities function as lower bounds for $\nicefrac{(T_{E}+T_{F})}{L}$. These RHSs are in agreement with two lower bounds on the DTB for cache-only schemes (see \eqref{eq:lb_cache_only}).

%% file: content/appendixd.tex
\section{Optimal Rate Allocation for Achievabability in Serial Transmission}
\label{app:up_bound_ser_ach}

In this section, we state the optimal rate allocation parameters at corner points $B_1,B_2$ and $C_1$ when solving the rate maximization problem \eqref{eq:lin_opt_problem} under serial transmission.

\begin{table}[!htbp]
\centering
\begin{tabular}{|c||c|c|c|}
\hline
Regimes $\mathcal{R}_{B_2}$ & $n_{d1}+n_{d3}\geq n_{d2}\geq n_{d3}\geq n_{d1}$ & $2n_{d3}\geq n_{d2}\geq n_{d1}\geq n_{d3}$ & $n_{d1}\geq n_{d2}\geq n_{d3},2n_{d3}\geq n_{d2}$ 
\\ \hline 
$R_{1,p}^{w,*}=R_{1,p}^{u,*}$ & $n_{d2}-n_{d3}$ &  $n_{d2}-n_{d3}$ & $n_{d2}-n_{d3}$
\\ \hline
$R_{c}^{u,*}$ & $\frac{n_{d3}-n_{d1}}{2}$ &  $0$ & $0$
\\ \hline
$R_{2,p}^{u,*}$ & $0$ & $0$ & $0$
\\ \hline 
$R_{\text{IN}}^{v,*}$  & $n_{d1}+n_{d3}-n_{d2}$ & $2n_{d3}-n_{d2}$ & $2n_{d3}-n_{d2}$
\\ \hline
$R_{c}^{v,*}$  & $\frac{2n_{d2}-n_{d1}-n_{d3}}{2}$ & $n_{d2}-n_{d3}$ & $n_{d2}-n_{d3}$
\\ \hline
$R_{\text{IN}}^{n,*}$  & $n_{d1}+n_{d3}-n_{d2}$ & $2n_{d3}-n_{d2}$ & $2n_{d3}-n_{d2}$
\\ \hline
$l_1^{*}$ & $0$ & $0$ & $n_{d1}-n_{d2}$ 
\\ \hline
$l_2^{*}$ & $2n_{d2}-n_{d1}-n_{d3}$ & $2n_{d2}-n_{d1}-n_{d3}$ & $n_{d2}-n_{d3}$
\\ \hline
$l_3^{*}$ & $0$ & $n_{d1}-n_{d3}$ & $n_{d2}-n_{d3}$ 
\\ \hline
$l_4^{*}$ & $0$ & $0$ & $n_{d1}-n_{d2}$
\\ \hline\hline
$\bar{L}^{*}$  & $\frac{n_{d1}+n_{d3}}{2}$ & $n_{d3}$ & $n_{d3}$ 
\\ \hline
$\Delta_{\text{LB}}^{\prime}(\mathbf{n})$  & $\frac{2}{n_{d1}+n_{d3}}$ & $\frac{1}{n_{d3}}$ & $\frac{1}{n_{d3}}$
\\ \hline
$\mu^{\prime}(\mathbf{n})$  & $2-\frac{2n_{d2}}{n_{d1}+n_{d3}}$ & $2-\frac{n_{d2}}{n_{d3}}$ & $2-\frac{n_{d2}}{n_{d3}}$
\\ \hline
\end{tabular}
\caption{\small Rate allocation parameters for corner point $B_2$ at fractional cache size $\mu^{\prime}(\mathbf{n})$ in all Class \rom{1} SCL channel regimes.}
\label{tab:rate_alloc_param}
\end{table}
\begin{table}[!htbp]
\centering
\begin{tabular}{|c||c|c|c|c|}
\hline
\multirow{2}{*}{Regimes $\mathcal{R}_{B_2}$} & \multirow{2}{*}{$n_{d1}\geq n_{d3}\geq n_{d2}$} & \multicolumn{2}{c|}{$n_{d3}\geq n_{d2}\geq n_{d1}$} & \multirow{2}{*}{$n_{d1}+n_{d2}\geq n_{d3}\geq n_{d1}\geq n_{d2}$} 
\\ \cline{3-4}
& & $n_{d1}+n_{d3}\geq 2n_{d2}\geq n_{d3}$ & $2n_{d2}\geq n_{d1}+n_{d3}\geq n_{d3}$ & \\ \hline
$R_{1,p}^{w,*}=R_{1,p}^{v,*}$ & $n_{d3}-n_{d2}$  & $n_{d3}-n_{d2}$ & $n_{d3}-n_{d2}$ & $n_{d3}-n_{d2}$
\\ \hline
$R_{c}^{u,*}$ & $0$ & $n_{d3}-n_{d2}$ & $\frac{n_{d3}-n_{d1}}{2}$ & $n_{d3}-n_{d1}$  
\\ \hline
$R_{2,p}^{u,*}$ & $n_{d3}-n_{d2}$ & $0$ & $0$ & $n_{d1}-n_{d2}$
\\ \hline
$R_{\text{IN}}^{v,*}$ & $n_{d2}$ & $2n_{d2}-n_{d3}$ & $n_{d1}$ & $n_{d1}+n_{d2}-n_{d3}$ 
\\ \hline
$R_{c}^{v,*}$ & $0$ & $0$ & $\frac{2n_{d2}-n_{d1}-n_{d3}}{2}$ & $0$ 
\\ \hline
$R_{\text{IN}}^{n,*}$  & $0$ & $2n_{d2}-n_{d3}$ & $n_{d1}$ & $n_{d1}+n_{d2}-n_{d3}$
\\ \hline
$l_1^{*}$ & $n_{d1}-n_{d3}$ & $0$ & $0$ & $0$ 
\\ \hline
$l_2^{*}$ & $0$ & $n_{d3}-n_{d1}$ & $n_{d3}-n_{d1}$ & $n_{d3}-n_{d1}$
\\ \hline
$l_3^{*}$ & $0$ & $n_{d1}+n_{d3}-2n_{d2}$ & $0$ & $n_{d3}-n_{d1}$  
\\ \hline
$l_4^{*}$ & $n_{d1}-n_{d3}$ & $0$ & $0$ & $0$
\\ \hline\hline
$\bar{L}^{*}$  & $n_{d3}$ & $n_{d2}$ & $\frac{n_{d1}+n_{d3}}{2}$ & $n_{d1}$
\\ \hline
$\Delta_{\text{LB}}^{\prime}(\mathbf{n})$  & $\frac{1}{n_{d3}}$ & $\frac{1}{n_{d2}}$ & $\frac{2}{n_{d1}+n_{d3}}$ & $\frac{1}{n_{d1}}$
\\ \hline
$\mu^{\prime}(\mathbf{n})$  & $1$ & $2-\frac{n_{d3}}{n_{d2}}$ & $2-\frac{2n_{d3}}{n_{d1}+n_{d3}}$ & $2-\frac{n_{d3}}{n_{d1}}$
\\ \hline
\end{tabular}
\caption{\small Rate allocation parameters for corner point $B_2$ at fractional cache size $\mu^{\prime}(\mathbf{n})$ in all Class \rom{1} WCL channel regimes.}
\label{tab:rate_alloc_param_WCL}
\end{table}  

\begin{table}[!htbp]
\centering
\setlength\tabcolsep{4pt}
\begin{minipage}{0.55\textwidth}
\centering
\begin{tabular}{|c||c|c|}
\hline
\multirow{2}{*}{Regimes $\mathcal{R}_{B_1}$} & \multirow{2}{*}{$n_{d1}\geq n_{d3}\geq 2n_{d2}$} & $n_{d1}+n_{d2}\geq n_{d3}\geq n_{d1}\geq n_{d2},$ \\ & & $n_{d3}\geq 2n_{d2}$ \\ \hline
$R_{1,p}^{w,*}=R_{1,p}^{v,*}$ & $n_{d3}-n_{d2}$ & $n_{d3}-n_{d2}$\\ \hline
$R_{c}^{u,*}$ & $n_{d2}$ & $n_{d2}$ \\ \hline
$R_{2,p}^{u,*}$ & $n_{d3}-2n_{d2}$ & $n_{d3}-2n_{d2}$ \\ \hline
$R_{\text{IN}}^{v,*}$  & $0$ & $0$ \\ \hline
$R_{c}^{v,*}$  & $0$ & $0$ \\ \hline
$R_{\text{IN}}^{n,*}$  & $0$ & $0$ \\ \hline
$l_1^{*}$ & $n_{d1}-n_{d3}$ & $0$ \\ \hline
$l_2^{*}$ & $0$ & $n_{d3}-n_{d1}$ \\ \hline
$l_3^{*}$ & $n_{d2}$ & $n_{d2}$ \\ \hline
$l_4^{*}$ & $n_{d1}+n_{d2}-n_{d3}$ & $n_{d1}+n_{d2}-n_{d3}$
\\ \hline\hline
$\bar{L}^{*}$  & $n_{d3}-n_{d2}$ & $n_{d3}-n_{d2}$
\\ \hline
$\Delta_{\text{LB}}^{\prime}(\mathbf{n})$  & $\frac{1}{n_{d3}-n_{d2}}$ & $\frac{1}{n_{d3}-n_{d2}}$
\\ \hline
$\mu^{\prime\prime}(\mathbf{n})$  & $\frac{n_{d3}-2n_{d2}}{n_{d3}-n_{d2}}$  & $\frac{n_{d3}-2n_{d2}}{n_{d3}-n_{d2}}$ 
\\ \hline
\end{tabular}
\caption{\small Rate allocation parameters for corner point $B_1$ at fractional cache size $\mu^{\prime\prime}(\mathbf{n})$}
\label{tab:rate_alloc_B_1} 
\end{minipage}%
\hfill
\begin{minipage}{0.43\textwidth}
\centering
\begin{tabular}{|c||c|}
\hline
\multirow{2}{*}{Regime $\mathcal{R}_{C_1}$} & \multirow{2}{*}{$n_{d3}\geq n_{d1}+n_{d2}\geq n_{d1}\geq n_{d2}$} \\ &  \\ \hline
$R_{1,p}^{w,*}=R_{1,p}^{v,*}$ & $n_{d3}-n_{d2}$ \\ \hline 
$R_{c}^{u,*}$ & $n_{d2}$ \\ \hline
$R_{2,p}^{u,*}$ & $n_{d1}-n_{d2}$ \\ \hline
$R_{\text{IN}}^{v,*}$  & $0$ \\ \hline
$R_{c}^{v,*}$  & $0$ \\ \hline
$R_{\text{IN}}^{n,*}$  & $0$ \\ \hline
$l_1^{*}$ & $0$ \\ \hline
$l_2^{*}$ & $n_{d3}-n_{d1}$ \\ \hline
$l_3^{*}$ & $n_{d2}$ \\ \hline
$l_4^{*}$ & $0$
\\ \hline\hline
$\bar{L}^{*}$  & $n_{d1}$
\\ \hline
$\Delta_{\text{LB}}^{\prime}(\mathbf{n})$  & $\frac{1}{n_{d1}}$
\\ \hline
$\mu^{\prime\prime\prime}(\mathbf{n})$  & $\frac{n_{d1}-n_{d2}}{n_{d1}}$
\\ \hline
\end{tabular}
\caption{\small Rate allocation parameters for corner point $C_1$ at fractional cache size $\mu^{\prime\prime\prime}(\mathbf{n})$} 
\label{tab:rate_alloc_C_1} 
\end{minipage}
\end{table}

%% file: content/appendixc.tex
\section{Lower Bound for Parallel Transmission}
\label{app:lw_bound_par}
In this section, we develop lower bounds on the DTB $\Delta_{\text{det}}^{*}(\mu,n_F,\mathbf{n})$ for the F-RAN with non-negative fronthauling capacity ($n_F\geq 0$) under a parallel fronthaul-edge transmission setting. We change the definition of $q$ in the LDM to $q=\max\{n_F,n_{d1},n_{d2},n_{d3}\}$. We use the same request pattern as in Appendix \ref{app:lw_bound_ser} and develop two lower bounds on the delivery time $T_{P}$. For this purpose, we modify the bounds \eqref{eq:conv_1_nF} and \eqref{eq:conv_4_nF} by replacing $T_{E}$ with $T_{P}$ ($T_{E}\leftrightarrow T_{P}$). The bounds \eqref{eq:conv_2} and \eqref{eq:conv_3} on the delivery time $T_{P}$ remain valid. 

We modify bound \eqref{eq:conv_1_nF} as follows:      
\begin{align}\label{eq:conv_1_par} 
2L\quad=&\quad H\big(W_1,W_2\big)\nonumber \\ =&\quad I\Big(W_1,W_2;\mathbf{S}^{q-n_{F}}\mathbf{x}_{F}^{T_{P}},\mathbf{S}^{q-\max\{n_{d2},n_{d3}\}}\mathbf{x}_{2}^{T_{P}},S_1,S_2\Big)\nonumber
 \\&\quad+H\big(W_1,W_2|\mathbf{S}^{q-n_{F}}\mathbf{x}_{F}^{T_{P}},\mathbf{S}^{q-\max\{n_{d2},n_{d3}\}}\mathbf{x}_{2}^{T_{P}},S_1,S_2\big)\nonumber \\\stackrel{(a)}\leq&\quad T_{P}\big(n_{F}+\max\{n_{d2},n_{d3}\}\big)+\mu L+L\epsilon_L.
\end{align} 
Similarly to \eqref{eq:conv_4_nF}, we obtain the bound
\begin{align}\label{eq:conv_4_par} 
L\quad=&\quad H\big(W_1\big)=H\big(W_1|W_2\big)\nonumber \\ =&\quad I\Big(W_1;S_1,S_2,\mathbf{S}^{q-n_{F}}\mathbf{x}_{F}^{T_{P}},\mathbf{S}^{q-n_{d2}}\mathbf{x}_{2}^{T_{P}}|W_2\Big)
 +H\Big(W_1|S_1,S_2,\mathbf{S}^{q-n_{F}}\mathbf{x}_{F}^{T_{P}},\mathbf{S}^{q-n_{d2}}\mathbf{x}_{2}^{T_{P}},W_2\Big)\nonumber\\ \stackrel{(a)}\leq &\quad T_{P}\big(n_{F}+n_{d2}\big)+\mu L+L\epsilon_L,
\end{align} 
where (a) follows by applying the exact same bounding technique as in \eqref{eq:conv_1} and \eqref{eq:conv_4}. Rearranging the expressions \eqref{eq:conv_1_par} and \eqref{eq:conv_4_par} and letting $L\rightarrow\infty$ such that $\epsilon_L\rightarrow 0$, yields the following two lower bounds on the DTB for the parallel fronthaul-edge transmission setting.
\begin{subequations}
\begin{alignat}{2}
\frac{T_P}{L}&\geq\frac{2-\mu}{n_{F}+\max\{n_{d2},n_{d3}\}},\label{eq:lb_par_1}\\
\frac{T_P}{L}&\geq \frac{1-\mu}{n_{F}+n_{d2}}.\label{eq:lb_par_2}
\end{alignat}   
\end{subequations} 
Combining bounds \eqref{eq:conv_2}, \eqref{eq:conv_3}, \eqref{eq:lb_par_1} and \eqref{eq:lb_par_2} on the DTB for the parallel fronthaul-edge transmission case finalizes the proof.

%% file: content/appendixe.tex
\section{Optimal Rate Allocation for Achievabability in Parallel Transmission}
\label{app:up_bound_par_ach}

In this section, we state the optimal rate allocation parameters at corner points $A_1$ and $A_2$ when solving the rate maximization problem \eqref{eq:lin_opt_problem_par} under parallel transmission.

\begin{table}[!htbp]
\scriptsize
\centering
\begin{tabular}{|c||c|c|c|c|c|c|}
\hline
Regimes $\mathcal{R}_{A_1}$ & \multicolumn{2}{c|}{$n_{d1}+n_{d3}\geq n_{d2}\geq n_{d3}\geq n_{d1}$} & \multicolumn{2}{c|}{$2n_{d3}\geq n_{d2}\geq n_{d1}\geq n_{d3}$} & \multicolumn{2}{c|}{$n_{d1}\geq n_{d2}\geq n_{d3},2n_{d3}\geq n_{d2}$}
\\ \hline
Time $t$ & $t_1$ & $t_2$ & $t_1$ & $t_2$ & $t_1$ & $t_2$ 
\\ \hline
$R_{1,p}^{w,*}=R_{1,p}^{u,*}$ &  $n_{d2}-n_{d3}$ &  $n_{d2}-n_{d3}$ &  $n_{d2}-n_{d3}$ &  $n_{d2}-n_{d3}$ &  $n_{d2}-n_{d3}$ &  $n_{d2}-n_{d3}$
\\ \hline
\multirow{2}{*}{$R_{1,c}^{u,*}$} &  $\big(n_{F,\max}(\mathbf{n})-$ &  \multirow{2}{*}{$0$} &  $\big(n_{F,\max}(\mathbf{n})-$ &  \multirow{2}{*}{$0$} &  $\big(n_{F,\max}(\mathbf{n})-$ &  \multirow{2}{*}{$0$}
\\ & $n_{F}\big)^{+}$ & & $n_{F}\big)^{+}$ & & $n_{F}\big)^{+}$ &
\\ \hline 
$R_{2,c}^{u,*}$  & $n_{d3}-n_{d1}$ & $0$ & \multicolumn{2}{c|}{$0$} & \multicolumn{2}{c|}{$0$}
\\ \hline
$R_{2,p}^{u,*}$ & \multicolumn{6}{c|}{$0$}
\\ \hline
$R_{\text{IN}}^{v,*}$ & \multicolumn{6}{c|}{$\min\big\{n_F,n_{F,\max}(\mathbf{n})\big\}$}
\\ \hline
\multirow{2}{*}{$R_{1,c}^{v,*}$} &  \multirow{2}{*}{$0$} & $\big(n_{F,\max}(\mathbf{n})-$ & \multirow{2}{*}{$0$} & $\big(n_{F,\max}(\mathbf{n})-$ & \multirow{2}{*}{$0$} & $\big(n_{F,\max}(\mathbf{n})-$ \\ & & $n_{F}\big)^{+}$ & & $n_{F}\big)^{+}$ & & $n_{F}\big)^{+}$
\\ \hline 
$R_{2,c}^{v,*}$  & $n_{d2}-n_{d3}$ & $n_{d2}-n_{d1}$ & \multicolumn{2}{c|}{$n_{d2}-n_{d3}$} & \multicolumn{2}{c|}{$n_{d2}-n_{d3}$}
\\ \hline
$R_{\text{IN}}^{n,*}$ & \multicolumn{6}{c|}{$\min\big\{n_F,n_{F,\max}(\mathbf{n})\big\}$}
\\ \hline
$l_1^{*}$ &  \multicolumn{2}{c|}{$0$} & \multicolumn{2}{c|}{$0$} & \multicolumn{2}{c|}{$n_{d1}-n_{d2}$} 
\\ \hline 
$l_2^{*}$  &  \multicolumn{2}{c|}{$2n_{d2}-n_{d1}-n_{d3}$} &  \multicolumn{2}{c|}{$2n_{d2}-n_{d1}-n_{d3}$} & \multicolumn{2}{c|}{$n_{d2}-n_{d3}$}   
\\ \hline
$l_3^{*}$ &  \multicolumn{2}{c|}{$\big(n_{F,\max}(\mathbf{n})-n_{F}\big)^{+}$} & \multicolumn{2}{c|}{$n_{d1}-n_{d3}+\big(n_{F,\max}(\mathbf{n})-n_{F}\big)^{+}$} & \multicolumn{2}{c|}{$n_{d2}-n_{d3}+\big(n_{F,\max}(\mathbf{n})-n_{F}\big)^{+}$}
\\ \hline
$l_4^{*}$ & \multicolumn{2}{c|}{$0$} & \multicolumn{2}{c|}{$0$} & \multicolumn{2}{c|}{$n_{d1}-n_{d2}$}
\\ \hline
\hline
$n_{F,\max}(\mathbf{n})$ & \multicolumn{2}{c|}{$n_{d1}+n_{d3}-n_{d2}$} & \multicolumn{2}{c|}{$2n_{d3}-n_{d2}$} & \multicolumn{2}{c|}{$2n_{d3}-n_{d2}$} \\ \hline
$\tilde{L}^{*}$ & \multicolumn{2}{c|}{$\min\big\{n_{d2}+n_{F},n_{d2}+n_{F,\max}(\mathbf{n})\big\}$} & \multicolumn{2}{c|}{$\min\big\{n_{d2}+n_{F},n_{d2}+n_{F,\max}(\mathbf{n})\big\}$} & \multicolumn{2}{c|}{$\min\big\{n_{d2}+n_{F},n_{d2}+n_{F,\max}(\mathbf{n})\big\}$} \\ \hline
$\Delta_{\text{det}}^{*}(0,n_F,\mathbf{n})$ & \multicolumn{2}{c|}{$\frac{2}{\min\big\{n_{d2}+n_{F},n_{d2}+n_{F,\max}(\mathbf{n})\big\}}$} & \multicolumn{2}{c|}{$\frac{2}{\min\big\{n_{d2}+n_{F},n_{d2}+n_{F,\max}(\mathbf{n})\big\}}$} & \multicolumn{2}{c|}{$\frac{2}{\min\big\{n_{d2}+n_{F},n_{d2}+n_{F,\max}(\mathbf{n})\big\}}$} \\ \hline
\end{tabular}
\caption{\small Rate allocation parameters for corner point $A_1$ at fractional cache size $\mu=0$ in all Class \rom{1} SCL channel regimes.}
\label{tab:rate_alloc_I_a_R1_R3}
\end{table}
\begin{table}[!htbp]
\scriptsize
\centering
\begin{tabular}{|c||c|c|c|c|}
\hline
\multirow{2}{*}{Regimes $\mathcal{R}_{A_1}$} & \multicolumn{4}{c|}{$n_{d3}\geq n_{d2}\geq n_{d1}$} 
\\ \cline{2-5}
& \multicolumn{2}{c|}{$n_{d1}+n_{d3}\geq 2n_{d2}\geq n_{d3}$} & \multicolumn{2}{c|}{$2n_{d2}\geq n_{d1}+n_{d3}\geq n_{d3}$} \\ \hline
Time $t$ & $t_1$ & $t_2$ & $t_1$ & $t_2$ 
\\ \hline
$R_{1,p}^{w,*}=R_{1,p}^{v,*}$ & \multicolumn{4}{c|}{$n_{d3}-n_{d2}$}
\\ \hline
$R_{1,c}^{u,*}$ & $(n_{F,\max}(\mathbf{n})-n_F)^{+}$ & $0$ & $(n_{F,\max}(\mathbf{n})-n_F)^{+}$ & $0$
\\ \hline
$R_{2,c}^{u,*}$  & \multicolumn{2}{c|}{$n_{d3}-n_{d2}$} & $n_{d3}-n_{d2}$ & $n_{d2}-n_{d1}$
\\ \hline
$R_{2,p}^{u,*}$ &  \multicolumn{4}{c|}{$0$}
\\ \hline
$R_{\text{IN}}^{v,*}$ & \multicolumn{4}{c|}{$\min\big\{n_F,n_{F,\max}(\mathbf{n})\big\}$}
\\ \hline
$R_{1,c}^{v,*}$ & $0$ & $(n_{F,\max}(\mathbf{n})-n_F)^{+}$ & $0$ & $(n_{F,\max}(\mathbf{n})-n_F)^{+}$  
\\ \hline
$R_{2,c}^{v,*}$ & \multicolumn{2}{c|}{$0$} & $2n_{d2}-n_{d1}-n_{d3}$ & $0$  
\\ \hline
$R_{\text{IN}}^{n,*}$ & \multicolumn{4}{c|}{$\min\big\{n_F,n_{F,\max}(\mathbf{n})\big\}$}
\\ \hline
$l_1^{*}$ &  \multicolumn{4}{c|}{$0$} 
\\ \hline 
$l_2^{*}$  &  \multicolumn{4}{c|}{$n_{d3}-n_{d1}$} 
\\ \hline
$l_3^{*}$ &  \multicolumn{2}{c|}{$n_{d1}+n_{d3}-2n_{d2}+(n_{F,\max}(\mathbf{n})-n_F)^{+}$} & \multicolumn{2}{c|}{$(n_{F,\max}(\mathbf{n})-n_F)^{+}$}
\\ \hline
$l_4^{*}$ & \multicolumn{4}{c|}{$0$}
\\ \hline
\hline
$n_{F,\max}(\mathbf{n})$ & \multicolumn{2}{c|}{$2n_{d2}-n_{d3}$} & \multicolumn{2}{c|}{$n_{d1}$}
\\ \hline
$\tilde{L}^{*}$ & \multicolumn{2}{c|}{$\min\big\{n_{d3}+n_{F},n_{d3}+n_{F,\max}(\mathbf{n})\big\}$} & \multicolumn{2}{c|}{$\min\big\{n_{d3}+n_{F},n_{d3}+n_{F,\max}(\mathbf{n})\big\}$} \\ \hline
$\Delta_{\text{det}}^{*}(0,n_F,\mathbf{n})$ & \multicolumn{2}{c|}{$\frac{2}{\min\big\{n_{d3}+n_{F},n_{d3}+n_{F,\max}(\mathbf{n})\big\}}$} & \multicolumn{2}{c|}{$\frac{2}{\min\big\{n_{d3}+n_{F},n_{d3}+n_{F,\max}(\mathbf{n})\big\}}$} \\ \hline
\end{tabular}
\caption{\small Rate allocation parameters for corner point $A_1$ at fractional cache size $\mu^{\prime}=0$ in Class \rom{1} WCL channel regimes for $n_{F}\geq 0$}
\label{tab:rate_alloc_I_a_R51_R52}
\end{table}
\newcolumntype{C}{>{\centering\arraybackslash}p{11.0em}}
\begin{table}[!htbp]
\centering
\scriptsize
\begin{tabular}{|c||C|C|C|C|}
\hline
Regimes $\mathcal{R}_{A_1}$ & \multicolumn{2}{c|}{$n_{d1}\geq n_{d3}\geq n_{d2}$} & \multicolumn{2}{c|}{$n_{d1}+n_{d2}\geq n_{d3}\geq n_{d1}\geq n_{d2}$} 
\\ \hline
Time $t$ & $t_1$ & $t_2$ & $t_1$ & $t_2$ 
\\ \hline
$R_{1,p}^{w,*}=R_{1,p}^{v,*}$ & \multicolumn{2}{c|}{$n_{d3}-n_{d2}$} & \multicolumn{2}{c|}{$n_{d3}-n_{d2}$}
\\ \hline
$R_{1,c}^{u*}$  & $\min\big\{n_{d2},(n_{d3}-n_F)^{+}\big\}$ & $\big(n_{d3}-n_{d2}-n_F\big)^{+}$ & $\min\big\{n_{d2},\max\{n_{d1}-n_{F},n_{d3}-n_{d1}\}\big\}$ & $\max\big\{n_{d3}-n_{d2}-n_F,n_{d3}-n_{d1}\big\}$
\\ \hline
$R_{2,c}^{u*}$  & \multicolumn{2}{c|}{$0$} & \multicolumn{2}{c|}{$0$}
\\ \hline
$R_{2,p}^{u,*}$ & $\min\{n_{d3}-n_{d2},n_{F}\}$ & $\min\Big\{\big(\max\{n_{d3}-2n_{d2},n_F-n_{d2}\}\big)^{+},n_{d3}-n_{d2}\Big\}$ &  $\min\{n_{F},n_{d1}-n_{d2}\}$ & $\min\Big\{\big(\max\{n_{d3}-2n_{d2},n_F-n_{d1}-n_{d2}+n_{d3}\}\big)^{+},n_{d1}-n_{d2}\Big\}$
 \\ \hline
$R_{\text{IN}}^{v,*}$ & $\big(\min\{n_{d2}+n_F-n_{d3},n_{d2}\}\big)^{+}$ & $\min\{2n_{d2}-n_{d3}+n_{F},n_{d2},n_{F}\}$ & $\big(\min\{n_{d2}+n_F-n_{d1},n_{d1}+n_{d2}-n_{d3}\}\big)^{+}$ & $\min\{2n_{d2}-n_{d3}+n_{F},n_{d1}+n_{d2}-n_{d3},n_{F}\}$
\\ \hline
$R_{1,c}^{v,*}$ & $0$ & $\big(\min\{2n_{d2}-n_{d3},n_{d2}-n_{F}\}\big)^{+}$ & $0$ & $\big(\min\{2n_{d2}-n_{d3},n_{d1}+n_{d2}-n_{d3}-n_{F}\}\big)^{+}$
\\ \hline
$R_{2,c}^{v,*}$ & \multicolumn{2}{c|}{$0$} & \multicolumn{2}{c|}{$0$}
\\ \hline
$R_{\text{IN}}^{n,*}$ & $\big(\min\{n_{d2}+n_F-n_{d3},n_{d2}\}\big)^{+}$ & $\min\{2n_{d2}-n_{d3}+n_{F},n_{d2},n_{F}\}$ & $\big(\min\{n_{d2}+n_F-n_{d1},n_{d1}+n_{d2}-n_{d3}\}\big)^{+}$ & $\min\{2n_{d2}-n_{d3}+n_{F},n_{d1}+n_{d2}-n_{d3},n_{F}\}$
\\ \hline
$l_1^{*}$ & \multicolumn{2}{c|}{$n_{d1}-n_{d3}$} & \multicolumn{2}{c|}{$0$}
\\ \hline
$l_2^{*}$  & \multicolumn{2}{c|}{$0$} & \multicolumn{2}{c|}{$n_{d3}-n_{d1}$}
\\ \hline
$l_3^{*}$ & $\min\{n_{d2},(n_{d3}-n_{F})^{+}\}$ & $\big(\max\{n_{d3}-n_{d2}-n_{F},n_{d2}-n_{F}\}\big)^{+}$ & $\min\big\{n_{d2},\max\{n_{d1}-n_{F},n_{d3}-n_{d1}\}\big\}$ & $\max\{n_{d3}-n_{d2}-n_{F},n_{d3}-n_{d1},n_{d2}-n_{F}\}$
\\ \hline
$l_4^{*}$ & $n_{d1}-n_{d3}+\big(n_{d3}-n_{d2}-n_{F}\big)^{+}$ & $\max\big\{\min\big\{n_{d1}+n_{d2}-n_{d3},n_{d1}-n_F,n_{d1}-n_{d2}\big\},n_{d1}-n_{d3}\big\}$ & $\big(n_{d1}-n_{d2}-n_F\big)^{+}$ & $\big(\min\{n_{d1}+n_{d2}-n_{d3},n_{F,\max}(\mathbf{n})-n_{F},n_{d1}-n_{d2}\}\big)^{+}$
\\ \hline\hline
$n_{F,\max}(\mathbf{n})$ & \multicolumn{2}{c|}{$n_{d3}$} & \multicolumn{2}{c|}{$2n_{d1}-n_{d3}$}
\\ \hline
$\tilde{L}^{*}$ & \multicolumn{2}{c|}{$\min\big\{n_{d3}+n_{F},n_{d3}+n_{F,\max}(\mathbf{n})\big\}$} & \multicolumn{2}{c|}{$\min\big\{n_{d3}+n_{F},n_{d3}+n_{F,\max}(\mathbf{n})\big\}$} \\ \hline
$\Delta_{\text{det}}^{*}(0,n_F,\mathbf{n})$ & \multicolumn{2}{c|}{$\frac{2}{\min\big\{n_{d3}+n_{F},n_{d3}+n_{F,\max}(\mathbf{n})\big\}}$} & \multicolumn{2}{c|}{$\frac{2}{\min\big\{n_{d3}+n_{F},n_{d3}+n_{F,\max}(\mathbf{n})\big\}}$} \\ \hline
\end{tabular}
\caption{\small Rate allocation parameters for corner point $A_1$ at fractional cache size $\mu=0$ for $n_{F}\geq (n_{d3}-2n_{d2})^{+}$. If $(n_{d3}-2n_{d2})^{+}=0$, we describe the remaining Class \rom{1} WCL channel regimes. Otherwise, if $(n_{d3}-2n_{d2})^{+}\geq 0$, the rate allocation parameters represent Class \rom{2} channel regimes.}
\label{tab:rate_alloc_I_a_R4_R62}
\end{table}
\begin{table}[!htbp]
\centering
\scriptsize
\begin{tabular}{|c||c|c|}
\hline
Regimes $\mathcal{R}_{A_2}$ & $n_{d1}\geq n_{d3}\geq 2n_{d2}$ & $n_{d1}+n_{d2}\geq n_{d3}\geq n_{d1}\geq n_{d2},n_{d3}\geq 2n_{d2}$
\\ \hline
Time $t$ & $[t_1:t_2]$ & $[t_1:t_2]$
\\ \hline
$R_{1,p}^{w,*}=R_{1,p}^{v,*}$ & $n_{d3}-n_{d2}$ & $n_{d3}-n_{d2}$
\\ \hline
$R_{1,c}^{u,*}$  & $0$ & $0$
\\ \hline
$R_{2,c}^{u,*}$  & $n_{d2}$ & $n_{d2}$
\\ \hline
$R_{2,p}^{u,*}$ & $\min\big\{n_F,n_{F,\text{IM}}(\mathbf{n})\big\}$ & $\min\big\{n_F,n_{F,\text{IM}}(\mathbf{n})\big\}$
\\ \hline
$R_{\text{IN}}^{v,*}$ & $0$ & $0$
\\ \hline
$R_{1,c}^{v,*}$ & $0$ & $0$ 
\\ \hline
$R_{2,c}^{v,*}$ & $0$ & $0$ 
\\ \hline
$R_{\text{IN}}^{n,*}$ & $0$ & $0$ 
\\ \hline
$l_1^{*}$ & $n_{d1}-n_{d3}$ & $0$  
\\ \hline 
$l_2^{*}$  & $0$ & $n_{d3}-n_{d1}$   
\\ \hline
$l_3^{*}$ & $n_{d2}$ & $n_{d2}$
\\ \hline
$l_4^{*}$ & $\max\big\{n_{d1}-n_{d2}-n_F,n_{d1}+n_{d2}-n_{d3}\big\}$ & $\max\big\{n_{d1}-n_{d2}-n_F,n_{d1}+n_{d2}-n_{d3}\big\}$
\\ \hline
\hline
$n_{F,\text{IM}}(\mathbf{n})$ & $n_{d3}-2n_{d2}$ & $n_{d3}-2n_{d2}$ \\ \hline
$\tilde{L}^{*}$ & $2\min\big\{n_{d2}+n_{F},n_{d2}+n_{F,\text{IM}}(\mathbf{n})\big\}$ & $2\min\big\{n_{d2}+n_{F},n_{d2}+n_{F,\text{IM}}(\mathbf{n})\big\}$ \\ \hline
$\Delta_{\text{det}}^{*}(0,n_F,\mathbf{n})$ & $\frac{1}{\min\big\{n_{d2}+n_{F},n_{d2}+n_{F,\text{IM}}(\mathbf{n})\big\}}$ & $\frac{1}{\min\big\{n_{d2}+n_{F},n_{d2}+n_{F,\text{IM}}(\mathbf{n})\big\}}$ \\ \hline
\end{tabular}
\caption{\small Rate allocation parameters for corner point $A_2$ at fractional cache size $\mu^{\prime}=0$ in all Class \rom{2} channel regimes for $n_{F}\leq n_{d3}-2n_{d2}$.}
\label{tab:rate_alloc_I_b_R4_R61_prime_low_fronthaul}
\end{table}
\begin{table}[!htbp]
\centering
\scriptsize
\begin{tabular}{|c||c|}
\hline
Regimes $\mathcal{R}_{A_2}$ & $n_{d3}\geq n_{d1}+n_{d2}\geq n_{d1}\geq n_{d2}$
\\ \hline
Time $t$ & $[t_1:t_2]$ 
\\ \hline
$R_{1,p}^{w,*}=R_{1,p}^{v,*}$ & $n_{d3}-n_{d2}$
\\ \hline
$R_{1,c}^{u,*}$  & $0$
\\ \hline
$R_{2,c}^{u,*}$  & $n_{d2}$
\\ \hline
$R_{2,p}^{u,*}$ & $\min\big\{n_F,n_{F,\max}(\mathbf{n})\big\}$
\\ \hline
$R_{\text{IN}}^{v,*}$ & $0$
\\ \hline
$R_{1,c}^{v,*}$ & $0$
\\ \hline
$R_{2,c}^{v,*}$ & $0$
\\ \hline
$R_{\text{IN}}^{n,*}$ & $0$
\\ \hline
$l_1^{*}$ & $0$ 
\\ \hline 
$l_2^{*}$  & $n_{d3}-n_{d1}$  
\\ \hline
$l_3^{*}$ & $n_{d2}$
\\ \hline
$l_4^{*}$ & $\big(n_{F,\max}(\mathbf{n})-n_F\big)^{+}$
\\ \hline
\hline
$n_{F,\max}(\mathbf{n})$ & $n_{d1}-n_{d2}$ \\ \hline
$\tilde{L}^{*}$ & $2\min\big\{n_{d2}+n_{F},n_{d2}+n_{F,\max}(\mathbf{n})\big\}$ \\ \hline
$\Delta_{\text{det}}^{*}(0,n_F,\mathbf{n})$ & $\frac{1}{\min\big\{n_{d2}+n_{F},n_{d2}+n_{F,\max}(\mathbf{n})\big\}}$ \\ \hline
\end{tabular}
\caption{\small Rate allocation parameters for corner point $A_2$ at fractional cache size $\mu^{\prime}=0$ in all Class \rom{3} channel regimes for $n_F\geq 0$.}
\label{tab:rate_alloc_I_b_C_62}
\end{table}

%% file: main.bbl
\begin{thebibliography}{10}
\providecommand{\url}[1]{#1}
\csname url@samestyle\endcsname
\providecommand{\newblock}{\relax}
\providecommand{\bibinfo}[2]{#2}
\providecommand{\BIBentrySTDinterwordspacing}{\spaceskip=0pt\relax}
\providecommand{\BIBentryALTinterwordstretchfactor}{4}
\providecommand{\BIBentryALTinterwordspacing}{\spaceskip=\fontdimen2\font plus
\BIBentryALTinterwordstretchfactor\fontdimen3\font minus
  \fontdimen4\font\relax}
\providecommand{\BIBforeignlanguage}[2]{{%
\expandafter\ifx\csname l@#1\endcsname\relax
\typeout{** WARNING: IEEEtran.bst: No hyphenation pattern has been}%
\typeout{** loaded for the language `#1'. Using the pattern for}%
\typeout{** the default language instead.}%
\else
\language=\csname l@#1\endcsname
\fi
#2}}
\providecommand{\BIBdecl}{\relax}
\BIBdecl

\bibitem{Kakar}
J.~Kakar, S.~Gherekhloo, Z.~H. Awan, and A.~Sezgin, ``{Fundamental Limits on
  Latency in Cloud- and Cache-Aided HetNets},'' \emph{IEEE International
  Conference on Communications}, May 2017.

\bibitem{cisco}
Cisco, ``{The Zettabyte Era: Trends and Analysis},'' Tech. Rep., 2014.

\bibitem{Bastug}
E.~Bastug, M.~Bennis, and M.~Debbah, ``Living on the edge: The role of
  proactive caching in {5G} wireless networks,'' \emph{IEEE Communications
  Magazine}, vol.~52, no.~8, pp. 82--89, Aug 2014.

\bibitem{Wang12}
X.~Wang, A.~V. Vasilakos, M.~Chen, Y.~Liu, and T.~T. Kwon, ``A survey of green
  mobile networks: Opportunities and challenges,'' \emph{Mob. Netw. Appl.},
  vol.~17, no.~1, pp. 4--20, Feb. 2012.

\bibitem{Wang14}
X.~Wang, M.~Chen, T.~Taleb, A.~Ksentini, and V.~C.~M. Leung, ``Cache in the
  air: exploiting content caching and delivery techniques for {5G} systems,''
  \emph{IEEE Communications Magazine}, vol.~52, no.~2, pp. 131--139, February
  2014.

\bibitem{Andrews13}
J.~G. Andrews, ``Seven ways that hetnets are a cellular paradigm shift,''
  \emph{IEEE Communications Magazine}, vol.~51, no.~3, pp. 136--144, March
  2013.

\bibitem{Shanmugam13}
K.~Shanmugam, N.~Golrezaei, A.~G. Dimakis, A.~F. Molisch, and G.~Caire,
  ``Femtocaching: Wireless content delivery through distributed caching
  helpers,'' \emph{IEEE Transactions on Information Theory}, vol.~59, no.~12,
  pp. 8402--8413, Dec 2013.

\bibitem{Malladi12}
D.~P. Malladi, ``Heterogeneous networks in {3G} and {4G},'' \emph{IEEE
  Communications Theory Workshop}, May 2012.

\bibitem{Lozano}
A.~Lozano, R.~W. Heath, and J.~G. Andrews, ``Fundamental limits of
  cooperation,'' \emph{IEEE Transactions on Information Theory}, vol.~59,
  no.~9, pp. 5213--5226, Sept 2013.

\bibitem{Soheil}
S.~Gherekhloo, A.~Chaaban, and A.~Sezgin, ``{Cooperation for Interference
  Management: A GDoF Perspective},'' \emph{IEEE Transactions on Information
  Theory}, vol.~62, no.~12, pp. 6986--7029, Dec 2016.

\bibitem{Checko}
A.~Checko, H.~L. Christiansen, Y.~Yan, L.~Scolari, G.~Kardaras, M.~S. Berger,
  and L.~Dittmann, ``Cloud {RAN} for mobile networks -- a technology
  overview,'' \emph{IEEE Communications Surveys Tutorials}, vol.~17, no.~1, pp.
  405--426, 2015.

\bibitem{Peng16}
M.~Peng, S.~Yan, K.~Zhang, and C.~Wang, ``Fog-computing-based radio access
  networks: issues and challenges,'' \emph{IEEE Network}, vol.~30, no.~4, pp.
  46--53, July 2016.

\bibitem{Azimi}
S.~M. Azimi, O.~Simeone, and R.~Tandon, ``Fundamental limits on latency in
  small-cell caching systems: An information-theoretic analysis,'' in
  \emph{2016 IEEE Global Communications Conference (GLOBECOM)}, Dec 2016, pp.
  1--6.

\bibitem{Liu2011}
Y.~Liu and E.~Erkip, ``Completion time in broadcast channel and interference
  channel,'' in \emph{Annual Allerton Conference on Communication, Control, and
  Computing (Allerton)}, Sept 2011, pp. 1694--1701.

\bibitem{Maddah-Ali2}
M.~A. Maddah-Ali and U.~Niesen, ``Fundamental limits of caching,'' \emph{Trans.
  on Info. Theory}, vol.~60, no.~5, pp. 2856--2867, May 2014.

\bibitem{Maddah_Ali}
------, ``Cache-aided interference channels,'' in \emph{IEEE ISIT}, June 2015,
  pp. 809--813.

\bibitem{avik}
A.~Sengupta, R.~Tandon, and O.~Simeone, ``Cache aided wireless networks:
  Tradeoffs between storage and latency,'' in \emph{CISS}, March 2016, pp.
  320--325.

\bibitem{Xu16}
F.~Xu, M.~Tao, and K.~Liu, ``Fundamental tradeoff between storage and latency
  in cache-aided wireless interference networks,'' \emph{CoRR}, vol.
  abs/1605.00203, 2016.

\bibitem{Hachem16}
\BIBentryALTinterwordspacing
J.~Hachem, U.~Niesen, and S.~N. Diggavi, ``Degrees of freedom of cache-aided
  wireless interference networks,'' \emph{CoRR}, vol. abs/1606.03175, 2016.
  [Online]. Available: \url{http://arxiv.org/abs/1606.03175}
\BIBentrySTDinterwordspacing

\bibitem{Amiri16}
M.~M. Amiri, Q.~Yang, and D.~G\"und\"uz, ``Coded caching for a large number of
  users,'' in \emph{2016 IEEE Information Theory Workshop (ITW)}, Sept 2016,
  pp. 171--175.

\bibitem{Wan16}
K.~Wan, D.~Tuninetti, and P.~Piantanida, ``On caching with more users than
  files,'' in \emph{IEEE International Symposium on Information Theory}, July
  2016, pp. 135--139.

\bibitem{Liu15}
A.~Liu and V.~K.~N. Lau, ``Exploiting base station caching in mimo cellular
  networks: Opportunistic cooperation for video streaming,'' \emph{IEEE
  Transactions on Signal Processing}, vol.~63, no.~1, pp. 57--69, Jan 2015.

\bibitem{Naderializadeh17}
N.~Naderializadeh, M.~A. Maddah-Ali, and A.~S. Avestimehr, ``Fundamental limits
  of cache-aided interference management,'' \emph{IEEE Transactions on
  Information Theory}, vol.~PP, no.~99, pp. 1--1, 2017.

\bibitem{Tandon}
R.~Tandon and O.~Simeone, ``Cloud-aided wireless networks with edge caching:
  Fundamental latency trade-offs in fog radio access networks,'' in \emph{IEEE
  ISIT}, July 2016, pp. 2029--2033.

\bibitem{SoheilTWC}
S.~Gherekhloo and A.~Sezgin, ``Latency-limited broadcast channel with
  cache-equipped helpers,'' \emph{IEEE Transactions on Wireless
  Communications}, no.~99, 2017.

\bibitem{Vahid17}
A.~Vahid, M.~A. Maddah-Ali, S.~Avestimehr, and Y.~Zhu, ``Binary fading
  interference channel with no {CSIT},'' \emph{IEEE Transactions on Information
  Theory}, vol.~PP, no.~99, pp. 1--1, 2017.

\bibitem{Yigit}
Y.~Ugur, Z.~H. Awan, and A.~Sezgin, ``Cloud radio access networks with coded
  caching,'' in \emph{WSA 2016}, March 2016, pp. 1--5.

\bibitem{Park}
S.~H. Park, O.~Simeone, and S.~Shamai, ``Joint optimization of cloud and edge
  processing for fog radio access networks,'' in \emph{IEEE ISIT}, July 2016.

\bibitem{Tao16}
M.~Tao, E.~Chen, H.~Zhou, and W.~Yu, ``Content-centric sparse multicast
  beamforming for cache-enabled cloud ran,'' \emph{IEEE Transactions on
  Wireless Communications}, vol.~15, no.~9, pp. 6118--6131, Sept 2016.

\bibitem{Chen16}
D.~Chen, S.~Schedler, and V.~Kuehn, ``Backhaul traffic balancing and dynamic
  content-centric clustering for the downlink of fog radio access network,'' in
  \emph{2016 IEEE 17th International Workshop on Signal Processing Advances in
  Wireless Communications (SPAWC)}, July 2016, pp. 1--5.

\bibitem{Avestimehr}
A.~S. Avestimehr, S.~N. Diggavi, and D.~N.~C. Tse, ``Wireless network
  information flow: A deterministic approach,'' \emph{IEEE Transactions on
  Information Theory}, vol.~57, no.~4, pp. 1872--1905, April 2011.

\bibitem{Anas}
A.~Chaaban and A.~Sezgin, ``{On the Generalized Degrees of Freedom of the
  Gaussian Interference Relay Channel},'' \emph{IEEE Transactions on
  Information Theory}, vol.~58, no.~7, pp. 4432--4461, July 2012.

\bibitem{Cover_2006}
T.~M. Cover and J.~A. Thomas, \emph{Elements of Information Theory}.\hskip 1em
  plus 0.5em minus 0.4em\relax Wiley, 2006.

\bibitem{Cover79}
T.~Cover and A.~E. Gamal, ``Capacity theorems for the relay channel,''
  \emph{IEEE Transactions on Information Theory}, vol.~25, no.~5, pp. 572--584,
  Sep 1979.

\bibitem{Willems82}
F.~M. Willems, ``Informationtheoretical results for the discrete memoryless
  multiple access channel,'' Ph.D. dissertation, Ph.D. dissertation, Katholieke
  Universiteit Leuven, Belgium, 1982.

\end{thebibliography}
